\documentclass[aip]{revtex4-1}
\usepackage[utf8]{inputenc}
\usepackage{float}
\usepackage{amsfonts}
\usepackage{amsmath}
\usepackage{amssymb}
\usepackage{caption}
\usepackage{subcaption}
\usepackage{ dsfont }
\usepackage{braket}
\usepackage{xcolor}

\usepackage{natbib}
\usepackage{graphicx}

\usepackage{graphicx}
\usepackage{dcolumn}
\usepackage{bm}

\usepackage[utf8]{inputenc}
\usepackage[T1]{fontenc}
\usepackage{mathptmx}
\usepackage{etoolbox}

\begin{document}
\title{Quantum double structure in cold atom superfluids}
\author{Emil Génetay Johansen}
\author{Chris Vale}
\author{Tapio Simula}
\affiliation{$^1$Optical Sciences Centre, Swinburne University of Technology, Melbourne 3122, Australia}

\begin{abstract}

The theory of topological quantum computation is underpinned by two important classes of models. One is based on non-abelian Chern--Simons theory, which yields the so-called $\rm{SU}(2)_k$ anyon models that often appear in the context of electrically charged quantum fluids. The physics of the other is captured by symmetry broken Yang--Mills theory in the absence of a Chern--Simons term, and results in the so-called quantum double models. Extensive resources have been invested into the search for $\rm{SU}(2)_k$ anyon quasi-particles; in particular the so-called Ising anyons ($k=2$) of which Majorana zero modes are believed to be an incarnation. In contrast to the $\rm{SU}(2)_k$ models, quantum doubles have attracted little attention in experiments despite their pivotal role in the theory of error correction. Beyond topological error correcting codes, the appearance of quantum doubles has been limited to contexts primarily within mathematical physics, and as such, they are of seemingly little relevance for the study of experimentally tangible systems. However, recent works suggest that quantum double anyons may be found in spinor Bose--Einstein condensates. In light of this, the core purpose of this article is to provide a self-contained exposition of the quantum double structure, framed in the context of spinor condensates, by constructing explicitly the quantum doubles for various ground state symmetry groups and discuss their experimental realisability. We also derive analytically an equation for the quantum double Clebsch--Gordan coefficients from which the relevant braid matrices can be worked out. Finally, the existence of a particle-vortex duality is exposed and illuminated upon in this context.
\end{abstract}
\maketitle

\tableofcontents
\newpage
\section{Introduction}

As one ventures beyond the realm of classical phases of matter, the classification paradigm due to Landau \cite{landau1937theory} ceases to apply. Such quantum systems often exhibit phenomena of a long-range collective nature connected to a non-trivial underlying topological structure. Consequently, topology may serve as a better fingerprint to categorize such phases. Owing to its inherent relationship to the connectivity of space, the spin-statistics theorem generally breaks down on non-trivial topologies, thus clearing the way for more exotic particle species to emerge that may not be classified as bosons or fermions. Such peculiar particles exhibit fractional statistics and are therefore referred to as \emph{anyons} \cite{arovas1985statistical,wilczek1990fractional,chen1989anyon,leinaas1977theory} (\emph{any} as in \emph{any} statistics). Permutation of anyons will, in contrast to fermions and bosons, generally implement a more complex unitary transformation than a simple change of sign of the wavefunction.

While anyons are very interesting from a fundamental viewpoint, most research on the topic concerns quantum information processing \cite{1982IJTP...21..467F,2010qcqi.book.....N}. A strong interest in anyons was sparked when Kitaev suggested that they may hold the key to the realization of fault tolerant quantum computation \cite{kitaev2009topological}. The prospects for quantum computation are tantalizing, but in order to not succumb to decoherence, the effects of environmental noise must be considered. This is naturally addressed in a quantum computer based on anyons. Owing to the intrinsic properties of topology, qubit states based on anyons possess a natural shield against various unwanted interactions with the environment. Error-immune computers of this kind are known as \emph{topological quantum computers} (TQC) \cite{kitaev2009topological,field2018introduction,pachos2012introduction,nayak2008non,wang2010topological}. 

Anyons may be realised in planar fermionic fluids, such as an electron gas in the fractional quantum Hall effect \cite{PhysRevB.61.10267} which effectively are governed by Chern--Simons theory \cite{witten1989quantum,1992PhR...213..179I,1999tald.conf..177D}. The Chern--Simons term is responsible for a type of charge-flux attachment, akin to that in the Aharanov--Bohm experiment \cite{aharonov1959significance,berry1984quantal}, which further results in quasi-particles with fractional statistics. These models known as $\rm{\rm{SU}(2)_k}$ models \cite{2013PhRvB..87w5120G,2020arXiv200810790G} where the parameter $\rm{k}$ is an integer referring to a particular deformed representation of $\rm{SU}(2)$. A strong interest for Majorana quasi-particle zero modes has been developed in recent years since they are believed to realise the $\rm{SU}(2)_2$ Ising anyons \cite{fan2010braid,bombin2010topological,2020arXiv200810790G}. Majorana zero modes have been predicted to emerge in the vortex quasi-particle spectrum of chiral p-wave superfluids such has in $^3$He superfluids or cold atom Fermi gases \cite{2007PhRvL..98a0506T,2001PhRvL..86..268I,PhysRevB.61.10267,PhysRevB.97.104501,Gurarie2007a,Mizushima2008a}. They are also expected to emerge in solid state systems such as in fractional quantum Hall fluids with filling fraction $\nu = 5/2$ \cite{kasahara2018majorana,zuo2016detecting,PhysRevLett.94.166802}, in superconducting-semiconducting nanowires \cite{2019NatCo..10.5128Z,2019arXiv191104512P,2019arXiv190706497B,stanescu2013majorana,Jiang2011a,2018NatRM...3...52L,livanas2019alternative,sato2016majorana}, and in the vortex cores of certain topological superconductors \cite{Sang2022a}.

Moreover, anyons are also believed to emerge in certain bosonic systems, such as Bose--Einstein condensates (BECs) \cite{1995Sci...269..198A}, as a result of spontaneous breaking of the initial continuous gauge symmetry to a discrete residual subgroup. Models based on spontaneously broken gauge symmetries are known as \emph{quantum doubles} \cite{1995hep.th...11201D,kitaev2009topological,gould1993quantum,kassel1995drinfeld,2006JMP....47j3511D}. A quantum double can be viewed as an emergent low-temperature symmetry algebra where the group structure is ``doubled" by combining it with its Fourier dual. Mathematically they constitute, just like the $\rm{SU}(2)_k$ models, examples of so-called \emph{quantum groups} \cite{kassel1995drinfeld,Drinfeld1988QuantumG}. In particular, quantum group symmetry emerge when the degrees of freedom pertaining to (generalized) electric charges interact with those of (generalized) magnetic fluxes. Consequently, the particle content of an anyon model is labelled by the irreducible representations of the pertinent quantum group. This is in contrast to conventional quantum field theory where the particles are labelled by those of an undeformed group. For the sake of completeness, a detailed exposition of the particular quantum group relevant to this work is provided in Appendix \ref{HA} and is also thoroughly discussed in e.g. \cite{1995hep.th...11201D}.

The quantum double structure has been studied extensively from a mathematical perspective, yet little efforts have been made to reconcile it with real physical systems. It is not well known that low-temperature phases of spinor BECs \cite{2007PhRvL..98j0401S,2010arXiv1001.2072K,2012PhRvA..85e1606L,makela2003topological,hall2016tying,liu2020interlocked,PhysRevLett.100.180403} with discrete residual symmetry are underpinned by a quantum double structure. Hence, this work primarily seeks to introduce the concept of quantum doubles to the cold atoms community and to provide concrete examples demonstrating how the various components of the quantum double structure might materialize in spinor BECs. There are mainly two types of quantum double excitations referred to as \emph{fluxons} and \emph{chargeons}. While the fluxons, as per homotopy theory, correspond to quantized vortices, the physical interpretation of chargeons is not that clear. By illustrating that fluxons and chargeons are in fact Fourier duals of one another, we find that the chargeons may be associated with delocalized waves. In particular, by conducting an analysis of the normal modes, the chargeons appear to be represented as spin rotations and spin waves, also known as \emph{magnons} \cite{1956PhRv..102.1217D,1940PhRv...58.1098H,RevModPhys.30.1,2009PhRvB..80b4420B}. This agrees with results obtained from numerical simulations carried out in \cite{mawson2018braiding} where spin waves were observed as remnants after fusion events involving non-abelian vortices. The fact that spinor BECs can be routinely produced in the laboratories\cite{PhysRevA.64.053602,PhysRevLett.81.5109,2001PhRvA..64b4702R,PhysRevLett.92.040402,2006NJPh....8..152W} brings further justification supporting the study of non-abelian anyons in such systems. Before moving on we wish to highlight that the field theory descriptions of anyons in fermionic and bosonic systems are not always of a Chern--Simons and Yang--Mills type, respectively. Indeed there exist systems that are based on spin-$1/2$ particles which are underpinned by quantum doubles, for example Kitaev's toric code \cite{kitaev2009topological} and honeycomb model \cite{2006AnPhy.321....2K}. Similarly there are systems whose constituent particles are bosons and are described by Chern--Simons theory. A Chern--Simons gauge gravity is an example of the latter \cite{1988NuPhB.311...46W,2023arXiv230516016G}.

\section{Chern-Simons anyons}

Before discussing the quantum double structure in spinor BECs, we describe anyons in fermionic systems of a Chern--Simons type for the purpose of providing a more cohesive perspective on anyons. In addition, the classification scheme for topological defects is described, which is later applied to spinor BECs in Section~\ref{spin01}-\ref{spin2} to derive their topological charge.

\subsection{From Aharonov--Bohm effect to topological quantum field theory}
\label{abtqft}

\begin{figure}[htp]
    \centering
    \includegraphics[width = 0.25 \linewidth]{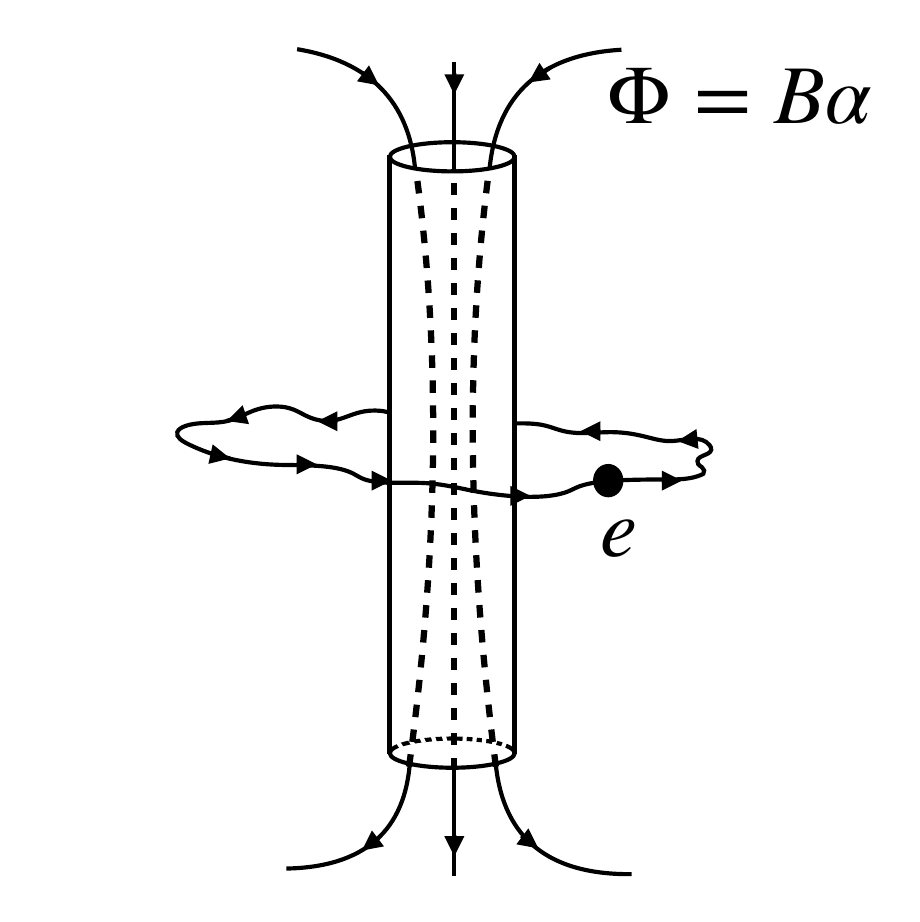}
    \caption{The Aharonov--Bohm experiment. An electron $e$ is encircling a magnetic flux tube $\Phi$, where B stands for magnetic field strength and $\alpha$ is the cross-sectional area.}
    \label{ab}
\end{figure}

Let us consider the Aharonov--Bohm set-up \cite{aharonov1959significance}. In this experiment, an electron encircles an (effectively) infinitely long tube of magnetic flux, see Fig. \ref{ab}, that is piercing through three-dimensional space. As the electron traverses around the flux-tube, the wave-function accumulates a complex phase due to the non-zero vector (gauge) potential originating from the flux.

Letting $A^{\mu}=(\phi/c,{\mathbf{A}})$ denote the electromagnetic 4-potential, where $\phi$ is the electrostatic potential, $c$ is the speed of light, and ${\mathbf{A}}$ is the magnetic vector potential, then the electric and magnetic fields, respectively, can be computed as
\begin{equation}
  {\mathbf{E}} = -\nabla \phi
  \hspace{1cm} {\rm and}\hspace{1cm}
  {\mathbf{B}} = \nabla \times {\mathbf{A}}.
\end{equation}

\begin{figure}[htp!]
    \centering
    \includegraphics[width = \textwidth]{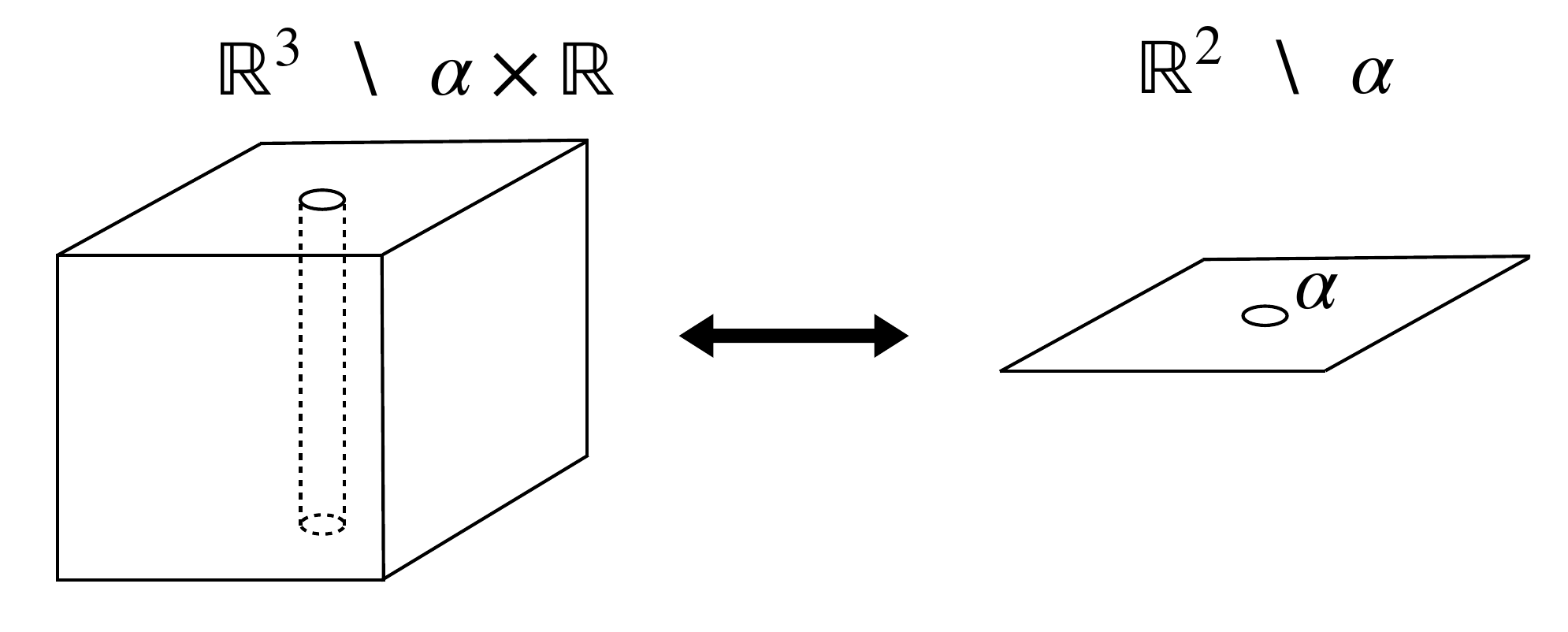}
    \caption{Topological equivalence of the spaces $\mathds{R}^3 \backslash \alpha \times \mathds{R}$ and $\mathds{R}^2 \backslash \alpha$: the space $\mathds{R}^3 \backslash \alpha \times \mathds{R}$ may be continuously deformed vertically to the plane $\mathds{R}^2 \backslash \alpha$.}
    \label{puncture}
\end{figure}

In (3+1)-dimensions the electromagnetic action is given by
\begin{equation}
    S[A^{\mu}] = \frac{1}{2} \int{d^4 x ({\mathbf{E}}^2 - {\mathbf{B}}^2 + {\mathbf{A}} \cdot {\mathbf{J}} + \frac{\phi \rho}{c}})
\end{equation}
where $\rho$ is the electric charge density and ${\mathbf{J}} = \rho \boldsymbol{v}$ is the current density. Variation of this action yields the sourceful Maxwell's equations of electrodynamics. The phase acquired by the electron as it encircles the flux may be computed using Feynman's path integral picture. From the electrons perspective, all fields in the action are zero except the vector potential so the only contribution to the phase is given by 
\begin{equation}
    e^{i \int{d^4x S[A^{\mu}]}} = e^{\frac{i}{\hbar c}\int{d^4x {\mathbf{A}}\cdot {\mathbf{J}}}} = e^{\frac{i e}{\hbar c}\oint_{\partial {\mathcal A}}{{\mathbf{A}} \cdot {d{\bf l}}}}.
\end{equation}
Applying Stokes' theorem the contour integral may be converted into a surface integral enclosed by the loop
\begin{equation}
\gamma=
    \frac{e}{\hbar c}\oint_{\partial {\mathcal A}}{{\mathbf{A}} \cdot {d{\bf l}}}_{\gamma} = \frac{e}{\hbar c} \iint_{\mathcal A} {\mathbf{B}} \cdot {d{\bf S}} =  \frac{e}{\hbar c} \iint_{\alpha} {\mathbf{B}} \cdot {dS} = \frac{e}{\hbar c} \Phi = \frac{e \alpha B}{\hbar c},
    \label{fluxtot}
\end{equation}
where ${\mathcal A}$ is the surface enclosed by the boundary $\partial {\mathcal A}$ and $\Phi$ is the magnetic flux through the cross section $\alpha$. This phase is an example of a \textit{Berry's phase} \cite{berry1984quantal}. Note that the contour $\partial {\mathcal A}$ is arbitrary so the only thing of relevance is the flux through the enclosed surface and not the particular shape of the contour. The value of the path integral only depends on whether the flux was enclosed by the loop or not. Note that the existence of the gauge field ${\bf A}$ can be attributed to the invariance under the gauge transformation
\begin{equation}
    {\mathbf{A}} \longrightarrow {\mathbf{A}} + \nabla \theta(t, {\bf r}),
    \label{gauge}
\end{equation}
since we can always add a non-zero field $\nabla \theta(t,{\bf r})$ in the exterior to the flux tube as ${\bf{B} = \nabla \times {\bf{A}}}=0$, for all functions $\theta (t,{\bf r})$. Fundamentally, what distinguishes the two scenarios when the flux tube is, and when it is not, encircled by the electron is the topology of the parameter space. Since the flux tube is effectively infinite, it will pierce through the entire space, thus rendering the space topologically non-trivial. From a mathematical perspective, the classification of the space is transitioning from being simply to multiply connected, see Fig. \ref{puncture}. 

Geometrically, the emerging $\rm{U}(1)$ transformation is defining a gauge connection in parameter space, according to which the state transforms as it is parallel transported along some path. In the language of differential geometry, this type of intrinsic tangent space with group structure constitutes a principal fiber bundle \cite{singer1982differential,trautman1980fiber}, and as we shall see, we may generalize this concept further to more complex spaces, which will allow for the implementation of topological quantum computation protocols. The magnetic field can also be regarded as the curvature of parameter space since the field strength tensor is the analogue of the Riemannian curvature tensor in the context of gauge theory. A non-zero magnetic field is thus inducing a non-trivial holonomy as the particle is parallel transported around the flux tube. Now, let us descend to the two dimensional plane. One of the key differences between two and three spatial dimensions is that the curl of a vector is a scalar, thus implying that the magnetic field $B^i = \epsilon^{i j k} \partial_j A_k$ is a scalar in the plane. This transition opens new doors since the most general Maxwell Lagrangian now accept new terms. In particular, Poincaré and gauge symmetry is respected by the additional term
\begin{equation}
    S_{\rm CS} = \frac{k}{4 \pi} \int d^3x \epsilon^{\mu \nu \rho} a_{\mu}\partial_{\nu} a_{\rho},
    \label{Lcs}
\end{equation}
where $k$ is a coupling constant. This Chern--Simons term has played a pivotal role in the development of our understanding of two-dimensional quantum physics, and in particular, the quantum Hall effects. Here $a^{\mu}$ is an emergent statistical gauge field originating from the punctures caused by the flux lines (vortices). Just like the flux tube penetrates the space in the Aharonov--Bohm experiment, topological excitations may puncture the system in two dimensions, which gives rise to an analogous phenomenon, see Fig. \ref{puncture}. Mathematically, what this entails is that the space no longer is simply connected since any closed loop around the puncture are non-contractible. In three dimensions, however, the loop can get wrapped around the hole through the third dimension, after which it can be contracted to a single point. In terms of exchange operators, we must therefore enforce that two exchanges is equal to doing nothing, i.e. $\hat{P}^2 = I$, which implies that the eigenvalue of $\hat{P}$ can only be $\pm 1$ which correspond to fermions $(-1)$ and bosons $(+1)$. But when the loop is non-contractible, the eigenvalue of $\hat{P}$ may be \textit{any} complex phase $e^{i \theta}$, which is why such quasi-particle excitations are known as anyons. Note that only the overall topology is relevant here and not the particular shape of the loop, which is reflected in Eq.~\eqref{Lcs} by the fact that it is independent of the metric. This is one of the defining properties of \textit{topological field theories} \cite{witten1988topological,atiyah1988topological,witten1989quantum}. 

\subsubsection{Charge-flux attachment}
\label{cf-attachment}
Chern--Simons theory alone only generates trivial equations of motion, but if we, for instance, couple the statistical gauge field $a_{\mu}$ to a matter current $J^{\mu}$ the total Lagrangian is given by
\begin{equation}
    S_{\rm CS+matter} = \frac{k}{4 \pi} \int d^3x (\epsilon^{\mu \nu \rho} a_{\mu}\partial_{\nu} a_{\rho} + a_{\mu}J^{\mu}).
    \label{Lcsm}
\end{equation}
Working out the Euler--Lagrange equations for this action yields a very interesting result \cite{dunne1999aspects}. The Gauss's law 
\begin{equation}
    \nabla \cdot {\mathbf{E}} + k B = \rho
\end{equation}
is modified to include an additional term. The physical interpretation of this equation is that in two dimensions, the charge distribution $\rho$ does not only give rise to a diverging electric field, but to a scalar magnetic field as well. If we further integrate over the charge distribution and convert the $\nabla \cdot {\mathbf{E}}$ integral to a surface integral we see that it vanishes at large distance scales since the ${\mathbf{E}}$ decays as $r^{-1}$. Consequently, we are left with
\begin{equation}
    \int d^2 x \rho = k \int d^2 x B,
\end{equation}
or equivalently
\begin{equation}
    Q = k \Phi,
    \label{cfattachment}
\end{equation}
where $Q$ is the total charge and $\Phi$ the magnetic flux. Thus every charged particle carries magnetic flux and one can not have one without the other. This is exactly what happens in quantum Hall fluids \cite{1987RvMP...59..781Y,2003PhT....56h..38A}. As the electrons are subjected to increasingly strong external magnetic fields they become trapped to tight cyclotron orbits around the magnetic flux lines and the Chern--Simons term begins to dominate the physics of the system. Note the similarity to the Aharanov--Bohm experiment where the electron couples to the flux tube via the gauge potential and acquires a phase that depends on the magnetic flux. In the quantum Hall systems the electrons couple to the flux quanta via the Chern--Simons gauge field, and similarly, the phase acquired by the wave function depends on the filling fraction $\nu = N_e / N_{\phi}$, which is the ratio of the number of electrons $N_e$ to the number of flux quanta $N_{\Phi}$. Different values of $\nu$ correspond to different anyon models determined by the Chern--Simons coupling constant $k$. For instance, the $\nu = 5/2$ system is expected to host the long sought-after Ising anyons, which we shall discuss further in Section \ref{isingtqc}. Moreover, systems with $\nu = 12/5$ are predicted to be inhabited by the so-called Fibonacci anyons \cite{2017PhRvB..95k5136M,2014PhRvL.113w6804V,xia2004electron,komijani2019kondo}, which set the golden standard for topological quantum computation. In Section \ref{DGT} we shall see that discrete gauge theories give rise to a similar phenomenon described above without the additional Chern--Simons term added. The mathematical structures describing such models are known as quantum doubles, which is the main focus of our discussion.

\subsection{Symmetry classification of topological particles}
\label{symclass}
\begin{figure}[htp!]
    \centering
    \includegraphics[width = \textwidth]{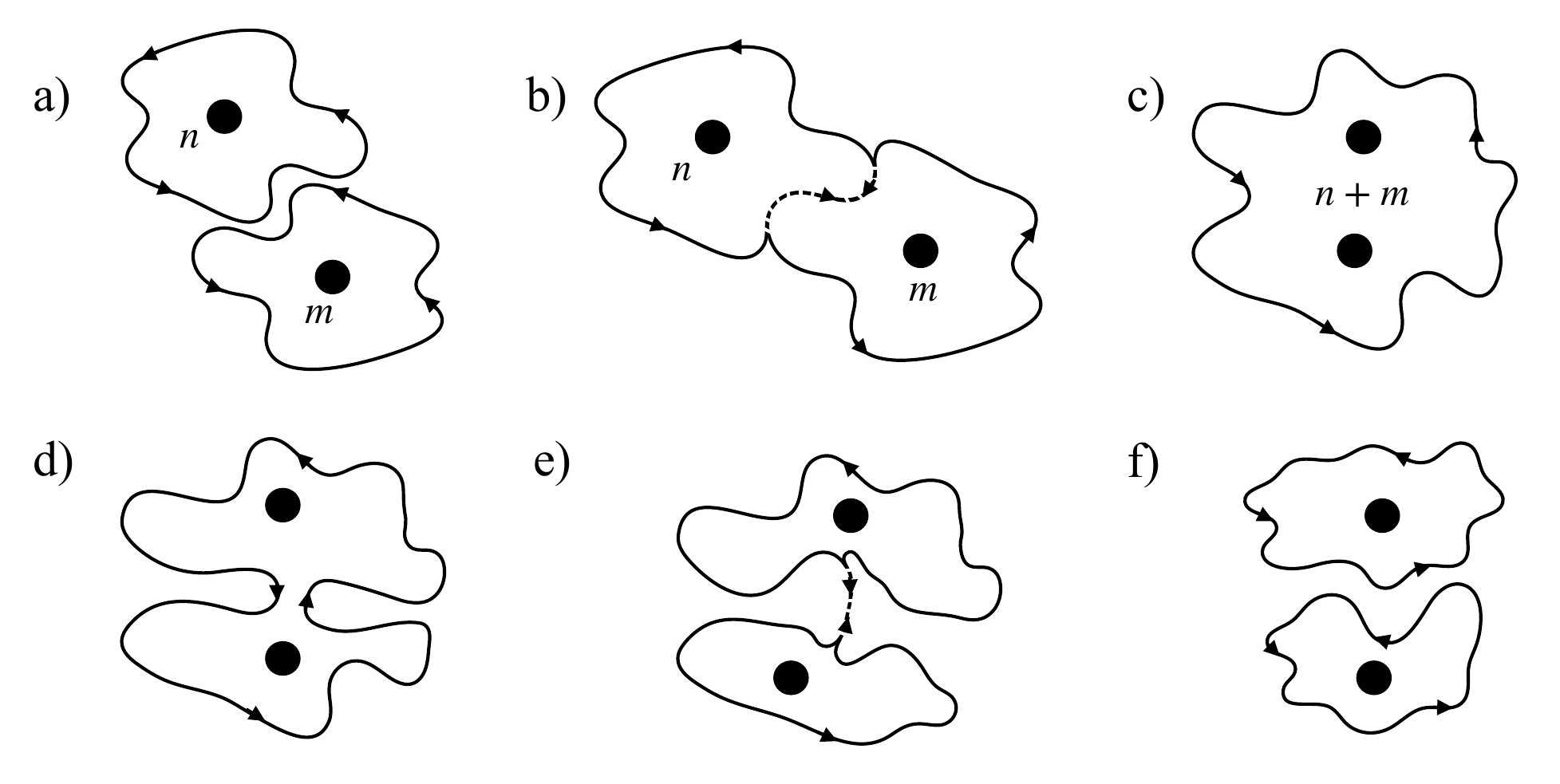}
    \caption{a)-c) Connection of two loops corresponding to winding $n$ and $m$ resulting in a single loop corresponding to winding $n+m$. d)-f) The reversed process of a)-c) where a single loop corresponding to winding $n+m$ splits into two distinct loops corresponding to windings $n$ and $m$, respectively.}
    \label{winding}
\end{figure}

The emergent charge-flux attachment concept discussed in the previous section is unique to (2+1)-dimensions. Imagine then that there are $n$ flux tubes enclosed by a loop. 
The loop may be continuously deformed such that it can be decomposed into $n$ distinct loops, each of which encircles one tube of magnetic flux, see Fig. \ref{winding} d)-f). This means that the total flux $\alpha B$ calculated in Eq.~\eqref{fluxtot} now is $n\Phi$. The integer $n$ counts the phase winding, which is why it is known as the \emph{winding number}, and as illustrated in Fig.~\ref{winding}, it must be a topological invariant. Similarly, we illustrate in Fig.~\ref{winding} a)-c) how loops can be attached, which allows for the combined flux to be deduced. 

\begin{figure}[htp!]
    \centering
    \includegraphics[width =\linewidth]{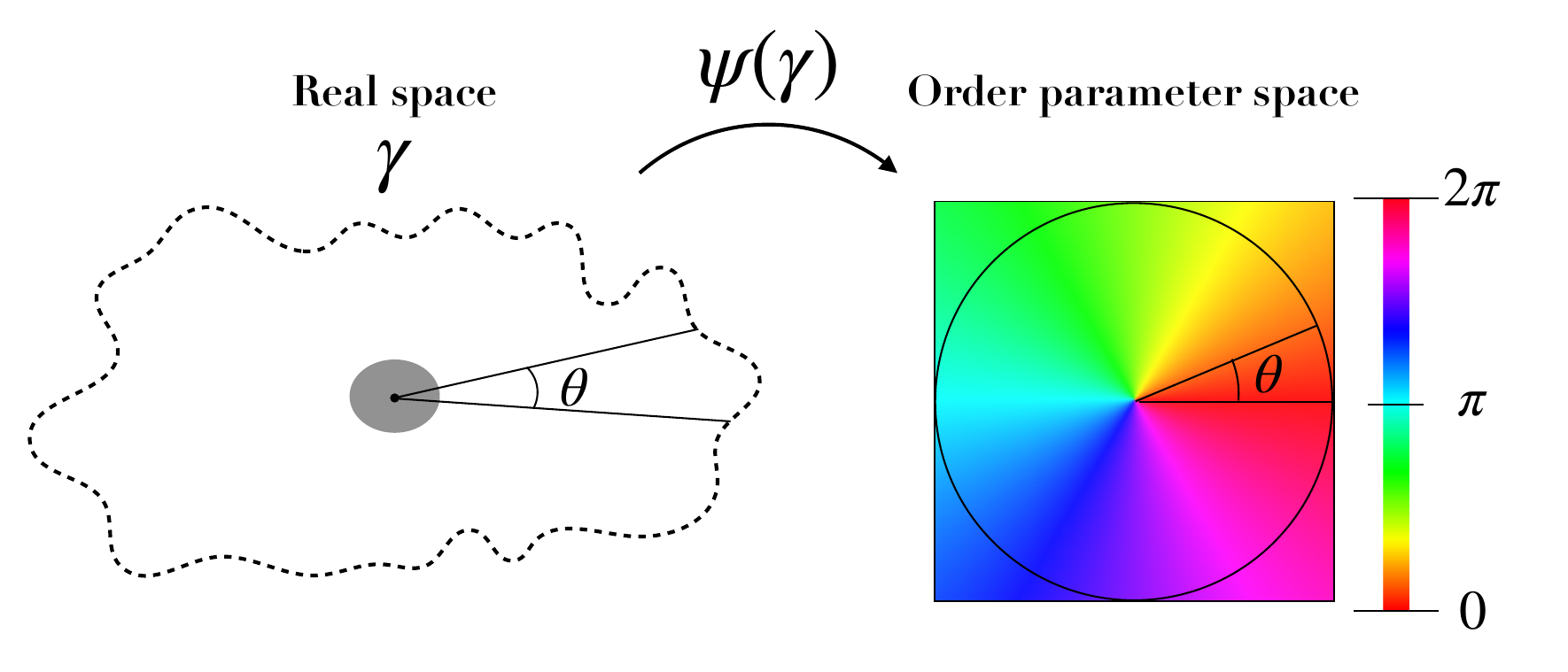}
    \caption{Phase winding of the wave function as a topological quasi-particle excitation with flux is encircled. The wave function $\psi(r)$ accumulates a $\rm{U}(1)$ phase $e^{i\theta}$ (right) as the topological defect is encircled (left), as a function of the path $\gamma$.}
    \label{phase}
\end{figure}

The properties of flux tubes in a $\rm{U}(1)$ theory reflect that of the group of integers under addition $(\mathds{Z},+)$ since the winding numbers are simply added together as the flux tubes are encircled. Hence, we shall use this group to classify the fluxons. The generic algebraic tool for classifying topological excitations is known as \textit{homotopy theory} \cite{hu1959homotopy,whitehead2012elements}. In this framework the classification is carried out by calculating the eigenvalues of loop operators corresponding to the encirclement $\gamma$ of excitations. These loop operators are known as \emph{Wilson loops} $W_{\gamma}$ and are defined as \cite{PhysRevD.10.2445}
\begin{equation}\label{wilson}
    W_{\gamma} = {\rm Tr}[\mathcal{P} \oint_{\gamma} A_{\mu} dx^{\mu}],
\end{equation}
where $A_{\mu}$ is the gauge field and $\mathcal{P}$ denotes path ordering. This is precisely equivalent to Eq.~\eqref{fluxtot} where we computed the phase in the Aharonov--Bohm experiment. Thus, Wilson loops represent the observables of a topological quantum field theory, since they measure the topological charge. Let us restrict to point-like excitations of a planar $\rm{U}(1)$ theory and employ the first order homotopy group $\pi_1$, see Appendix~\ref{FG} for definition, also known as the fundamental group, over the $\rm{U}(1) \simeq S^1$ manifold.This yields $\pi_1 (\rm{U}(1)) \simeq \pi_1 (S^1) \simeq \mathds{Z}$, since one can only encircle a circle $S^1$ an integer number of times, see Fig.~\ref{phase}. Considering the total winding of two fluxons $|n\rangle$ and $|m\rangle$, where $n,m \in \mathds{Z}$, their joint state is given by
\begin{equation}
    |n\rangle \otimes |m\rangle = |n+m\rangle.
\end{equation}
The group of integers possesses abelian structure, which is why the flux-charge composites considered here are abelian anyons. However, we may also consider more general gauge theories in which the topology of the gauge group is more complicated and for a gauge group $G$, the fundamental group $\pi_1(G)$ need not be isomorphic to the group of integers. Contrary to bosons and fermions, if two flux-charge composites in such a theory were to be exchanged twice, the wave function may not return to its original state as an arbitrary phase factor $\hat{P}^2 = e^{i \theta}$ may be acquired.

The wave function of such "fractional" particles therefore transforms as
\begin{equation}
    \Psi ({\bf r},t) \longrightarrow \psi({\bf r},t) e^{i \theta}.
    \label{U1}
\end{equation}
The commutative property of the phase $\theta$ can be traced back to the Chern--Simons action in Eq.~\eqref{Lcsm}, which is an abelian gauge theory. Therefore, more interesting theories can be constructed with a ground state degeneracy by promoting the Chern--Simons field $a_{\mu}$ of the $\rm{U}(1)$ theory to a $\rm{SU}(N)$ gauge field similar to the one in Yang--Mills theory, which gives rise to a non-abelian theory. Non-abelian Chern--Simons theories result from the action \cite{witten1989quantum,1992PhR...213..179I,1999tald.conf..177D}
\begin{equation}
    \label{CSNA}
    S_{\rm CS,non-abelian} = \frac{k}{4 \pi} \int d^3x\epsilon^{\mu \nu \rho} \frac{2}{3} a_{\mu}[a_{\nu},a_{\rho}].
\end{equation}
Contrary to the $\rm{U}(1)$ theory, the state vector is now allowed to rotate in a higher dimensional space when being parallel transported, which gives rise to a matrix valued phase factor $e^{i \Theta_{ij}}$. Just like the integer flux-charge composites are classified by the group of integers $(\mathds{Z},+)$ under addition, non-abelian anyons can be classified by a non-abelian homotopy group $\pi_1 (\rm{SU}(N))$ over $\rm{SU}(N)$. The global state of two such anyons can thus be expressed as
\begin{equation}
    |g_i\rangle \otimes |g_j\rangle = |g_i \circ g_j\rangle,
\end{equation}
where $g_i , g_j \in \rm{SU}(2)$ are the group elements that are labelling the two anyons and $\circ$ denotes the group operation. Such a vector space is generally decomposible into blocks since $G \otimes G$ might have multiple irreducible representations. There is one particle species per irreducible representation so the outcome when combining, or fusing, two non-abelian anyons is generally indefinite. In this section we have primarily focused on Chern--Simons anyons but the concept of homotopy can be extended to arbitrary anyon models, such as the spinor BEC-based quantum doubles that will be the subject of discussion in the Section~\ref{afb}. Next we shall consider a specific instance of non-abelian Chern--Simons theory, which underpins the so-called Ising anyon model.

\subsection{Ising anyon $\rm{SU}(2)_2$ topological quantum computer}
\label{isingtqc}
\begin{figure}
    \centering
    \includegraphics[width=\textwidth]{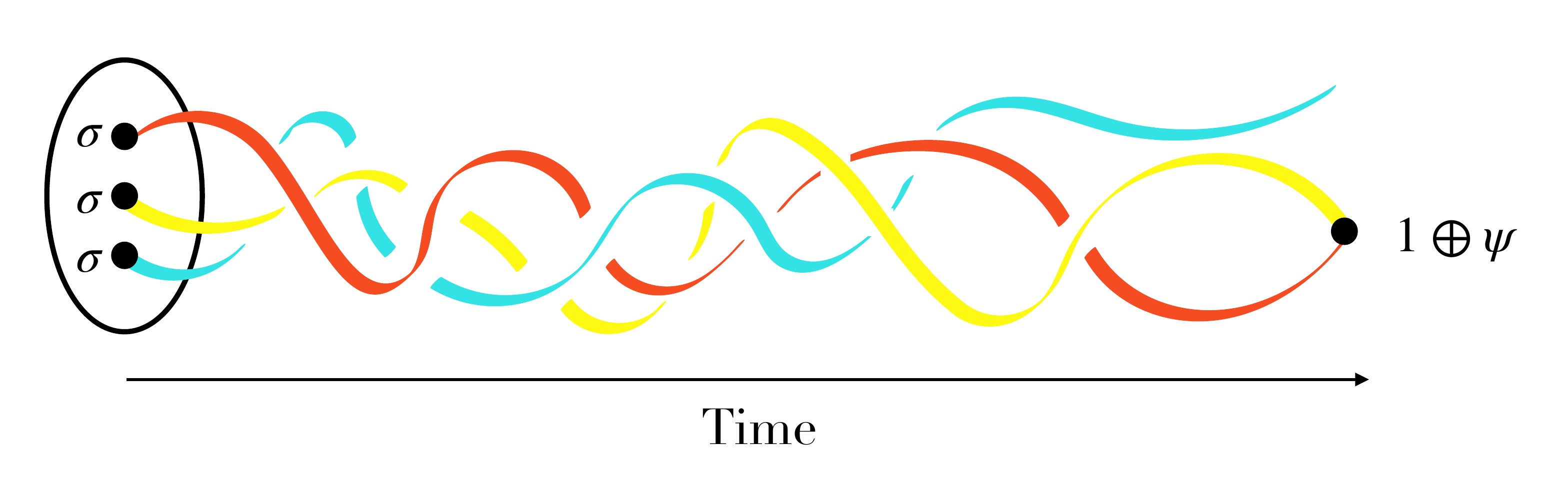}
    \caption{Braiding three Ising anyons. As the positions of the anyons are permuted in the plane their world lines form a braid in space-time. When the braiding is performed two of the anyons are fused to measure the state of the topological qubit.}
    \label{ising}
\end{figure}

A particular interest for quasi-particles with non-abelian exchange symmetry is steadily increasing in the quantum information community. This is due to the belief that they potentially can be utilized as quantum memories, which are well guarded from environmental decoherence by topological equivalence. Here the indefinite fusion products of non-abelian anyons constitute the qubit states in which information can be encoded. The simplest, and perhaps also the most experimentally viable model, is based on Majorana fermion zero modes, which are believed to realize so-called Ising anyons \cite{fan2010braid}. Amplitudes of processes involving Ising anyon are governed by a particular instance of non-abelian Chern--Simons theory. Specifically, the $\rm{\rm{SU}(2)_k}$ theory with $k=2$, where the integer coupling constant $k$ in Eq.~\eqref{CSNA} is also known as the level of the theory. Each level $k$ give rise to a unique theory pertaining to a particular deformation of the full $\rm{\rm{SU}(2)}$ representation space. The integer $k$ plays the role of a deformation parameter so theories labelled by different $k$ values host different types of anyons with different topological charges. Ising anyons in particular have the topological charges $\mathds{1, \sigma}$ and $\psi$ (anyons), which obey the following fusion rules
\begin{align}
    \sigma \otimes \sigma = \mathds{1} \oplus \psi\\
    \sigma \otimes \psi = \sigma\\
    \psi \otimes \psi = \mathds{1},
\end{align}
where $\mathds{1}$ denotes the vacuum, $\sigma$ is known as the Ising anyon and $\psi$ as the $\psi$-anyon. Level $k=2$ non-abelian Chern--Simons theory is believed to bear relevance to fractional quantum Hall fluids with filling fraction $\nu = 5/2$, which is why Majorana fermion quasi-particles have been predicted to emerge in such systems \cite{kasahara2018majorana,zuo2016detecting,PhysRevLett.94.166802}. They are also believed to be found in topological superconductor-semiconductor nanowires \cite{2019NatCo..10.5128Z,2019arXiv191104512P,2019arXiv190706497B,stanescu2013majorana,Jiang2011a,2018NatRM...3...52L,livanas2019alternative,sato2016majorana} and chiral $p$-wave paired Fermi superfluids \cite{2007PhRvL..98a0506T,2001PhRvL..86..268I,2007PhRvL..98a0506T,PhysRevB.97.104501,Gurarie2007aa,Mizushima2008a}. In a topological quantum computer based on Ising anyons the fusion product $\sigma \otimes \sigma = \mathds{1} \oplus \psi$ constitutes the qubit where the two possible measurement outcomes correspond to $\mathds{1}$ (the vacuum) and $\psi$ (the $\psi$-anyon). In order to process the information encoded in the fusion product, and thus to carry out computation, logic gates must be implemented. The special trait possessed by non-abelian anyons that distinguish them from bosons and fermions is that by interchanging them, an implementation of a braid group representation is realized, as opposed to that of an abelian permutation group. These transformations may then be used as logic gates. In Fig.~\ref{ising} a computational process is illustrated where a set of Ising anyons are created from the vacuum and then braided according to the quantum circuit one wishes to implement. Finally, the anyons are brought together, or fused, to read out the outcome of the computation. Since any deformation of the braid that does not change the topology leave the transformation invariant, the braids are protected from external noise that otherwise would cause decoherence and the computation to fail. 
In a one-qubit system there are three anyons participating in the computation and thus two braid matrices implementing the braiding of the first with the second anyon, and the second with the third, respectively. The braid matrices in the Ising anyon model are 
\begin{equation}
    \sigma_1 = R =e^{i\frac{\pi}{8}}\begin{pmatrix}
    -1 & 0\\
    0 & -i
    \end{pmatrix} \;\;\;\;{\rm and}\;\; \  \
    \sigma_2 = F RF^{-1} = \frac{e^{-i \frac{4 \pi}{8}}}{\sqrt{2}} \begin{pmatrix}
    1 & i\\
    i & 1
\end{pmatrix}.
\end{equation}

However, this model is computationally non-universal, meaning that the braid group generated from $\sigma_1$ and $\sigma_2$ is finite and is thus only capable of implementing a finite set of unitary rotations of the Bloch sphere. In order to be able to implement every logic gate we need a braid group of infinite order that is generating a topologically dense cover in $\rm{\rm{SU}(2)}$, so that any rotation of the Bloch sphere can be realized. Nevertheless, universality can be achieved by supplementing the set of braids with an additional conventional operation at the cost of sacrificing the complete topological protection \cite{2015PhRvA..92a2301L,2003PhRvA..67b2315M}. Such a quantum computer is therefore to be regarded as a hybrid as arbitrary computational processes will still rely on error correcting protocols to some extent. Next, we shall redirect our attention towards bosonic systems and their quantum double structure in phases with spontaneously broken symmetry.

\section{Quantum double anyons}\label{afb}

In Section \ref{abtqft} we described how two-dimensional electrodynamics with interactions governed by continuous $\rm{U}(1)$ gauge symmetry give rise to an Aharonov--Bohm like phase. The emergence of this phase is due to the punctures caused by the vortex-like excitations (fluxons), which thus play the role of flux tubes. We also saw that a modified Gauss's law emerged as a consequence of the additional Chern--Simons term, which lead to a flux-charge attachment, and thus to fractional statistics. Topological physics can also be realized in the absence of such a term via the Higg's mechanism, since the gauge bosons of the theory may acquire mass thus resulting in a topological field theory, as the non-topological interactions are rendered short range.

\subsection{The general picture}

\subsubsection{Discrete gauge theory in (2+1)-D}
\label{DGT}

Here we shall consider bosonic condensates which have undergone spontaneous symmetry breaking from a continuous symmetry group $G$ to a finite discrete subgroup $H$, where each subgroup $H$ serves as a signature of a low-temperature phase. We refer to this group as the isotropy group of the condensate as it does not alter the physical properties of the system. However, since all cosets of the isotropy group possess the same structure, the full order parameter manifold $\mathcal{M}$ is defined by the quotient space $\mathcal{M} = G/H$. Just like in Section~\ref{symclass}, where it was described how the fluxons can be classified by means of the first homotopy group over the order parameter manifold $\rm{U}(1)$, here the excitations are classified according to $\pi_1 (G/H)$. But if $H$ is discrete and $G/H$ is simply connected, they are connected by an isomorphism so $\pi_1 (G/H) \simeq H$, which entails that all information about the excitations is encoded in $H$, so that the elements of $H$ may be used for labeling the excitations. However, the exact classification is slightly more subtle since charge conservation must be respected. That is, if we regard two fluxons $h_i,h_j \in H$ and bring the $h_i$ fluxon around the $h_j$ one, it returns in a transformed state $h_i'$. Flux conservation thus enforces the condition $h_i' h_j = h_j h_i$, which further entails that the state must transform under conjugation $h_i' = h_j h_i h_j^{-1}$. Conjugation defines an equivalence relation on the group so the fluxons should therefore be classified according to the conjugacy class partitioning, which consequently implies that all particles labelled by an element within the same conjugacy class are indistinguishable. This phenomenon is known as \emph{flux metamorphosis} \cite{bais1980flux, 1992PhLB..280...63A}, and if the residual group $H$ is non-commutative, there are generally more than one element in each conjugacy class which together span a multi-dimensional Hilbert space inhabited by a non-abelian fluxon.
 
 \subsubsection{Higg's mechanism and symmetry breaking}
 
 Following Refs.~\cite{bais1980flux,1995hep.th...11201D}, we may model the dynamics of a planar bosonic system supporting topological excitations by means of a (2+1)-dimensional Yang--Mills--Higgs Lagrangian 
\begin{equation}
    S_{\rm YMH} = \int d^4x \mathcal{L}_{\rm YMH} = \int d^3x [-\frac{1}{4} F^{\mu \nu} F_{\mu \nu} + (\partial^{\mu} \Phi_{\mu})^* \partial^{\mu} \Phi_{\mu}-V(|\Phi|)],
    \label{action}
\end{equation}
where $F_{\mu \nu} = \partial_{\mu}A_{\nu}-\partial_{\nu}A_{\mu} + i [A_{\mu},A_{\nu}]$ is the Yang--Mills field strength tensor with $A_{\mu}$ being the vector potential and $\Phi$ is the Higgs field. As the temperature is lowered below the critical value the system undergoes symmetry breaking and the Higgs potential takes on a vacuum expectation value that is only invariant under the residual subgroup $H \subset G$. Contrary to Chern--Simons electrodynamics without matter, see Section \ref{cf-attachment}), the dynamics governed by this Lagrangian does indeed encompass some interesting physics. The fluxons are introducing non-trivial monodromy to the otherwise flat gauge connection, due to the non-zero components of the curvature tensor $F^{\mu \nu}$ at the location of the fluxon. We shall see later on in Sections \ref{spin01}-\ref{spin2} that fluxons in a spinor BEC correspond to quantized vortices. For a gauge theory with a discrete gauge group, the gauge bosons, as a consequence of the Higg's mechanism, acquire mass in the symmetry breaking process, thus rendering the interactions short ranged. The gauge fields would therefore require huge amounts of energy to get excited, or curved, which is the origin of the topological invariance since the only contribution to the curvature is that coming from the inside of the fluxon. The monodromy, which is measured by the Wilson operator, is therefore the gauge invariant observable of such a theory. In addition to the fluxons, we may introduce matter to the action in Eq.~\eqref{action}, to which we couple the gauge field. This can be achieved by adding a current term $j^{\mu}A_{\mu}^a$ and adopt a covariant derivative via minimal coupling to the gauge field $A_{\mu}^a$. Hence, gauge symmetry enforces the covariant derivative to take on the form
\begin{equation}
    \partial_{\mu} \longrightarrow \mathcal{D}_{\mu} = \partial_{\mu} +ig T_a A_{\mu}^a,
    \label{covariant}
\end{equation}
 where $g$ is the coupling constant and $T_a$ is a generator of $G$. The matter fields transform according to various unitary irreducible representations (UIRs) $\Gamma_i$ of $H$. Similarly to the $\rm{\rm{U}(1)}$ bundle in Fig.~\ref{phase}, we are now dealing with an $H$ bundle where $H$ is attached to each point in space, acting on the matter field according to its UIRs. Such fields are called \emph{chargeons}.

\begin{figure}
    \includegraphics[width=0.7\textwidth]{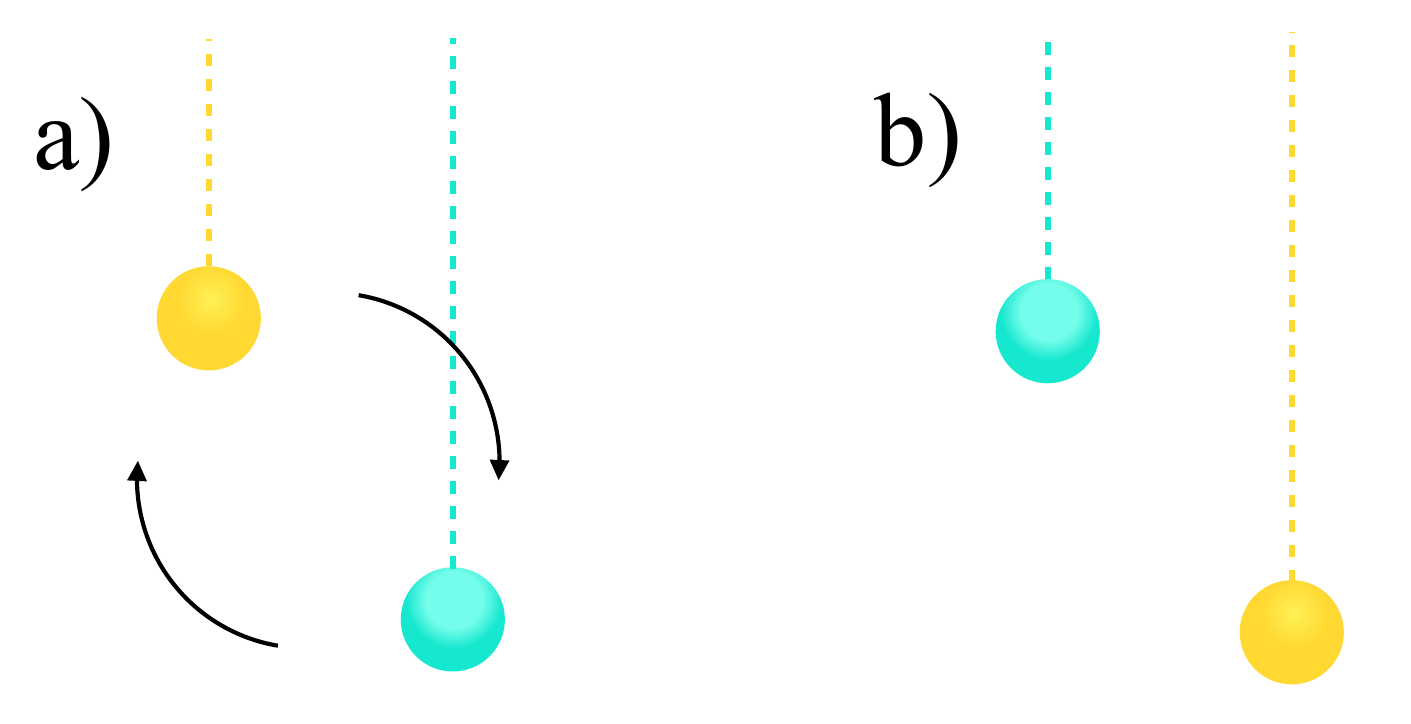}
    \caption{Exchange of two anyons showing the state (a) before and (b) after the exchange, such that the yellow crosses the Dirac string attached to the turquoise one. The turquoise one never crosses the yellow Dirac string so the entire gauge transformation is picked up by the yellow anyon.}
    \label{dirac}
\end{figure}

Inheriting the nomenclature adopted in the Aharonov--Bohm experiment, the fluxons play the role of flux tubes while the matter fields correspond to electric charges, or \textit{chargeons}, which are arranged according $\Gamma_i$. If a chargeon labelled by some UIR $\Gamma_i$ encircles a fluxon labelled by some element $h_j$ the Aharonov--Bohm phase is given by $\Gamma_i(h_j)$. Thus, due to the minimal coupling introduced in the covariant derivative in Eq.~\eqref{covariant}, we have a notion of charge-flux attachment akin to that in Eq.~\eqref{cfattachment}. Such composite objects are called \textit{dyons}. However, the labelling of the dyons is subtle and requires a more careful analysis. Following \cite{1993PhRvD..48.4821L}, we shall consider the Aharonov--Bohm experiment with two dyons where one of them is hidden between two slits in a plate placed in front of a screen. The flux part of the two dyons is labelled by $h_i$ and $h_j$, respectively, so if the first dyon goes through the left slit, the flux part of the dyon between the slits transforms according to $h_j \longrightarrow h_i h_j h_i^{-1}$. However, if the first dyon is goes through the right slit, the second one sitting between the slits is left invariant, i.e. $h_j \longrightarrow h_j$, and consequently, we have asymmetry between the two beams. This is due to the fact that there is a Dirac string connecting the fluxon-antifluxon pairs, which is crossed only when passing the fluxon on one of its sides, see Fig.~\ref{dirac}. 
Owing to this correlation, we only have constructive interference if $h_i$ and $h_j$ commute since the $h_i$ will slip through so that $h_j \longrightarrow h_i h_j h_i^{-1} = h_j h_i h_i^{-1} = h_j$. This implies that the charge attached to $h_j$ must transform under UIRs of elements that commute with $h_j$. Such a set of elements always possesses group structure and is known as the centralizer group $Z(h_j)$ of $h_j$. In conclusion, the adequate labelling of the dyons is given by the conjugacy classes $C_i$ partitioning $H$ (the fluxon part) and the UIRs of the centralizers of the conjugacy classes $\Gamma_j(Z(C_i))$ (the chargeon part). We have now dissected the structure of the quantum double construction, which revealed that fluxon Hilbert spaces are spanned by the elements of the conjugacy class considered. That is, a generic fluxon state is a coherent superposition $\sum_i c_i \ket{h_i} \in \mathds{C}[H]$, where $h_i$ are elements within the same conjugacy class and $c_i$ are coefficients belonging to the field $\mathds{C}$. Moreover, the chargeon Hilbert space is a space of functions $\mathfrak{F}(H)$ on $H$ (the centralizer UIRs), which means that the full quantum double Hilbert space is given by their tensor product $\mathfrak{F}(H) \otimes \mathds{C}[H]$. Before we move on and discuss the quantum double algebra in more detail, we shall provide an example of a simple quantum double model based on the group $\mathds{Z}_N$, which will help us to motivate the various components of the structure.

\subsection{$\mathds{Z}_N$ lattice electrodynamics -- a simple example}
\label{ZN}
\begin{figure}
\centering
\includegraphics[width=\textwidth]{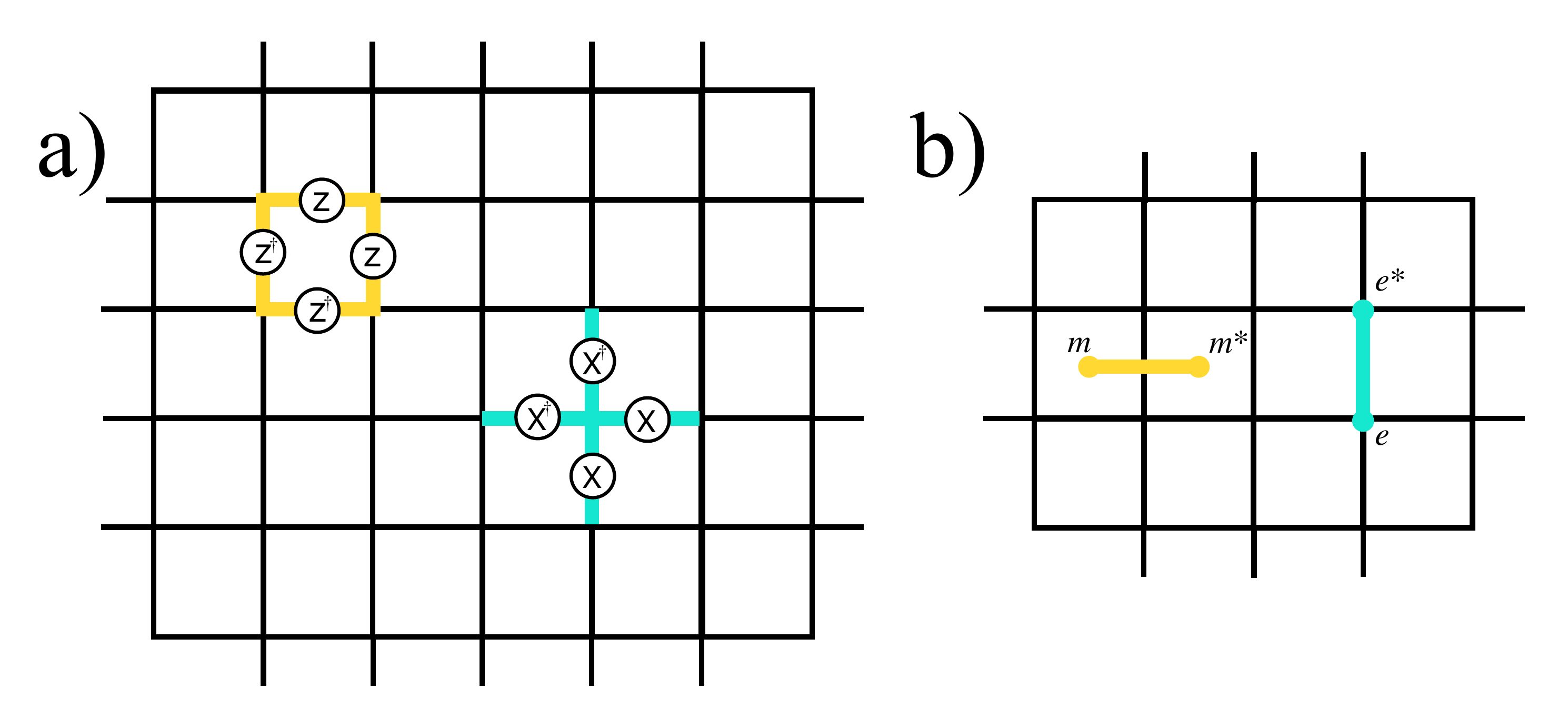}
\caption{a) A $\mathds{Z}_N$ lattice with the plaquette and star operators highlighted in yellow and turquoise, respectively. b) Creating particle anti-particle pairs on the plaquettes (fluxons) and on the sites (chargeon), by the application of an $X$ and a $Z$ operator.}
\label{lattice}
\end{figure}

Let us consider a $2D$ lattice with a continuous group $G$ ($\rm{U}(1)$ for instance) that is spontaneously broken down to $\mathds{Z}_N$ so that there are $N$ degrees of freedom per site. This is an abelian model where the gauge field is inducing transformations in the group $\mathds{Z}_N$ \cite{2012PhRvB..86p1107Y,1992PhRvB..45.5737D,2013JSMTE..10..024B}. In such a theory measurements of charges is still possible through the Aharonov--Bohm effect, despite the electric fields vanish globally as the $\rm{U}(1)$ symmetry is broken. That is, Gauss's law breaks down globally, since the photons have become massive and decay. Thus, the particles are only interacting via topology. The Hamiltonian can be constructed from two distinct types of four-body interactions given by the ``plaquette" operators $A_{\square_i}=Z_k^{\dagger} Z_{k+1} Z_{l}^{\dagger} Z_{l+1}$ and the ``star" operators $B_{+_i}=X_k X_{k+1}^{\dagger} X_{l}^{\dagger} X_{l+1}$, see Fig.~\ref{lattice},
\begin{equation}
    H = -J_{\square}\sum_{\square_i} A_{\square_i} - J_{+}\sum_{+_{j}} B_{+_j},
\end{equation}
where $J_{\square}$ and $J_{+}$ are coupling constants and $Z$ and $X$ are Weyl matrices defined as
\begin{equation}
   Z = 
 \begin{pmatrix}
  1 & 0 & 0 & \cdots & 0 \\
  0 & \omega & 0 & \cdots & 0 \\
  0 & 0 & \omega^2 & \cdots & 0 \\
  \vdots  & \vdots & \vdots & \ddots & \vdots  \\
  0 & 0 & 0 & \cdots & \omega^{N-1}
 \end{pmatrix}
  \ \ {\rm and} \ \  X = 
 \begin{pmatrix}
  0 & 0 & 0 & \cdots & 1 \\
  1 & 0 & 0 & \cdots & 0 \\
  0 & 1 & 0 & \cdots & 0 \\
  \vdots  & \vdots & \vdots & \ddots & \vdots  \\
  0 & 0 & \cdots & 1 & 0
 \end{pmatrix}.
\end{equation}
These obey the following algebraic relations
\begin{equation}
    X^N=Z^N=I,\ \ Z^{\dagger}=Z^{N-1}, \ \ X^{\dagger}=X^{N-1}, \ \ XZ=\omega ZX,
    \label{alg}
\end{equation}
where $\omega = e^{i\frac{2 \pi}{N}}$ is a unitary phase quantized by $N$. Note that these matrices reduce to the standard Pauli $X$ and $Z$ if we set $N=2$.
A nice property of this model is that all terms in the Hamiltonian commute since there are always two overlapping $X$ and $Z$ operators for the adjacent plaquettes and stars, and all other terms commute trivially since they do not overlap and thus occupy different blocks in their matrix representations. Consequently, all terms simultaneously diagonalize the Hamiltonian, which further entails that the ground state $|\Psi_0\rangle$ corresponds to the maximum eigenvalue of all plaquette and star operators, i.e. $A_{\square} |\Psi_0\rangle = |\Psi_0\rangle$ and $B_{+} |\Psi_0\rangle = |\Psi_0\rangle$. The excitations are thus created by violating these conditions. For instance, we can create a particle anti-particle pair on two adjacent plaquettes by applying an $X$ operator on the edge they share since $A_{\square} (X^k |\Psi_0 \rangle) = \omega^{\pm k} X^k (A_{\square} |\Psi_0 \rangle) = \omega^{\pm k} (X^k |\Psi_0 \rangle)$ due to the algebra defined in Eq.~\eqref{alg}, where $\pm$ refers to the eigenvalues corresponding to the two plaquattes, $+$ for the particle and $-$ for the anti-particle. Similarly, we can create pairs at the center of the stars as $B_{+} (Z^l |\Psi_0 \rangle) =  \omega^{\pm l} (Z |\Psi_0 \rangle)$. Note that if we consider an arbitrary state $|l \rangle$, the $X$ and $Z$ operators act according to
\begin{equation}
   X^p |q\rangle = | q + p \mod  N \rangle \ \ {\rm and} \ \ Z^n | m \rangle = \omega^{nm} |m \rangle.
    \label{em}
\end{equation}
This looks familiar from the previous section. It appears that the $X$ operator is rotating the state like a gauge transformation similar to the chargeon and that $Z$ is pulling out a phase like the fluxon. We can thus view plaquettes occupied by fluxons as magnetic flux tubes, which the chargeons interact with via the $\mathds{Z}_N$ gauge potential. If there are no fluxons present the gauge potential is everywhere flat and therefore there is no gauge transformation implemented on the chargeons. Further, the excitations can be moved around by applying string operators, which create and annihilate excitations in a sequence which effectively translates the particle in space. This allows us to perform braiding and fusion with the chargeons and fluxons. In fact, the chargeons live on the Fourier dual to the lattice inhabited by the fluxons, which means that their trajectories should be interpreted in reciprocal space. This insight will help us to interpret the real physical objects realizing the chargeons. A discussion on the interpretation of chargeons in spinor BECs is provided in Section \ref{NMA}. 

\begin{figure}
\centering
\includegraphics[width=\textwidth]{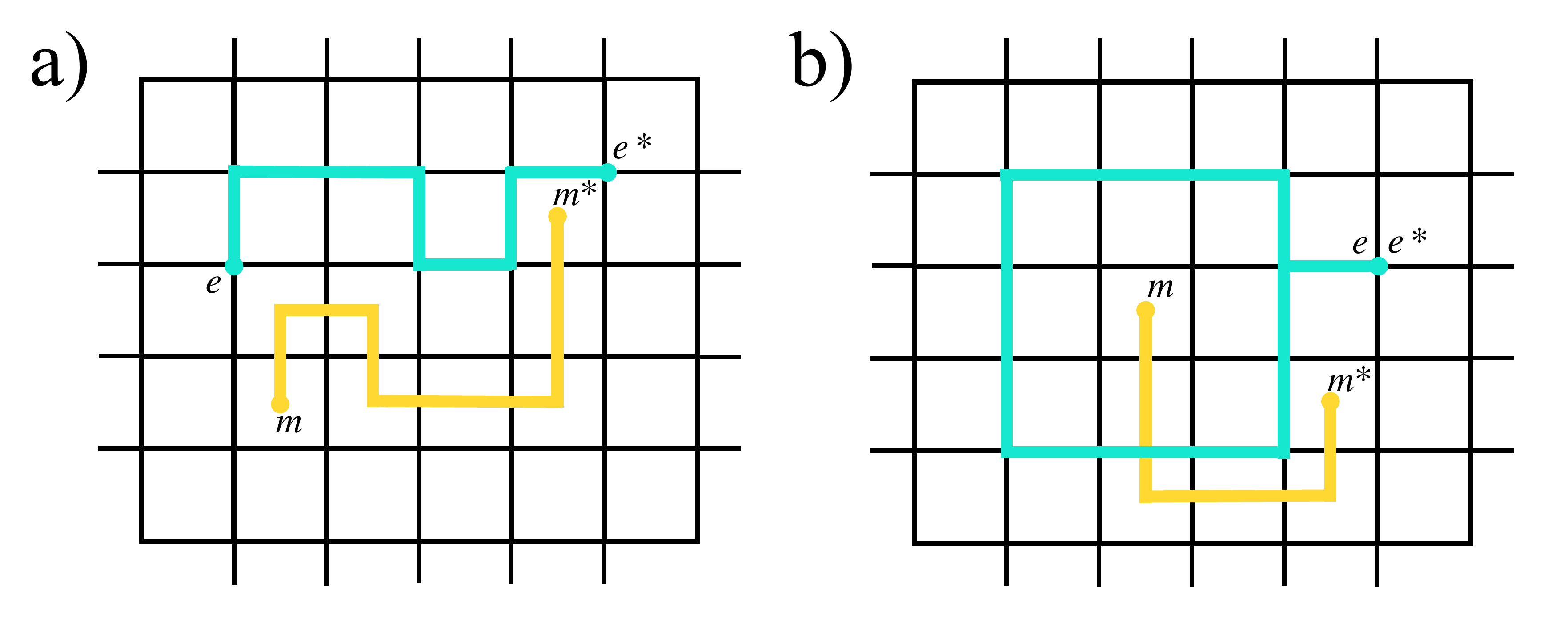}
\caption{a) An anti-fluxon $m^*$ is transported and fused with a anti-chargeon $e^*$, thus forming a dyon $m^* e^*$. b) A chargeon $e$ is brought around a fluxon and then fused with its anti-partner $e^*$.}
\label{strings}
\end{figure}

We may also consider composite objects called dyons. In fact, this is the most generic excitation as a chargeon simply is a dyon with trivial fluxon part, and similarly, a fluxon, simply is a dyon with trivial chargeon part. We concluded earlier that an arbitrary dyon is labelled by a conjugacy class $C_i$ of the symmetry group (the fluxon part) and an UIR of the centralizer of this conjugacy class $\Gamma_j (Z(C_i))$.  This is the essence of the quantum double structure. The structure is ``doubled" via a Fourier duality, which allows for a unified description of the two excitation spectra. A dyon can be viewed as a fusion product of a fluxon and a chargeon and since the fluxons correspond to the plaquette operators $A_{\square}$, which project out the flux and the chargeons to the gauge transformations $B_{+}$, the dyons must be classified according to the UIRs $\Lambda(A_{\square}B_{+})$. All fusion rules in this model are presented in Eq.~\eqref{fusion} where the powers belong to $\mathds{Z}_N$. For a graphical illustration, see Fig.~\ref{strings} a) where the string operators are depicted, which enables fusion of an anti-fluxon and and anti-chargeon.

\begin{equation}
e^q \otimes e^p=e^{q+p}, \ \ m^j \otimes m^l=m^{j+l}, \ \ e^k \otimes m^l=\epsilon^{(k,l)}
\label{fusion}
\end{equation}
\begin{equation*}
e^i\otimes \epsilon^{(k,l)}=\epsilon^{(i+k,l)},\ \ m^j\otimes \epsilon^{(k,l)}=\epsilon^{(k,j+l)}, \ \ \epsilon^{(i,j)} \otimes \epsilon^{(k,l)}= \epsilon^{(i+k,j+l)}
\end{equation*}

We can also derive the monodromy by creating string operators that, for instance, are bringing an $e$ around an $m$, as in Fig.~\ref{strings} b). Note that the loop corresponds to the same operation as the product of the plaquettes it encircles. All of the plaquettes return eigenvalue $1$ except the one on which the fluxon is residing. That is, $S_{\rm loop} | \Psi \rangle = \prod_{\square} A_{\square} |\Psi\rangle = A_{\boxdot} |\Psi\rangle = \omega |\Psi\rangle$. The eigenvalue of a braiding process where an $e$ is braided twice with an $m$ is thus given by $\omega$. This is, in fact, a discrete version of Stokes' theorem where the plaquette inhabited by the fluxon carries circulation so that the loop can be smoothly deformed around this plaquette, without affecting the outcome. As shown in Fig.~\ref{strings} b), the Dirac string connecting $m$ and $m^*$ is crossed once independent of the deformation of the loop, which results in a phase $\omega$ due to the relation between $X$ and $Y$ in Eq.~\eqref{alg}. We may thus conclude that topological equivalence is naturally encoded in this model. The plaquette operator $A_{\boxdot}$ acting on the fluxon can be regarded as a measurement operator that is projecting out the flux. In electromagnetism this is equivalent to $\nabla \times {\textbf{A}} = {\textbf{B}}$ and the action of the vertex operator thus represents something akin to Gauss's law $\nabla \cdot \nabla \phi = -\frac{\rho}{\epsilon_0}$. Generally, the flux is measured by $Z$ operators while the $X$ operators correspond to gauge transformations. Hence, the dyons, which can be regarded as the elementary objects of the quantum double structure, must be arranged according to the UIRs implementing the action of a flux measurement followed by a gauge transformation.

\subsubsection{Chargeon-fluxon duality}
\label{emdual}
Before discussing the algebraic structure of the quantum double we shall investigate some aspects regarding the excitations in this model. Since $\mathds{Z}_N$ is an abelian group the only element in the conjugacy class of each element is the element itself since $h_i h_j h_i^{-1} = h_j h_i h_i^{-1} = h_j \ \ \forall h_i,h_j \in \mathds{Z}_N$. Hence, we have $|\mathds{Z}_N|=N$ distinct fluxons in this theory. 
Specifically, if we let $\omega = e^{i \frac{2 \pi}{N}}$ be the generator of $\mathds{Z}_N$, we have the following fluxons
\begin{equation}
m^0 = \mathds{1}, \ \ m^1 = \omega, \ \ m^2 = \omega^2,..., \ \ m^{N-1} = \omega^{N-1}.
\label{fluxons}
\end{equation}
We can also establish that the centralizer of each element must be the whole group, since all elements in the group commute. Moreover, since the generator $1$ must be mapped onto the identity generator in the image of $\Gamma_i$, we know that $\Gamma_n(1)$ must be an $N$'th root of unity such that $(\Gamma_n(1))^N = 1 \ \ \forall n$. This fixes the number of UIRs to $N$ and the maps are given by $\Gamma_n(m) = e^{i \frac{2mn \pi}{N}} = \omega^{nm}$ where $m \in \mathds{Z}_N$. These UIRs are exactly the elements representing the fluxons, which means that we can interpret the fluxons and chargeons as duals of one another. The fluxons thus perceive the chargeons as flux tubes in the same way the chargeons experience the fluxons as flux tubes, to which they couple via the gauge field. Furthermore, for the special case $N=2$ we have that the operators $X$ and $Z$ are hermitian (the standard Pauli matrices), that is $X=X^\dagger$ and $Z=Z^\dagger$, which implies that they must square to the identity so that all excitations are their own anti-particles. Hence, in addition to the $e \leftrightarrow m$ symmetry, we also have $e \leftrightarrow e^*$ and $m \leftrightarrow m^*$ symmetry in the $\mathds{Z}_2$ model. In section~\ref{C2} we shall see an explicit example of a physical system realizing this particular structure.

\subsection{The quantum double construction}
\label{QD}
\begin{figure}[htp!]
    \centering
    \includegraphics[width=0.4\textwidth]{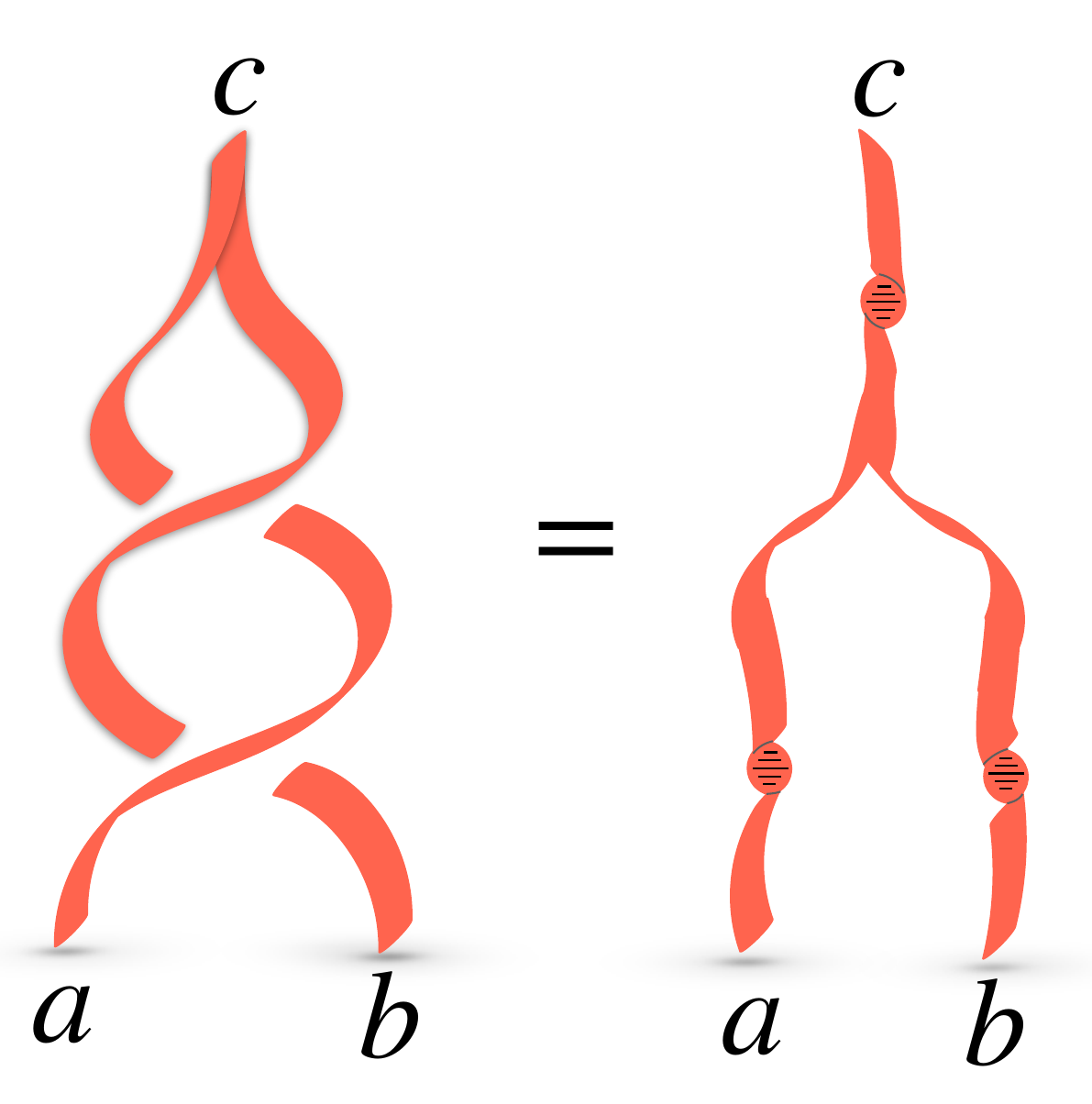}
    \caption{Pictorial illustration of the generalized spin-statistics connection.}
    \label{spinstat}
\end{figure}

As illustrated above, a quantum double structure comprises two elementary excitations called fluxons ($m$) and chargeons ($e$), and together they form a composite object, the dyon, which can be classified according to UIRs $\Pi(P_h g)$ implementing a gauge transformation $g$ followed by a flux measurement $P_h$. These elements constitute the building blocks of the quantum double of a group. In the example provided in the preceding section, the quantum double is given by $\mathfrak{D}(\mathds{Z}_N) =\{P_{h_i} g_j\}^{N-1}_{i,j = 0}$.
Now, due to flux metamorphosis and the property $P_h P_{h'}=P_h \delta_{h,h'}$ of the projection operators, we enforce the following multiplication rule
\begin{equation}
    P_h g P_{h'} g' = P_h P_{gh'g^{-1}} g g' = \delta_{h,g h' g^{-1}} P_h g g',
\end{equation}
where $h,h',g,g' \in G$. The image of a representation of a quantum double element has an action on the space $V^{\Gamma}_C = V_C \otimes V^{\Gamma}$ formed by the conjugacy classes $C$ (the fluxons) and the UIRs of the corresponding centralizers $\Gamma$, with basis $|h^C_i,u^{\Gamma}_j \rangle$ where $i = 0,1,.., |C|$ and $j = 0,1,..,\dim(\Gamma)$. The action of the image of $\Pi(P_h g)$ can be thought of as performing a gauge transformation $g$ subsequently followed by a flux measurement $P_h$ which yields \cite{1992PhLB..280...63A,1995hep.th...11201D}
\begin{equation}
\label{QDaction}
    \Pi_C^{\Gamma}(P_h g) | h_i^C, u^{\Gamma}_j \rangle = \delta_{h, g h_i g^{-1}} | g h_i^C g^{-1}, \Gamma(g^*) u^{\Gamma}_j \rangle,
\end{equation}
where $g^*= a_k g a_i^{-1}$ and $a_k$ is defined via $h_k = g h_i g^{-1}$, where $a_i \in H$ is the element that maps the conjugacy class representative element $h_*^C \in C$ to $h^C_i = a_i h^C_* a_i^{-1} \in C$. For a more detailed discussion the reader may consult Appendix~\ref{QDrep} or e.g. \cite{1992PhLB..280...63A,1995hep.th...11201D}. Now, if we tensor two quantum double representations (dyons), we find the fusion product of the two particles by decomposing the reducible tensored representation into its irreducibles subblocks. For instance, consider particle $a$ and $b$ with fusion channel $c$
\begin{equation}
    \Pi_a \otimes \Pi_b = \bigoplus_c N_{ab}^c \Pi_c,
\end{equation}
each of which can be formed in $N_{ab}^c$ unique ways. A noteworthy subtlety of the above fusion rule is that if we let $\Pi_a = \Pi^{1}_{C}$ and $\Pi_b = \Pi^{1}_{C^{-1}}$ represent a pure flux-antiflux pair, their fusion outcome might still contain charges. This can be explained by virtue of the fact that the representation $\Pi_a \otimes \Pi_b$ might still be reducible which signals that there may still be surviving charge quantum numbers, even if the fluxons annihilate one another. In particular, if the flux $C$ is brought around a distant flux $A$, then it returns as $ACA^{-1}$ so that $C^{-1}ACA^{-1} \neq I$ when it is fused with the antiflux $C^{-1}$. These quantum numbers are labelling delocalized charges that are not attached to the positions of the fluxes and are often referred to as \emph{Cheshire charges} (inspired by the Cheshire cat in Alice in wonderland who suddenly vanished while leaving the reminiscent of its grin behind). The integer $N_{ab}^c$ can be calculated by applying the projection operator that is mapping the reducible tensored representation $\Pi_a \otimes \Pi_b$ onto the irreducible orthogonal $\Pi_c$ blocks. That way one can disassemble the vector space corresponding to $\Pi_a \otimes \Pi_b$ into its irreducibles where the multiplicity of each irreducible is given by \cite{1995hep.th...11201D}
\begin{equation}
    N_{ab}^c = \frac{1}{|G|} \sum_{g,h \in G} {\rm Tr}[\Pi_c(P_h g)] {\rm Tr}[\Pi_a \otimes \Pi_b(\Delta (P_h g))]^*,
\end{equation}

where $\Delta$ is the coproduct map defined in Appendix~\ref{HA}. However, a more convenient way to compute the multiplicities is to employ the so-called modular S-matrix and the Verlinde formula \cite{1988NuPhB.300..360V}
\begin{equation}
\label{verlinde}
    N_{ab}^{c} = \sum_{D,\delta} \frac{S_{AD}^{\alpha \delta} S_{BD}^{\beta \delta} S_{CD}^{\gamma \delta}}{S_{eD}^{0 \delta}}.
\end{equation}
The S-matrix is computed as
\begin{equation}
\label{Smatrix}
    S^{\Gamma \Lambda}_{AB} = \frac{1}{|H|} \sum_{h_A \in C^A, h_B \in C^B} {{\rm Tr}[\Gamma(g_A^{-1} h_B g_A)]}^* {\rm Tr}[\Lambda(g_B^{-1} h_A g_B)]^*,
\end{equation}
where the capital letters denote the conjugacy classes and $\Gamma$ and $\Lambda$ the centralizer UIRs. Finally, we are also interested in the topological spin of the particles since it provides information regarding statistics and braiding. Considering \emph{Shur's lemma} \cite{burnside} which states that if a complex valued matrix $A$ commutes with the image of some UIR $\Gamma$, then the matrix $A$ must be represented as a complex matrix that is proportional to the identity in the representation $\Gamma$, i.e. ${\rm Rep}(A) = e^{i \theta_\Gamma(A)} \mathds{1}_{\Gamma(A)}$. Given some anyon model we may compute the phases corresponding to each of the particles in the model and then store them in a matrix by stacking them on the diagonal. The resulting matrix is known as the modular T-matrix, which together with the S-matrix discussed above, span the modular group ${\rm SL}(N,\mathds{F})$ of conformal transformations which encodes the conformal structure of anyon models. Note that this fits into the quantum double construction as the flux $h$ of a dyon, per definition, commutes with its centralizer $Z(h)$. Thus, considering some UIR $\Gamma$ of $Z(h)$, we may establish a generalized spin-statistics connection \cite{1982PhRvL..49..957W,1988CMaPh.116..127F} $e^{i 2 \pi s}=e^{i 2 \pi \theta_{\Gamma(h)}}$. This relation can be understood from a graphical perspective by considering the twisting of the world lines of two anyons as they are braided around one another. Here it is useful to think of the world lines as ribbons, see Fig.~\ref{spinstat}, as the twisting becomes more apparent.
Fig.~\ref{spinstat} is a pictorial illustration of the topological equivalence between the double interchange $\mathcal{R}^2$ of $a$ and $b$ (this is exactly the monodromy illustrated in Fig.~\ref{phase} and Fig.~\ref{strings} b)) and the twisting of their world lines. From this we can straightforwardly deduce the monodromy matrix elements as
\begin{equation}
\label{monomat}
    [\mathcal{R}^2]_{ab}^c = e^{i 2 \pi (s_c-s_a-s_b)},
\end{equation}
where $s_i$ are the topological spins of the particles. We shall return to this in Section \ref{spin01} and \ref{spin2} and calculate the spins explicitly for various excitations in spinor BECs. 

\subsection{Vortex chromodynamics picture}\label{vqcd}

The action described in Eq.~\eqref{action} is generic and here we offer an interpretation of it in the context of a spinor BEC. A superfluid can be used for reproducing the physics of relativistic electrodynamics \cite{popov1973quantum} by defining the superfluid velocity ${\boldsymbol v}_s =(\hbar/m)\nabla \theta$ as the electric field and the mass density of particles $\rho_s = m|\Psi|^2$ as the magnetic field, jointly forming a Maxwellian field strength tensor $F_{\mu \nu}$. Alternatively, the phase change $\partial_t \theta$ may be assigned to the role of magnetic field to avoid nonzero vacuum magnetisation \cite{2020PhRvA.101f3616S}. Building on this idea, it is conceivable that a superfluid comprised of spinful particles may reproduce a generic Yang--Mills theory \cite{peskin2018introduction}. Indeed, it was demonstrated in \cite{2018PhRvA..97b3613E,tylutki2016confinement,cirac2010cold} that certain aspects of quantum chromodynamics (QCD), such as quark confinement, can be simulated in BECs with spin. Other forms of gauge theories, such as that involving an emergent Ising gauge field \cite{2003AnPhy.308..692Z}, have also been studied using BECs comprised of spinful particles. By adopting a mean-field description (see Appendix~\ref{mft}), a Yang--Mills field strength tensor may be constructed by applying the Madelung transformation $\Psi^a(\textbf{r},t) = |\psi^a(\textbf{r},t)| e^{i \theta^a(\textbf{r},t)}$ to the spinor components. This allows one to separate the phase and the density which can  be interpreted as Yang--Mills field strength and gravity, respectively. However, if we consider a static phase, the full field strength $F^a_{\mu \nu}$ is completely defined by the spin current (phase gradient)\cite{2010arXiv1001.2072K}\\
\begin{equation}
\label{spincurrent}
    [{\boldsymbol v}_s]^{a} = \frac{\hbar}{2Mni} \sum_{m,m'=-S}^S [f_a]_{m m'}\Big(\psi_m^*(r,t)(\nabla \psi_{m'}(r,t))-(\nabla \psi_{m}^*(r,t))\psi_{m'}(r,t)\Big),
\end{equation}
where $a$ denotes the components of the spin. In this picture, the particles can be thought of as color charged ``quarks" which are interacting via a generalized electric field --- the ``strong" force. However, if we instead consider a viewpoint in which one particle is a charged analogue quark, and a second particle is a generalized flux tube, then the interpretation of Eq.~\eqref{spincurrent} is that it corresponds to a non-abelian gauge field $A^a_{\mu}$, such that instead $F^a_{\mu \nu} = \partial_{\mu}A^a_{\nu}-\partial_{\nu}A^a_{\mu} + ig [A^a_{\mu},A^a_{\nu}]$, where $g$ is the coupling. Since the superfluid velocity field arises due to the non-vanishing of the phase gradient $\nabla \theta^a$, the Aharonov--Bohm phase acquired as the vortex is encircled is given by the contour integral of $A_{\mu}^a = [\boldsymbol{v}_s]^a$. Furthermore, the vorticity is given by $\boldsymbol{\omega} = \nabla \times \boldsymbol{v}_s$ allowing it to be interpreted as the magnetic part of the field strength. If the vorticity, or field strength, is only non-vanishing inside of the vortex core the resulting interaction is topological. In the Gross--Pitaevskii equation (GPE) \cite{2010arXiv1001.2072K} which governs the dynamics of a BEC, the covariant derivative encoding the interactions mediated by $A_{\mu}^a = [\boldsymbol{v}_s]^a$ is obtained by transforming to a frame co-moving with the vortex. The Laplacian in the kinetic term can then be expressed as
$
    \nabla^2 \longrightarrow (\nabla + \kappa^a[\boldsymbol{v}_s]^{a})^2,
$
so the charge $g$ corresponds to the generalised circulation quantum $\kappa^a$. This term is responsible for the interactions within the quantum double as it encodes the response of the vortex due to the motion of the surrounding atoms. Note also that this Laplacian is taking on exactly the same form as that giving you the kinetic energy of an electron coupled to a magnetic field in e.g. the quantum Hall effects.
In analogy with QCD, we may therefore say that the particles of the theory carry color charge which is preserved under the interactions mediated by $A^a_{\mu}$. This quantum number can thus be identified as the topological spin, or generalised quantum of circulation, of the particles defined in the preceding subsection. The vortices represent defects which carry a non-zero field strength to which the chargeons couple topologically via $A^a_{\mu}$ (or the reversed if implementing the S-matrix duality which is to be discussed in the following subsections). A generic dyonic state can, in analogy with Gell-Mann's eightfold way, thus be regarded as a superposition in an $N$-tuplet color space, where $N = |C| \times \dim(\Gamma)$. Interestingly, as demonstrated in \cite{genetay2023topological}, by choosing the density profile of the system in a specific way, the gravity-like force arising from the quantum pressure can be implemented topologically. In particular, arbitrary topological phase rotations can be achieved which may be employed as additional quantum logic gates. Hence, such a vortex QCD plus gravity may result in a universal platform for topological quantum computation.

\subsection{Anyons in Spin-0 and Spin-1 BECs}\label{spin01}
In a spinor BEC, the order parameter representing the phase under consideration is not a scalar but a spinor. The order parameter wave function of a spin-$F$ condensate belongs to the complex vector space $\mathcal{H} = \mathds{C}^{2S+1}$ which is isomorphic to the real space $\mathds{R}^{4S+2}$, i.e. $\mathcal{H} = \mathds{C}^{2S+1} \simeq \mathds{R}^{4S+2}$, where $S$ is the spin. This entails that the order parameter is a map onto a surface isomorphic to a sphere, that is, it maps onto the manifold $\mathcal{M} \simeq S^{4S+2-1} = S^{4S+1}$. Consequently, we can conclude that the full symmetry group of the system is $\textrm{SO}_S (3)$ accompanied by phase invariance, i.e. $\textrm{SO}_S (3) \times \rm{U}(1)$. Here we consider two types of condensates with spin-$0$ and spin-$1$ degrees of freedom, respectively, which are both governed by abelian theories. In the $S=0$ case, we know that the corresponding wave function is a scalar $\Psi (r,\theta) = |\Psi(r,\theta)|e^{i \theta}$. Moreover, if we break the $\rm{U}(1)$ symmetry, we know that each subgroup is inheriting the commutative property, more precisely each subgroup must be isomorphic to a $\mathds{Z}_N$ group, thus giving rise to a quantum double akin to the lattice model discussed in Section \ref{ZN}. In the spin-$1$ case we have three low-temperature inert states $(1,0,0)^T, (0,0,1)^T$ and $(0,1,0)^T$, which correspond to the two ferromagnetic states and the polar state, respectively. The symmetry groups of these states are given by rotations about the axes parallel to the states so that $\textrm{SO}_S (3) \times \rm{U}(1)$ reduces to $\textrm{SO}(2)_{S_z} \times \rm{U}(1)$ for the ferromagnetic states, and $D_{\infty} \times \rm{U}(1)$ for the polar state. Here $D_{\infty}$ denotes the infinite dihedral group which also includes reflection symmetry, as opposed to $\textrm{SO}(2)_{S_z}$. As for discrete symmetries in these phases, we can conclude that we only have one corresponding to the reflection part of the $D_{\infty}$ symmetry, which can be represented by the cyclic group $C_2$. 

A convenient way to deduce the symmetries is to consider a graphical representations of the spinorial states, such as the spherical harmonics or the Majorana star representation \cite{2010arXiv1001.2072K}. Note that since these groups are abelian, spin-0 or a spin-1 condensates are not able to support non-abelian anyons as their natural excitations. This fact can be established from the analysis outlined in Section \ref{symclass}, where we discussed how the topological excitations in planar systems can be classified according to the first homotopy group $\pi_1 (G/H)$ of the coset space $G/H$ generated by the subgroup $H$.

\subsubsection{The quantum double $\mathfrak{D}(C_2)$ of $C_2$}
\label{C2}
All elements in an abelian group constitute their own conjugacy class, and the centralizers are given by the whole group. For instance, $C_2 = \{I,r\}$ has two elements: one identity $I$ and one $\pi$-rotation $r$, which are the two conjugacy classes of the group, and consequently the centralizer has two UIRs, the $1D$ trivial representation $\Lambda_{\rm sym}$ and the $1D$ asymmetric representation $\Lambda_{\rm asym}$. The character table for this model is provided in Table~\ref{C2table}. We thus conclude that there are four abelian anyons in the model corresponding to the UIRs of the quantum double element $P_h g$. In particular, we have a vacuum  sector $\mathds{1}$, one fluxon $m$, one chargeon $e$ and one dyon $\epsilon$. The fusion rules can thus be summarized as
\begin{align}
    m \otimes m =e \otimes e = \epsilon \otimes \epsilon = \mathds{1}\\
    e \otimes m = \mathds{1} \oplus \epsilon.
\end{align}
Note that this model is equivalent to the $\mathds{Z}_2$ quantum double model which is given by $N=2$ instance of the model discussed in Section~\ref{ZN}. As for braiding, we can conclude that the fluxon $m$ as well as the chargeon $e$ are both bosons. This fact is evident from the character table Table~\ref{C2table}, from which we can also deduce that the dyon $\epsilon$ is a fermion. Moreover, while the bosons have trivial self-braiding, winding the chargeon around he fluxon (or the fluxon around the chargeon) yields a factor $-1$, which can be derived from the generalized spin-statistics theorem in Eq.~\eqref{monomat}. The topological spins $s$ can be deduced from the character table as $\chi/d = e^{i 2 \pi s}$, where $\chi$ is the character and $d$ is the dimension of the corresponding UIR. This relationship can be derived from the definition of the Wilson loop defined in Eq.~\eqref{wilson}, which is nothing but a measurement of the monodromy illustrated in Fig.~\ref{phase}, which is equivalent to the character values of the representation under consideration. Reading off the character table, we may establish that $s_{\mathds{1}} = s_m = s_e =0$ and $s_{\epsilon} = \frac{1}{2}$ as expected for bosons and fermions.
\begin{table}
\centering
\caption{Character table of the group $C_2$.}
\label{C2table}
\begin{tabular}{ |c|c|c| } 
 \hline
  & $I$ & $r$ \\ 
 \hline
  $\Lambda_{\rm sym}$ & 1 & 1 \\ 
 \hline
 $\Lambda_{\rm asym}$ & 1 & -1 \\ 
 \hline
\end{tabular}
\end{table}
We next compute the S-matrix of the model which we briefly encountered in Section \ref{QD}, as well as the T-matrix, which together generate the modular group ${\rm SL}(2,\mathds{Z})$. Applying Eq.~\eqref{Smatrix} yields the matrix
\begin{equation}
    S_{C_2} =  \frac{1}{2} \begin{pmatrix}
    1 & 1 & 1 & 1\\
    1 & 1 & -1 & -1\\
    1 & -1 & 1 & -1\\
    1 & -1 & -1 & 1
    \end{pmatrix}.
\end{equation}
For abelian theories, the interpretation of this matrix is simple. The elements $S_{C_2}^{ij}$ represent the phase acquired when exchanging particles $i$ and $j$, where the indices run through the charges $\mathds{1},e,m,\epsilon$. Moreover, the T-matrix is given by stacking the twist factors $e^{i 2 \pi s}$ of the anyons on the diagonal which yields the T-matrix
\begin{equation}
    T_{C_2} =  \begin{pmatrix}
    1 & 0 & 0 & 0\\
    0 & 1 & 0 & 0\\
    0 & 0 & 1 & 0\\
    0 & 0 & 0 & -1
    \end{pmatrix}.
\end{equation}

\subsubsection{Particle-vortex duality in $\mathfrak{D}(C_2)$}

Interestingly, as shown in \cite{1999JPhA...32.8539K}, the modular S-matrix can be interpreted as a generalization of the quantum Fourier transform based on the representation theory of the quantum double. In fact, this is exactly the transformation that is implementing the chargeon-fluxon duality discussed in Section \ref{ZN}, where the chargeons live on the direct lattice and the fluxons live on the reciprocal lattice. To illustrate this explicitly, all we have to show is that the set of characters corresponding to the chargeon transform, under the action of $S$, into the set of characters corresponding to the fluxon, since these sets form an orthonormal basis for their respective Hilbert space. We first need the representations labelling the two anyons which can be worked out by virtue of Eq.~\eqref{QDaction}, which defines the quantum double action. Using $\Pi^e = \Pi^{\rm asym}_I$ to denote the representation corresponding to the chargeon $e$ and $\Pi^m = \Pi^{\rm sym}_r$ to denote that of the fluxon $m$, then the image of their $1D$ quantum double representations, given the action defined in Eq.~\eqref{QDaction}, is given by Table~\ref{reptable}.

\begin{table}
\centering
\caption{The elements in the image of the representation $\Lambda$.}
\label{reptable}
\begin{tabular}{ |c|c|c|c|c| } 
 \hline
  & $P_I I$ & $P_I r$ & $P_r I$ & $P_r r$\\ 
 \hline
  $\Pi^{asym}_I (P_h g)$ & 1 & -1 & 0 & 0 \\ 
 \hline
 $\Pi^{sym}_r (P_h g)$ & 0 & 0 & 1 & 1 \\ 
 \hline
\end{tabular}
\end{table}
Since these representations are one-dimensional they yield the characters directly since the characters are computed as the trace of the elements in the representation image. In light of this, we may denote by $\chi_e = (1,-1,0,0)^T$ the set of characters for the chargeon and by $\chi_m = (0,0,1,1)^T$ the set of characters for the fluxon. It can then be shown that indeed $S \chi_e = \chi_m$, which proves that $S$ is implementing a duality between the two anyons. Moreover, if we work out the characters corresponding to the dyon $\epsilon$, we have that $\chi_{\epsilon} = (0,0,1,-1)^T$, and if we act with $S$ we find that the state is left invariant, i.e. $S \chi_{\epsilon} = \chi_{\epsilon}$, as expected since $\epsilon$ is a fluxon-chargeon composite so it must be left untouched if we swap the constituent anyons $e \leftrightarrow m$. This duality is exactly the same as the one discussed in Section \ref{emdual} for $N=2$. Here we took a different route via the representation theory and the S-matrix and arrived at the same result, without any direct knowledge about the structure of the Hamiltonian, other than its symmetries. Next, we shall consider spin-2 systems which exhibit phases with non-abelian symmetry groups that may be capable of topological quantum computation.

\subsection{Anyons in Spin-2 BECs}
\label{spin2}
Unlike the spin-0 and spin-1 condensates, the spin-2, and in fact any spin $F\geq 2$, exhibit phases whose order parameters are invariant under non-abelian groups. The order parameters can be worked out from the mean-field theory described in Appendix~\ref{mft}. In the unbroken phase, such order parameters have rotational $\textrm{SO}(3)$ symmetry as well as phase invariance $\textrm{U}(1)$, which results in the full symmetry $\textrm{SO}(3) \times \textrm{U}(1)$, but if we consider instead the simply connected special unitary representation $\textrm{SU}(2)$, the group structure is given by $ G = \rm{SU}(2) \times \textrm{U}(1)$. If this symmetry is broken to a subgroup $H \subset G$, the full order parameter manifold is given by the coset space formed by taking the quotient $G/H$, i.e. $\mathcal{M} = \textrm{SU}(2) \times \textrm{U}(1) / H$. Here we will direct our attention towards two particular ground state phases, the binary tetrahedral phase and the biaxial-nematic phase. The particle content of the emerging anyon models can be derived by extending the analysis carried out in the abelian case, but as we shall see, these structures are much more complex which gives rise to vast particle spectra. The topological spins of the particles are, again, derived from the character theory in the same way as in the previous section.

\subsubsection{Binary tetrahedral phase and its quantum double $\mathfrak{D}(T^*)$}

The order parameter of the binary tetrahedral phase is invariant under the action of the group $H = T^* \in \rm{SU}(2)$, which is the simply connected double cover of $T \in \rm{SO}(3)$, and is given by $\Psi = \frac{1}{2} (i,0,\sqrt{2},0,i)^T$ \cite{2010arXiv1001.2072K}. The tetrahedral group comprises the symmetries of a tetrahedron and has 12 elements in total formed by $pi/3$-rotations and reflections. The binary representation, however, has 24 elements due to the two-to-one double cover as each element $t \in T$ is mapped onto $t,\bar{t} \in T^*$, i.e. $f: \ \ t \longrightarrow \{ t, \bar{t}\}$. The structure of $T^*$ can be represented as $\{a,b,c | a^3=b^3=(ab)^2=-1,  \}$ \cite{2008arXiv0810.3225L}, and since $T^* \subset \rm{SU}(2)$, we may pick the elements $a = 1/2(\mathds{1} + i \sigma_x + i \sigma_y + i \sigma_z)$ and $b = 1/2(\mathds{1} + i \sigma_x + i \sigma_y - i \sigma_z)$ as generators. It is straightforward to show that these satisfy the relationships between the generators of the group. Unpacking the structure, we first note that they both generate disjoint 6-cycles $\mathds{Z}_6$ thus comprising in total 12 distinct elements individually, which further implies that they must form another 12 distinct elements when combined, since there are 24 elements in the group in total. Noting that $ab = i \sigma_x$ and $ba = i \sigma_y$, and that $i \sigma_y i \sigma_x = i \sigma_z$ and $a^4 = -a$ (and consequently $a^5 = -a^2$ where the same relations hold for $b$), we can conclude that all of the 24 elements can be generated from $\pm a=\pm \tilde{\sigma}$ (or $b$) and its product with $\pm i \sigma_i$ for $i \in \{x,y,z\}$. In particular, we have the following partitioning into conjugacy classes (CC) \cite{2007PhRvL..98j0401S}
\begin{align}
 {\rm CC}_1(T^*) &= \{ (I, n_w) \} \\
 {\rm CC}_2(T^*) &=\{ (-I, n_w) \} \notag\\
 {\rm CC}_3(T^*) &= \{ ( \pm i \sigma_x, n_w), ( \pm i \sigma_y, n_w), ( \pm i \sigma_z, n_w) \} \notag\\
 {\rm CC}_4(T^*) &= \{ (\Tilde{\sigma}, n_w + \frac{1}{3}), (-i \sigma_x \Tilde{\sigma}, n_w + \frac{1}{3}), (-i \sigma_y \Tilde{\sigma}, n_w + \frac{1}{3}), (-i \sigma_z \Tilde{\sigma}, n_w + \frac{1}{3}) \} \notag\\
 {\rm CC}_5(T^*) &= \{ (-\Tilde{\sigma}, n_w + \frac{1}{3}), (i \sigma_x \Tilde{\sigma}, n_w + \frac{1}{3}), (i \sigma_y \Tilde{\sigma}, n_w + \frac{1}{3}), (i \sigma_z \Tilde{\sigma}, n_w + \frac{1}{3}) \} \notag\\
 {\rm CC}_6(T^*) &= \{ (\Tilde{\sigma}^2, n_w + \frac{2}{3}), (-i \sigma_x \Tilde{\sigma}^2, n_w + \frac{2}{3}), (-i \sigma_y \Tilde{\sigma}^2, n_w + \frac{2}{3}), (-i \sigma_z \Tilde{\sigma}^2, n_w + \frac{2}{3}) \} \notag\\
 {\rm CC}_7(T^*) &= \{ (-\Tilde{\sigma}^2, n_w + \frac{2}{3}), (i \sigma_x \Tilde{\sigma}^2, n_w + \frac{2}{3}), (i \sigma_y \Tilde{\sigma}^2, n_w + \frac{2}{3}), (i \sigma_z \Tilde{\sigma}^2, n_w + \frac{2}{3}) \}\notag
\label{conjclassT}
\end{align}
where $n_w \in \mathds{Z}$ denotes the winding number and the $\frac{1}{3}$ fractions are due to the breaking of $\rm{U}(1)$ to $\mathds{Z}_3$. This particular partitioning can be derived either by brute force computation of all conjugates, or by e.g. exploiting the fact that all conjugate elements must have the same order, thus allowing one to group potential conjugate partners together. Then using the property $\pm i \sigma_i i\sigma_j i\sigma_i^{-1} = \pm i\sigma_j^{-1}$ where $i \neq j \in \{x,y,z\}$, one can further split the sets of equal-order elements into subsets that only contain mutual conjugates. In order to derive the the chargeon particle spectrum, we need to find the centralizers and their irreducible representations. ${\rm CC}_1(T^*)$ and ${\rm CC}_2(T^*)$ are both abelian which entails that the entire group $T^*$ must be their centralizer. First, let us see if there are any one-dimensional UIRs. Due to the geometry of the tetrahedron, we may conclude that it is invariant under $2 \pi/3$-rotations, about three distinct axes, i.e. $\mathds{Z}_3$. This group has a trivial UIR sending all elements to the identity, in addition to two representations which are obtained by permuting the trivial representation by the group generator $e^{i\frac{2 \pi}{3}}$. This must be all one-dimensional UIRs since $\mathds{Z}_3$ is abelian so that each of the three elements is its own conjugacy class, and consequently, there must be three UIRs $\Lambda^{T^*}_i$ ($i=4,5,6$), due to Burnsides's theorem \cite{burnside} in the theory of finite groups. Further, we must have a faithful two-dimensional complex representation since the binary tetrahedron is naturally embedded in $\mathds{C}^2$. We can construct yet another two two-dimensional UIRs by taking the tensor product of it with the two non-trivial one-dimensional UIRs, thus resulting in a total of three inequivalent two-dimensional representations which we shall denote by $\Lambda^{T^*}_i$ ($i=1,2,3$). Now, as a collorary of Burnside's theorem, the order of the group must be equal to the sum of the UIR dimensions squared. Thus far we have $3 \times 1^2 + 3 \times 2^2 = 15$ which implies that there must exist one three-dimensional representation since $\sqrt{24-15}=3$. In order to find this representation, we may exploit the fact $\rm{SU}(2)$ is the double cover of $\rm{SO}(3)$, and hence, there must exist a two-to-one map between $T^*$ and $T$. $T$ has, owing its embedding, a natural three-dimensional faithful representation $\Lambda^{T^*}_7$ which is simply permuting the corners of the tetrahedron in real $\mathds{R}^3$ space, so by composing this map with the double covering map, we find a three dimensional UIR for $T^*$. 

Instead of writing all of these matrices out explicitly as we did in the abelian $C_2$ case, we are simply providing the corresponding character tables as all essential information can be extracted from there. The characters of the one-dimensional UIRs are found trivially since the traces of them are the same as the group elements themselves. As for the two-dimensional representations, we pick one element from each conjugacy class in Eq.~\eqref{conjclassT} which we multiply by each of the one-dimensional UIRs and then take the trace, since the trace is invariant under conjugacy due to its invariance under cyclic permutations. Finally, the characters for the three-dimensional UIR can be found simply by defining a map $f$ that sends each Pauli matrix $\sigma_i$ in Eq.~\eqref{conjclassT} to the corresponding three-dimensional Lorentz rotation matrix $R_i$, since the Lorentz rotations $\rm{SO}(3)$ and $\rm{SU}(2)$ adhere to the same algebra. The characters for all of these seven representations can be found in Table~\ref{reptableT}.
\begin{table}
\caption{Character table for $\Lambda^{T^*}_i$, where $i=1,2,3$ are one-dimensional, $i=4,5,6$ are two-dimensional and $i=7$ is three-dimensional. Here $\omega = e^{i \frac{2 \pi}{3}}$.}
\centering
\label{reptableT}
\begin{tabular}{|c|c|c|c|c|c|c|c|} 
 \hline
  & ${\rm CC}_1(T^*)$ & ${\rm CC}_2(T^*)$ & ${\rm CC}_3(T^*)$ & ${\rm CC}_4(T^*)$ & ${\rm CC}_5(T^*)$ & ${\rm CC}_6(T^*)$ & ${\rm CC}_7(T^*)$\\ 
 \hline
  $\Lambda^{T^*}_1$ & 1 & 1 & 1 & 1 & 1 & 1 & 1\\  
 \hline
 $\Lambda^{T^*}_2$ & 1 & 1 & 1 & $\omega$ & $\omega$ & $\omega^*$ & $\omega^*$\\ 
 \hline
 $\Lambda^{T^*}_3$ & 1 & 1 & 1 & $\omega^*$ & $\omega^*$ & $\omega$ & $\omega$\\ 
 \hline
 $\Lambda^{T^*}_4$ & 2 & -2 & 0 & -1 & 1 & 1 & -1\\
 \hline
 $\Lambda^{T^*}_5$ & 2 & -2 & 0 & $-\omega^*$ & $\omega^*$ & $\omega$ & $-\omega$\\ 
 \hline
 $\Lambda^{T^*}_6$ & 2 & -2 & 0 & $-\omega$ & $\omega$ & $\omega^*$ & $-\omega^*$\\
 \hline
 $\Lambda^{T^*}_7$ & $3$ & $3$ & $-1$ & $0$ & $0$ & $0$ & $0$\\
 \hline
\end{tabular}
\end{table}
Again, the topological spins of the anyons in the spectrum can be deduced from the character table via the relation $e^{i 2 \pi s} = \chi/d$, where $s$ is the spin, $\chi$ is the character value and $d$ is the dimension. Reading off the first and the second column, corresponding to fluxons ${\rm CC}_1(T^*)$ and ${\rm CC}_2(T^*)$, we note that all characters are integers which means that they are all bosons. Next, we turn to the ${\rm CC}_3(T^*)$ conjugacy class which has $\mathds{Z}_4$ as centralizer group. This can be concluded by noting that all elements in this set are pure rotations around the $x,y$ and $z$ axis, respectively. Hence, the only elements commuting with a representative from this conjugacy class is a rotation about the same axis. Considering the group structure, there can only be 4 of those (generated by the element itself) which leads us to the conclusion that the centralizer is $\mathds{Z}_4$. Moreover, each element in $\mathds{Z}_4$ is its own conjugacy class since it is an abelian group, and consequently, as per Burnside's theorem, there must be four UIRs $\Gamma_i$ ($i=1,2,3,4$) which can be found simply by permuting the the trivial representation which send all elements to 1, by rotations of $\pi/2$. We thus obtain the character table presented in Table \ref{reptableZ4}.

\begin{table}
\caption{Character table for the four one-dimensions UIRs $\Gamma^{\mathds{Z}_4}_i$ ($i=1,2,3,4$) and $x$ is the generator of $\mathds{Z}_4$.}
\centering
\label{reptableZ4}
\begin{tabular}{|c|c|c|c|c|} 
 \hline
  & $1$ & $x$ & $x^2$ & $x^3$\\ 
 \hline
  $\Gamma^{\mathds{Z}_4}_1$ & 1 & 1 & 1 & 1\\  
 \hline
 $\Gamma^{\mathds{Z}_4}_2$ & 1 & $i$ & -1 & $-i$\\ 
 \hline
 $\Gamma^{\mathds{Z}_4}_3$ & 1 & $-1$ & 1 & $-1$\\ 
 \hline
 $\Gamma^{\mathds{Z}_4}_4$ & 1 & $-i$ & $-1$ & $i$\\
 \hline
\end{tabular}
\end{table}

The self-statistics, and hence the spin, can be read off from the second column in Table \ref{reptableZ4} which reveals that the $({\rm CC}_3,\Gamma^{\mathds{Z}_4}_1)$ dyon is a boson, the $({\rm CC}_3,\Gamma^{\mathds{Z}_4}_2)$ is a spin-$\frac{1}{4}$ particle, $({\rm CC}_3,\Gamma^{\mathds{Z}_4}_3)$ is a fermion and the $({\rm CC}_3,\Gamma^{\mathds{Z}_4}_4)$ must be a spin-$\frac{3}{4}$ particle. Finally, by the same argument we applied to deduce the $\mathds{Z}_4$ centralizer, we can establish that the centralizer of the remaining conjugacy classes ${\rm CC}_4(T^*)$, ${\rm CC}_5(T^*)$, ${\rm CC}_6(T^*)$ and ${\rm CC}_7(T^*)$ must be the 6-cycle $\mathds{Z}_6$ since $\tilde{\sigma}$ and $\tilde{\sigma} \sigma_i$ (where $i \in \{x,y,z \}$) have order 6. Again, owing to the abelian structure of $\mathds{Z}_6$, it must have six UIRs whose characters are provided in Table \ref{reptableZ6}.

\begin{table}
\centering
\caption{Character table for the six one-dimensions UIRs $\Gamma^{\mathds{Z}_6}_i$ ($i=1,2,3,4,5,6$) and $y$ is the generator of $\mathds{Z}_6$ and $\varphi = e^{i \frac{2 \pi}{6}}$.}
\label{reptableZ6}
\begin{tabular}{|c|c|c|c|c|c|c|} 
 \hline
  & $1$ & $y$ & $y^2$ & $y^3$ & $y^4$ & $y^5$\\ 
 \hline
  $\Gamma^{\mathds{Z}_6}_1$ & 1 & 1 & 1 & 1 & 1 & 1\\  
 \hline
 $\Gamma^{\mathds{Z}_6}_2$ & 1 & $\varphi$ & $\varphi^3$ & $-1$ & $\varphi^4$ & $\varphi^5$\\ 
 \hline
 $\Gamma^{\mathds{Z}_6}_3$ & 1 & $\varphi^2$ & $\varphi^4$ & $1$ & $\varphi^2$ & $\varphi^4$\\  
 \hline
 $\Gamma^{\mathds{Z}_6}_4$ & 1 & $-1$ & $1$ & $-1$ & $1$ & $-1$\\
 \hline
 $\Gamma^{\mathds{Z}_6}_5$ & 1 & $\varphi^4$ & $\varphi^2$ & $1$ & $\varphi^2$ & $\varphi^4$\\
 \hline
 $\Gamma^{\mathds{Z}_6}_6$ & 1 & $\varphi^5$ & $\varphi^4$ &$-1$ & $\varphi^4$ & $\varphi$\\
 \hline
\end{tabular}
\end{table}

Even here we find some interesting dyonic particles with fractional spin $s=\frac{1}{6},\frac{1}{3},\frac{2}{3},\frac{5}{6}$, which can be deduced from the second column. We have now worked out the entire particle spectrum of the quantum double of $T^*$ which is comprised of one vacuum, 6 pure fluxons, 6 pure chargeons and 29 dyons, thus amounting to a total of 42 distinguishable particles. The fusion rules of these anyons can be obtained by first computing the S-matrix according to Eq.~\eqref{Smatrix} and then employing e.g. the Verlinde equation in Eq.~\eqref{verlinde}. However, computing all such combinations would be a monstrous task so we will not present these here. The underlying principle is the same though as in the much simpler $\mathfrak{D}(C_2)$ anyon model. Interestingly, as already pointed out in Section \ref{QD}, fusing a fluxon with its anti-partner may result in a particle with Cheshire charge. This possibility stems from the reducibility of the tensored representation space of the two fluxons. If the resulting space has invariant subspaces, these subspaces correspond to multiplets, each of which is labelled by an invariant charge quantum number.

\subsubsection{Biaxial nematic phase and its quantum double $\mathfrak{D}(D^*_4)$}

The biaxial-nematic phase is represented by the order parameter $\Psi = \frac{1}{\sqrt{2}}(1,0,0,0,1)^T$ which is invariant under the action of the binary dihedral-4 group $D^*_4$. Consequently, the full order parameter manifold is given by $\mathcal{M} = \rm{U}(1) \times \rm{SU}(2)/D^*_4$ and the fluxons and chargeons are classified according to the conjugacy classes of $D^*_4$ and their centralizer UIRs, respectively. The group $D^*_4$ is the binary extension of $D_4$ whose structure can be represented as $D^*_4 = \{ a,b | a^{8}=1, b=a^2, bab^{-1}=a^{-1}\}$ \cite{gt}. In words, we have an 8-cycle $a$ about a primary axis, say the z-axis, and a 4-cycle $b$ which reverses $a$ under conjugation. In the $\rm{SU}(2)$ representation, we may pick $R_z(2\pi/8) = 1/\sqrt{2}(\mathds{1}+i \sigma_z)$ as the $\mathds{Z}_8$ generator, and one can easily verify that either $i \sigma_x$ or $i \sigma_y$ works as an $\rm{SU}(2)$ representation of $b$, so we may pick for instance $i \sigma_x$. Note that $i \sigma_x i \sigma_z = i \sigma_y$, meaning that all elements can be expressed as products of $i \sigma_x, i\sigma_y$ and $\tilde{\sigma} = 1/\sqrt{2}(\mathds{1}+i\sigma_z)$. Moreover, since $D^*_4$ has sixteen elements in total, and the $\mathds{Z}_8$ subgroup generated by $1/\sqrt{2}(\mathds{1}+i\sigma_z)$ has eight elements, the remaining eight can straightforwardly be found by acting with $\pm i \sigma_x$ and $\pm i\sigma_y$ on $\tilde{\sigma}$. These elements can be partitioned into seven conjugacy classes according to \cite{2007PhRvL..98j0401S}
\begin{align} 
 {\rm CC}_1(D^*_4) &= \{ (I, n_w) \} \\
 {\rm CC}_2(D^*_4) &= \{ (-I, n_w) \} \\
 {\rm CC}_3(D^*_4) &= \{ ( \pm i \sigma_x, n_w), ( \pm i \sigma_y, n_w) \} \\
 {\rm CC}_4(D^*_4) &= \{ (\pm i \sigma_z, n_w) \} \\
 {\rm CC}_5(D^*_4) &= \{ (\Tilde{\sigma}, n_w + \frac{1}{2}), (-i \sigma_z \Tilde{\sigma}, n_w + \frac{1}{2}) \} \\
 {\rm CC}_6(D^*_4) &= \{ (-\Tilde{\sigma}, n_w + \frac{1}{2}), (i \sigma_z \Tilde{\sigma}, n_w + \frac{1}{2}) \}  \\
 {\rm CC}_7(D^*_4) &= \{ (\pm i \sigma_x \Tilde{\sigma}, n_w + \frac{1}{2}), (\pm i \sigma_y \Tilde{\sigma}, n_w + \frac{1}{2}) \}
\label{conjclassD4}
\end{align}

The conjugacy class structure and the centralizer UIRs of this group can be found by applying the same set of arguments as in the $T^*$ case, so we are simply jumping straight to the character tables here. Again, the abelian fluxons ${\rm CC}_1(D_4^*)$ and ${\rm CC}_2(D_4^*)$ have the entire group $D_4^*$ as centralizer whose character table is provided in Table~\ref{reptableD4}.
\begin{table}
\caption{Character table for $\Lambda^{D^*_4}_i$, where $i=1,2,3,4$ are one-dimensional, $i=5$ is two-dimensional and $i=6,7$ are four-dimensional.}\label{reptableD4}
\begin{tabular}{|c|c|c|c|c|c|c|c|}
 \hline
  & ${\rm CC}_1(D^*_4)$ & ${\rm CC}_2(D^*_4)$ & ${\rm CC}_3(D^*_4)$ & ${\rm CC}_4(D^*_4)$ & ${\rm CC}_5(D^*_4)$ & ${\rm CC}_6(D^*_4)$ & ${\rm CC}_7(D^*_4)$\\ 
 \hline
  $\Lambda^{D^*_4}_1$ & 1 & 1 & 1 & 1 & 1 & 1 & 1\\  
 \hline
 $\Lambda^{D^*_4}_2$ & 1 & 1 & -1 & $1$ & $1$ & $1$ & $-1$\\ 
 \hline
 $\Lambda^{D^*_4}_3$ & 1 & 1 & 1 & $1$ & $-1$ & $-1$ & $-1$\\
 \hline
 $\Lambda^{D^*_4}_4$ & 1 & 1 & -1 & $1$ & $-1$ & $-1$ & $1$\\ 
 \hline
 $\Lambda^{D^*_4}_5$ & 2 & -2 & 0 & $0$ & $-2$ & $0$ & $0$\\
 \hline
 $\Lambda^{D^*_4}_6$ & $4$ & $-4$ & $0$ & $0$ & $2 \sqrt{2}$ & $-2 \sqrt{2}$ & $0$\\
 \hline
  $\Lambda^{D^*_4}_7$ & $4$ & $-4$ & $0$ & $0$ & $-2 \sqrt{2}$ & $2 \sqrt{2}$ & $0$\\
 \hline
\end{tabular}
\end{table}
The dyons with flux corresponding to ${\rm CC}_1$ and ${\rm CC}_2$ are consequently all bosons since all of the characters in the first and second column are integers. The centralizer of ${\rm CC}_3(D^*_4)$ and ${\rm CC}_7(D^*_4)$ is given by $\mathds{Z}_4$, which coincides with the centralizer of ${\rm CC}_3(T^*)$, whose character table is already provided in Table~\ref{reptableZ4}. Finally, the centralizer for the remaining conjugacy classes ${\rm CC}_4(D_4^*)$, ${\rm CC}_5(D_4^*)$ and ${\rm CC}_6(D_4^*)$ are given by the cyclic abelian group $\mathds{Z}_8$ whose character table is provided in Table \ref{reptableZ8}.
\begin{table}
\caption{Character table for the six one-dimensions UIRs $\Gamma^{\mathds{Z}_8}_i$ ($i=1,2,3,4,5,6,7,8$) and $z$ is the generator of $\mathds{Z}_8$ and $\theta = e^{i \frac{2 \pi}{8}}$.}
\label{reptableZ8}
\begin{tabular}{|c|c|c|c|c|c|c|c|c|}
 \hline
  & $1$ & $z$ & $z^2$ & $z^3$ & $z^4$ & $z^5$ & $z^6$ & $z^7$\\ 
 \hline
  $\Gamma^{\mathds{Z}_8}_1$ & 1 & 1 & 1 & 1 & 1 & 1 & 1 & 1\\  
 \hline
 $\Gamma^{\mathds{Z}_8}_2$ & 1 & $-1$ & $1$ & $-1$ & $1$ & $-1$ & $1$ & $-1$\\ 
 \hline
 $\Gamma^{\mathds{Z}_8}_3$ & 1 & $i$ & $-1$ & $-i$ & $1$ & $i$ & $-1$ & $-i$\\  
 \hline
 $\Gamma^{\mathds{Z}_8}_4$ & 1 & $-i$ & $-1$ & $i$ & $1$ & $-i$ & $-1$ & $i$\\
 \hline
 $\Gamma^{\mathds{Z}_8}_5$ & 1 & $\theta$ & $\theta^2$ & $\theta^3$ & $\theta^4$ & $\theta^5$ & $\theta^6$ & $\theta^7$\\
 \hline
 $\Gamma^{\mathds{Z}_8}_6$ & 1 & $\theta^7$ & $\theta^6$ & $\theta^5$ & $\theta^4$ & $\theta^3$ & $\theta^2$ & $\theta$\\
 \hline
 $\Gamma^{\mathds{Z}_8}_7$ & 1 & $\theta^3$ & $\theta^6$ & $\theta$ & $\theta^4$ & $\theta^7$ & $\theta^2$ & $\theta^5$\\
 \hline
 $\Gamma^{\mathds{Z}_8}_8$ & 1 & $\theta^5$ & $\theta^2$ & $\theta^7$ & $\theta^4$ & $\theta$ & $\theta^6$ & $\theta^3$\\
 \hline
\end{tabular}
\end{table}

We deduce from these characters non-abelian dyons with fractional topological spins $s=\frac{1}{8},\frac{1}{4},\frac{3}{8},\frac{5}{8},\frac{3}{4},\frac{7}{8}$.

\subsubsection{A note on the quaternionic phase and its quantum double $\mathfrak{D}(Q^*_8)$}
\label{quaternion}

In a separate work \cite{2022QuIP...21...31G}, devoted entirely to the  $\mathfrak{D}(Q^*_8)$ model, we analysed the feasibility of employing the quaternionic phase of a spin-2 BEC as a TQC platform. This phase can be obtained from the $D^*_4$ one by reducing its rotational symmetry to a 4-cycle. For the sake of completeness, we are here outlining some key aspect of this model. One of the primary challenges of developing a theoretical model of a TQC based on the non-abelian phases considered in this work, is that the dimensionality of the fusion spaces are generally larger than two. This is problematic since we only need a two-level system to form a qubit, meaning that the redundant particles in the fusion outcomes will generally absorb amplitude and consequently cause information leakage \cite{2003PhRvA..67b2315M,PhysRevB.75.165310}. It is therefore of interest to find a low-temperature symmetry group whose quantum double contains fusion products that form two-level states. The quaternionic phase offers such a model if one consider a closed subset of the particles in the theory. 

A natural question that may arise is why we do not want to utilize all of the particles in a fusion outcome to represent a generic qudit. The issue with this idea, as thoroughly discussed in \cite{2022QuIP...21...31G}, is that one would need to have access to a group whose algebra is of a higher dimension. For instance, the braid group of a four-level qudit still only have two generators $\sigma_1$ and $\sigma_2$, which is insufficient to span an $\rm{SU}(4)$ Bloch sphere. Instead, by considering two qubits, which combined form a four-level system, the braid group has five generators, which may thus make universality more attainable. For instance, as already touched upon in subsection \ref{vqcd} and demonstrated in \cite{genetay2023topological}, a topological gravity may be simulated via the quantum pressure in the fluid, which results in additional topologically protected phase rotations which one may supplement any non-universal braid set with. Only one such additional phase gate would be required in order to make the $\mathfrak{D}(Q^*_8)$ universal. Another advantage of this model is that all fusion rules are multiplicity free. As already argued in \cite{2010JPhA...43M5205A}, the presence of multiplicities is complicating the calculation of the Clebsch--Gordan coefficients of the theory. This is also apparent in our derivation presented in Section~\ref{CGderivation} since the number of terms on the right hand side in Eq.\eqref{solution} depends on the multiplicity. Only in the multiplicity-free case is an explicit solution accessible. 


\subsection{Symmetry-based normal mode analysis}
\label{NMA}
As indicated by the Fourier duality connecting the fluxons and the chargeons, if one of the two is considered to be localized in real space, its dual must be localized in the reciprocal space, with respect to a generalized quantum Fourier transform (the S-matrix). While it is, as per homotopy theory, well established that the fluxons must map onto quantized vortices, what the physical incarnations of chargeons are is less clear. The purpose of this section is therefore to offer an interpretation of the chargeon degrees of freedom. We find, by means of a symmetry analysis, that these normal modes correspond to spin-rotations and spin-waves, or \textit{magnons}. These systems also have regular phonon excitations but those are naturally associated with the breaking of translational symmetry. Generally speaking, a system in equilibrium exhibits some sort of symmetry that, if broken, excites modes in the system. These solutions transform according to the various UIRs of the, possibly broken, symmetry group, and as such, furnish a basis for the vector spaces the matrices in images of the UIRs act on. Guided by these considerations, we may now proceed and work out the normal modes in a spinor BEC. 

\subsubsection{Tetrahedral phase -- an example}

\begin{figure}[htp!]
    \centering
    \includegraphics[width = \textwidth]{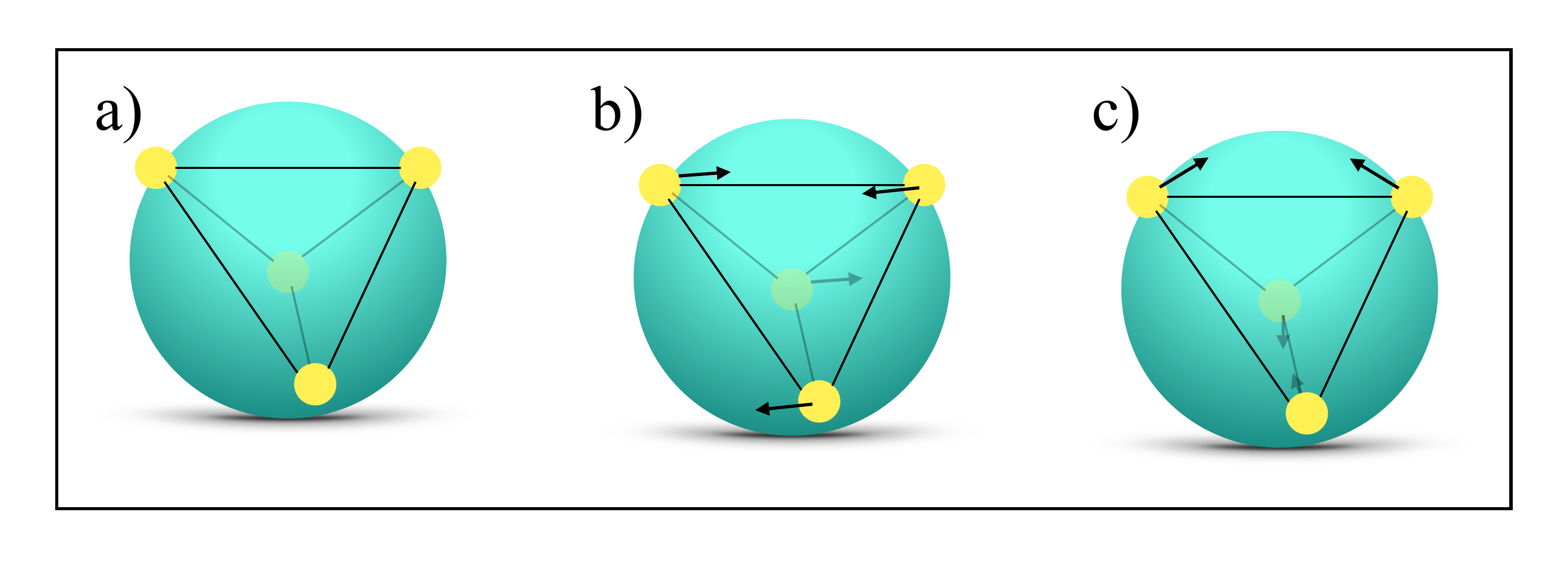}
    \caption{(a) Spin-nodes within the Majorana star representation of the tetrahedral phase (b) spin-rotation mode where the whole tetrahedron is simply rotating (c) spin-wave mode where the nodes oscillate out-of-phase (this mode is three-fold degenerate).}
    \label{spinwaves}
\end{figure}

Here we will study the tetrahedral phase to provide an explicit example. In particular we shall consider the regular tetrahedral group $T$ instead of its binary representation $T^*$ previously considered, since this group has a simpler geometric structure which will facilitate the discussion and analysis. To be more explicit in our formulation, we consider the Majorana star representation of the order parameter in which the spin nodes are located on the corners of a tetrahedron, see Fig.~\ref{spinwaves}. The problem of finding the modes of motion of such a system can be cast into the form of an eigenvalue problem 
\begin{equation}
\label{EOM}
    \hat{A} \bar{\theta} = \omega^2 \bar{\theta},
\end{equation}
where the matrix $\hat{A}$ describes the energy carried by the system and $\hat{\theta}$ is a displacement vector, i.e. the normal mode. Before we move on and work out these modes, we shall make pertinent remarks regarding the vector space formed by the modes. First, note that a tetrahedron has four corners to which four distinct spin nodes are attached. Secondly, note that each node can move in two perpendicular angular directions $\alpha$ and $\beta$ (Euler angles), and consequently, the system must have $4\times 2 = 8$ degrees of freedom in total. The resulting vector space constructed from the space of node labels and the vectors describing the displacements of the nodes must consequently have a tensor structure $V = V_{nodes} \otimes V_{direction}$. This eight-dimensional vector space has a basis of the form $\bar{\theta} = (\alpha_1,\beta_1,\alpha_2,\beta_2,\alpha_3, \beta_3,\alpha_4,\beta_4)^T$, where $\alpha_i$ and $\beta_i$ are the angular coordinates relative to the equilibrium position of each node. Due to the uniqueness of the solution to Eq.~\eqref{EOM}, given some initial state vector $\bar{\theta_i}$, we know that the evolution of the system is completely determined by these values at time $t=0$. Further, since the eigenvalue spectrum of $\hat{A}$ is invariant under $T$, we may consider the action of the image of the full eight-dimensional representation on $V$. Generally, this representation is, owing to the tensor structure of $V$, reducible, which implies that it can be constructed from a set of irreducible orthogonal subblocks
\begin{equation}
    \Gamma(T) = \bigoplus_i m_i \Gamma_i (T),
\end{equation}
where $m_i$ is the multiplicity of block $i$. Since the various eigenmodes $\bar{\theta}_i$ of $A$  are orthogonal and inhabit distinct blocks of $V$,  they must transform according to the UIRs of $T$. Hence, the action of $\Gamma(T)$ can be decomposed as
\begin{equation}
    \Gamma(T) \sum_i\bar{\theta}_i = m_1 \Gamma_1 (T) \bar{\theta}_1 + m_2 \Gamma_2 (T) \bar{\theta}_2 +...+ m_n \Gamma_n (T) \bar{\theta}_n.
\end{equation}
In light of these considerations, we may establish that the eigenvalues attached to each subspace $V_i$ cannot be altered by $\Gamma(T)$, that is to say
\begin{equation}
    (\Gamma_i(T) \hat{A}) \bar{\theta_i} = \Gamma_i (T) (\hat{A} \bar{\theta_i}) = \omega_i^2 (\Gamma_i (T) \bar{\theta_i}) = \omega_i^2 \bar{\theta'_i} \ \ \forall i.
\end{equation}

What we learn from this is that there is one unique eigenvalue $\omega_i^2$ attached to each UIR, and that this eigenvalue must be the same for $\bar{\theta}_i'$ and $\bar{\theta}_i$. Moreover, the dimension of the subspace $V_i$, and hence the degeneracy of $\omega_i^2$, must be equal to the dimension of the UIR $\Gamma_i$. We have now reduced the problem of finding the normal modes of the system to finding the various irreducible representations of the underlying symmetry group. Remarkably, the number of distinct eigenvalues and their degeneracies can be completely deduced from the representation theory, without any consideration of the physical parameters; the entire measurable spectrum of possible eigenvalues is classified by the symmetry of the system. Note that this is in complete agreement with the quantum double particle labelling. The fluxons are point particles labelled by the conjugacy classes of the group, and the chargeons are labelled by the various UIRs (or the reverse if we act with the S-matrix and consider the system in reciprocal space).

Again we consider the character table of $T$ as it contains all of the pertinent information required in order to work out the normal modes. A tetrahedron is constructed from four triangular faces, thus comprising $4 \times 3 = 12$ rotational symmetries in total. It has four conjugacy classes corresponding to the identity $I$, all $2 \pi/3$-rotations $R(2 \pi/3)$, all $4 \pi/3$-rotations $R(4 \pi/3)$, in addition to $\pi$-rotations $R(\pi)$. Let us now deduce the UIRs. First, we always have a trivial one-dimensional representation $\Lambda_1$ sending all elements to unity. Then, we must also have another one-dimensional representation $\Lambda_2$ sending all of the $2 \pi/3$-rotations to $e^{i \frac{2 \pi}{3}}$. We can further permute this representation giving us a map $\Lambda_3$ that sends all of the $2 \pi/3$-rotations to $e^{i \frac{4 \pi}{3}}$. Again, using the corollary to Burnside's theorem which states that the order of a finite group is equal to the sum of the representation dimensions squared, we can conclude that there must also exist a representation of dimension $\sqrt{12-1^2-1^2-1^2} = 3$. This is, of course, the three-dimensional representation acting on the $\mathds{R}^3$ space in which the tetrahedron is embedded. By virtue of the above considerations, in addition to the orthogonality condition enforced on the rows in a character table, we may construct the complete character table of the group which is presented in Table~\ref{tetrahedron}.
\begin{table}
\caption{Character table for the the one-dimensions UIRs $\Lambda_i$ ($i=1,2,3$) and the three dimensional one $\Gamma_4$.}
\begin{tabular}{|c|c|c|c|c|}
 \hline
  & $I$ & $R(2 \pi/3)$ & $R(4 \pi/3)$ & $R(\pi)$\\ 
 \hline
  $\Lambda_1$ & 1 & 1 & 1 & 1\\  
 \hline
 $\Lambda_2$ & 1 & $e^{i\frac{\pi}{3}}$ & $e^{i\frac{2 \pi}{3}}$ & $1$\\ 
 \hline
 $\Lambda_3$ & 1 & $e^{i\frac{2 \pi}{3}}$ & $e^{i\frac{\pi}{3}}$ & $1$\\ 
 \hline
 $\Lambda_4$ & $3$ & $0$ & $0$ & $-1$\\
 \hline
\end{tabular}
\label{tetrahedron}
\end{table}
Directing our attention to the fist column of Table~\ref{tetrahedron}, reveals that the characters add up to six, which means that the combined dimension of the UIRs is six since these characters correspond to the traces of the identity element. This means that there must be some multiplicities $m_i$ involved since the full representation $\Gamma(T)$ is eight-dimensional. All these multiplicities can be found by projecting the UIRs onto the full representation through the equation
\begin{equation}
    m_i = \frac{1}{|T|} \sum_{t \in T} \chi_i^* (t) \chi_{\textrm full} (t),
\end{equation}
where $\chi_i$ and $\chi_{\rm full}$ are the characters of the $i$'th UIR and the full representation, respectively, and $t\in T$. This is a somewhat cumbersome, yet straight forward, calculation to perform so we will not carry it out here as the normal modes can be found without any knowledge of these multiplicities. As already noted in the previous cases, the number of UIRs must be the same as the number of conjugacy classes, that is four, and consequently we must have four normal modes in total. This can be understood intuitively by considering the geometry of the tetrahedron. The three nodes can either rotate about the three axes or the nodes can oscillate in an out-of-phase fashion. Moreover, we see from the character table that the $\Lambda_3$ UIR must be three-dimensional since the character corresponding to the identity is three. From this we can infer that the corresponding mode must be three-fold degenerate as these degenerate modes span the irreducible subspace of $\Lambda_3$. Again, we can confirm this result through some simple geometrical considerations. The tetrahedron has four vertices in total so if one node is pairing up with one other node, the remaining two nodes are forced to pair up as well, but since the first node has three choices of partner, and this choice automatically fixes the choices of the remaining nodes, we can only have three inequivalent configurations of pair-vibrations. We have now, guided by the symmetry of the system, completely determined the normal mode spectrum as well as the degeneracies without any knowledge of the physical parameter values. We found three non-degenerate spin-rotation modes and one spin-wave mode with three-fold degeneracy corresponding to the out-of-phase oscillations of the nodes. The strategy deployed here is completely generic and can as such be applied to systems exhibiting any type of symmetry, to obtain the normal mode spectra.

\section{Vortex anyons in superfluid gases of cold atoms}
To draw a connection between the theoretical concepts and experiments, we briefly mention two closely related physical systems, superfluid Fermi gases and Bose--Einstein condensates, that may be able to host non-abelian anyons inside the cores of quantised vortices. We will first discuss abelian vortices, routinely created and observed in experiments, and how to manipulate them before contemplating their respective non-abelian extensions yet to be realised in the laboratory.

\subsection{Quantised vortices in cold atomic superfluids}

Consider a generic complex valued scalar order parameter 
\begin{equation}
\psi({\bf r}) = |\psi({\bf r})| e^{iS({\bf r})},
\end{equation}
which may be used for modeling, for instance, ${\rm U}(1)$ symmetric scalar BECs. The corresponding first homotopy group is $\mathbb Z$ so the topological excitations, called quantised vortices, that form the particles of this theory are labelled by an integer winding number $n\in \mathbb Z$. The superfluid velocity $\boldsymbol{v}_s = \hbar \nabla S / m^*$, where $S({\bf r})$ is the real valued phase of the order parameter and $m^*$ is the mass of the constituent particles comprising the superfluid. For a simple BEC $m^*=m$ where $m$ is the mass of a bosonic atom, while for a simple superfluid Fermi gas $m^*=2m$ due to Cooper pairing of fermionic atoms of mass $m$. The circulation of the superflow is quantised according to
\begin{equation}
\Gamma = \oint \boldsymbol{v}_s \cdot d{\bf l} = \kappa 2\pi n
\end{equation}
where $\kappa =h/m^*$ is the quantum of circulation such that for the typical $n=\pm 1$ vortices the phase $S({\bf r})$ of the superfluid has a $\pm 2\pi$ phase winding around the vortex core. Since such scalar vortices are the elements of an abelian group, they are not suitable candidates for topological quantum computation. However, they can provide laboratory demonstrations of all the required actions from vortex pair creation, braiding and fusion, required to perform braid based topological quantum information processing using the more exotic flavor of non-abelian vortices discussed in later subsections.  

Scalar vortices were created and observed in Bose--Einstein condensates in 1999 \cite{Matthews1999a} and in superfluid Fermi gas in 2005 \cite{Zwierlein2005a}. Since then, they have been routinely observed in numerous cold atom laboratories using a variety of techniques. Further details about vortex experiments in cold atom superfluids may be found for instance in Refs.~\cite{Anderson2010a,Fetter2009a,Simula2019a}.

In both bosonic and fermionic cases the vortices, if left to equilbrate, will typically arrange into a regular Wigner-crystalline triangular vortex lattice due to the repulsive 2D-Coulomb-like interaction between the vortices.

\subsection{Non-abelian vortices in a superfluid Fermi gas}
Here we briefly outline how the non-abelian anyons may emerge in the vortex cores of chiral p-wave paired superfluid Fermi gas. This situation is generic and believed to be relevant to other topological fermionic superfluids such as in liquid $^3$He and certain Type II superconductors \cite{Kopnin1991a,PhysRevB.61.10267,Gurarie2007a,2007PhRvL..98a0506T,Mizushima2008a,Li2022a,Masaki2023a}. 

A canonical transformation of spinless fermion fields to the Bogoliubov quasi-particle basis
\begin{equation}
\left( 
\begin{matrix}
\Psi({\bf r}) \\
\Psi^\dagger({\bf r})
\end{matrix}
\right)
=
\sum_q
\left( 
\begin{matrix}
u_q({\bf r}) & -v_q^*({\bf r}) \\
v_q({\bf r}) & u_q^*({\bf r})
\end{matrix}
\right)
\left( 
\begin{matrix}
\gamma_q\\
\gamma^\dagger_q
\end{matrix}
\right)
\end{equation}
enables the fermionic second-quantised Hamiltonian to be expressed as an effective field theory in terms of quasi-particles. The quasi-particle eigenstates with amplitudes $u_q,v_q$ and energies $E_q$ are described by the Bogoliubov-deGennes equation 
\begin{equation}\label{BdG}
\left( 
\begin{matrix}
\mathcal{L}({\bf r}) & \Delta({\bf r}) \\
\Delta^*({\bf r}) & -\mathcal{L}^*({\bf r})
\end{matrix}
\right)
\left( 
\begin{matrix}
u_q({\bf r})  \\
v_q({\bf r}) 
\end{matrix}
\right)
=E_q
\left( 
\begin{matrix}
u_q({\bf r}) \\
v_q({\bf r})
\end{matrix}
\right),
\end{equation}
where the structure of the operator $\mathcal{L}({\bf r})$ on the diagonal as well as the off-diagonal pair potential $\Delta({\bf r})$  depend on the specifics of the physical model considered. 
The BdG equations are particle hole symmetric such that for every eigenstate with energy $E$, there is another, negative energy $-E$, eigenstate obtained via the transformation $(u_{-E},v_{-E})^T = (v^*_{E},u^*_{E})^T $. This further implies that $\gamma^\dagger_E = \gamma_{-E}$, that is, creating a quasi-particle with energy $E$ is equivalent to annihilating a quasi-hole with energy $-E$.

In the case where the superfluidity emerges in a chiral p-wave paired state, the gap function for a single physical vortex in an axisymmetric system may be expressed as a vortex order parameter $\Delta({\bf r}) = |\Delta(r)|e^{in\theta}$, and the conclusions may be generalized to multiple vortices \cite{Mizushima2010a}. Under such circumstances the system possesses an exact zero energy solution $E_q=0$ with the quasi-particle probability density peaking inside the vortex. This yields the defining property of a Majorana quasi-particle
\begin{equation}
\gamma^\dagger _{E=0}= \gamma_{E=0}
\end{equation}
stating that a Majorana zero mode is its own antiparticle. The Majorana zero mode is a topologically protected zero energy Caroli--deGennes--Matricon (CdGM) quasi-particle mode familiar from ordinary type II superconductors where they occur with non-zero energies. Since this quasi-particle is trapped by the vortex core, it and the host vortex are sometimes used interchangeably and referred to as the Majorana fermion (quasi-particle).  

Majorana zero modes come in pairs. For odd numbers of vortices, one vortex mode is paired with a zero energy edge mode and the remaining even number of vortices form paired Majorana quasi-particle bound states. Considering two vortices hosting two "real" Majoranas $\gamma_1$ and $\gamma_2$, one can construct "complex" Dirac fermion operators $c = 1/2 (\gamma_1+i\gamma_2)$ and $c^\dagger = 1/2 (\gamma_1-i\gamma_2)$ characterized by the usual Fermi statistics. Taking a product of these relations yields
\begin{equation}
i\gamma_1\gamma_2 = 2 c^\dagger c -1
\end{equation}
which provides a pedagogical link between the quasi-particle creation and annihilation operators and the Majorana Ising anyon fusion rules 
\begin{equation}
\sigma\otimes\sigma={\bf 1}\oplus \psi,
\end{equation}
where $\sigma$ maps onto $\gamma$, $\bf 1$ corresponds to the vacuum $|0\rangle$ and $\psi$ corresponds to a regular fermion $|1\rangle$. The number operator $c^\dagger c$ may take two possible values $0,1$ corresponding to respective states $|0\rangle$ and $|1\rangle$ resulting in the two values for the Majorana parity $i\gamma_1\gamma_2 =\pm 1$. More rigorous analysis leads to the conclusion that the Majorana zero mode vortices indeed correspond to realisations of the ${\rm SU}(2)_2$ Ising anyons discussed in Sec.~\ref{isingtqc} \cite{nayak2008non}.

From theoretical perspective, vortices must be well separated to avoid quasi-particle tunnelling between vortex cores, which would cause energy splitting of the Majorana pairs. From an experimental perspective, the greatest unresolved challenge is to realise a suitable topological superfluid such as a spinless p-wave paired superfluid phase of a Fermi gas. Ultracold Fermi gases near a p-wave Feshbach resonance, have been studied experimentally \cite{Zhang2004,Schunck2005,Gaebler2007,Fuchs2008,Inada2008,Maier2010,Luciuk2016,Gerken2019} and offer a potential route to realising chiral p-wave superfluidity, however, they suffer from inelastic losses which are enhanced near the resonance \cite{Waseem2019} Lower-dimensional gases \cite{Levinsen2008,Kurlov2017,Muhammed2018,Fonta2020,Chang2020}, optical lattices \cite{Venu2023} and even nonequilibrium \cite{Foster2013} systems provide promising opportunities for studying Fermi gases with resonantly enhanced p-wave interactions and remain the subject of ongoing research.

\subsection{Non-abelian vortices in a superfluid Bose gas}
Here we briefly outline how non-abelian anyons may emerge in the vortex cores \cite{Kobayashi2009a,mawson2018braiding} of $F=2$ spinor Bose--Einstein condensates  \cite{UedaRMP,2010arXiv1001.2072K,2015arXiv151101575M}. A canonical transformation of bosonic fields to Bogoliubov quasi-particle basis
\begin{equation}
\left( 
\begin{matrix}
\Psi({\bf r}) \\
\Psi^\dagger({\bf r})
\end{matrix}
\right)
=
\sum_q
\left( 
\begin{matrix}
u_q({\bf r}) & v_q^*({\bf r}) \\
v_q({\bf r}) & u_q^*({\bf r})
\end{matrix}
\right)
\left( 
\begin{matrix}
\gamma_q\\
\gamma^\dagger_q
\end{matrix}
\right)
\end{equation}
is similar but has a sign difference in comparison to the fermionic case. In the case where the atoms condense in a state with a hyperfine spin $F$, the quasi-particle eigenmodes have $2F+1$ components such that for $F=2$ they are five component spinors. The BEC itself is an exact Nambu--Goldstone zero mode satisfying $E=0$ and $u=v^*$, where the energy is measured with respect to the chemical potential. It is therefore convenient to discuss the ground state topology in terms of the ground state order parameter $\Psi = [u_{-2},u_{-1},u_{0},u_{1},u_{2}]^T$, where the subscripts refer to the five spin-projections of the hyperfine spin. For instance, in the cyclic ground state phase, the state respects the symmetry of the non-abelian binary tetrahedral group and therefore the topological vortex excitations (fluxons) are determined by the conjugacy classes of the group with the quantum double construction leading to the additional chargeon and dyon degrees of freedom as clarified in Sec.~\ref{spin2}. 

In contrast to the fermionic case where the role of the Majorana vortices is merely to trap and host the core localised quasiparticles, which define the qubit of the non-abelian fusion algebra, in the quantum double construction of spinor BECs both the carrier vortices (fluxons) and the core localised quasiparticles (chargeons) participate in the fusion process.  As such, the Majorana zero modes may be viewed as the counterparts of pure chargeons of the quantum double.

Even the simplest non-abelian quantum double of $F=2$ BEC, realised by the quaternion phase \cite{2022QuIP...21...31G}, leads to fusion rules of higher complexity in comparison to the Ising anyon model, such as
\begin{equation}
\Sigma_x\otimes\Sigma_x={\bf 1}\oplus \rho_x\oplus {\bar \rho}_y\oplus \bar{\rho}_z,
\end{equation}
where the result of fusing two vortices (pure fluxons) $\Sigma_x$ may yield either a vacuum or three different types of pure chargeons. 
The physical interpretation of this fusion rule that defines a 4-qudit is thus that the two vortices annihilate leaving behind one of the four possible chargeons, including the trivial one.
This arises due to the fact that the order parameter around the vortex core is now a five component spinor such that there may exist multiple distinct ways spin currents and superflow mass currents can be combined along a path encircling a vortex core that nevertheless result in the same topological vortex invariant. This is to be contrasted with a simple vortex in a scalar BEC where the only degree of freedom along the path around the vortex is a single number, the phase S(r) of the order parameter, that uniquely defines the topology of the vortex flux.

\subsection{Experimental considerations regarding creation, fusion and braiding}

It is useful to first consider the most abundant laboratory system of a simple scalar BEC with a quantised abelian vortex fluxon. Such vortices are readily produced in research laboratories via a variety of methods \cite{Anderson2010a,Fetter2009a,Simula2019a}, including the so-called ``chopsticks'' method \cite{Samson2016a}, which enables deterministic production of vortices and control over their positions, thereby conceptually facilitating pair creation, braiding and fusion protocols for vortex anyons. Recently, this method has been utilised to develop a programmable vortex collider in a superfluid Fermi gas \cite{Kwon2021a}. In principle, these techniques can be extended to larger numbers of vortices using optical tweezer arrays, as being deployed for the storing Rydberg atom qubits in neutral atom quantum computers \cite{SaffmanRMP,Graham2022a,Bluvstein2022a}.

As the vortices are moved around during braiding, one potential concern is the effect of phonon excitations. Although phonons cannot change the topology of isolated vortices, it is in principle possible for a high-energy phonon to split into a vortex-antivortex pair via a Sauter--Schwinger-like pair creation process and such vortices could then accidentally become braided as the computation is carried out leading to quasiparticle poisoning of the computation. The probability of such processes should, however, be negligible at low enough temperatures achievable in the experiments. In the spinor BECs it is also conceivable that an absorption of a phonon by the vortex might affect the atomic probability distribution the different hyperfine spin levels of the core localised modes. This could potentially alter the amplitudes of the quantum double fusion rules. However, we are not aware of any works that would have addressed this issue.
\\
Regarding the readout of the fusion outcome, the greatest experimental challenge is to observe the vortex core localised quasiparticles, which has not been achieved to date even in the case of a scalar BEC. It was only recently discovered \cite{PhysRevA.104.L061301} that even in a scalar BEC a vortex core hosts a zero energy BdG quasiparticle mode known as a kelvon \cite{PhysRevLett.101.020402,PhysRevA.69.043617}. Such a kelvon can be regarded as a quantum mechanical equivalent of a classical Kelvin wave \cite{doi:10.1080/14786448008626897}. In a 3D scalar BEC kelvons have been indirectly observed \cite{PhysRevLett.90.100403} as they distort the shape of a vortex line into a helix. However, in 2D it is not clear how to experimentally resolve the zero energy kelvon, which can be viewed as a quantum depleted non-condensate atom that is trapped by the vortex core and as such is inseparable from the vortex fluxon. It may thus be tempting to interpret the kelvon as the charge of a quantum double anyon. Indeed, from our analysis we found that all pure fluxons are in fact bosons, while the dyons generally have fractional charge. Therefore the non-abelian anyons in a spinor BEC are always charge-flux composites. We may therefore conjecture that the chargeon part of a dyon corresponds to the spin mode of the kelvon attached to the vortex. This view is supported by numerical experiments carried out in \cite{mawson2018braiding} where spin waves were observed after a fusion event involving two vortices. Even if a vortex is annihilated by an anti-vortex, they may still have a charge that survives the fusion process. In a realistic model of TQC based on vortices in a spinor BEC one would then build the qubits from vortices and their kelvons. 
\\
Thus the technologies to create, braid, and fuse vortex anyons have been demonstrated in the case of abelian vortices. There are two remaining major problems that need to be resolved in order to achieve a complete experimental demonstration of a non-abelian vortex based topological qubit in a cold atom superfluid. The first is the stabilisation of a ground state whose order parameter would support non-abelian vortices as its natural excitations. For instance, it is not presently known if the non-abelian cyclic, biaxial nematic, or the quaternion phases can be stabilized as ground states \cite{UedaRMP}. 
The second, which has already been discussed, is the development of experimental techniques that allow probing the kelvon and to measure its charge, a task that is yet to be realised even for vortices in existing quasi-two-dimensional scalar BECs.

\section{Discussion}

We have aimed at providing a self-contained account of the emergent low-temperature quantum double structure in spinor BECs, a connection that is not well covered in the literature to date. It is well understood that spinor BECs may support non-abelian anyons in the form of quantized vortices (fluxons). However, as emphasized in this paper, this is just one side of the story. The existence of a particle-vortex duality, whose structure is made exact within the quantum double framework, implies that the description of vortices can be mapped onto a dual side, which describes a different type of particle (chargeons). We have shown that the chargeons belonging to the dual side correspond to spin waves and spin rotation modes. As illustrated, despite being non-abelian, the fluxons only have fractional mutual statistics whereas their self-statistics is bosonic. The same is true for the chargeons since their algebraic structure is, owing to the duality, equivalent. However, the quantum double structure also accommodate charge-flux composites which, as demonstrated, may indeed have fractional self-statistics. 

The strong interest in non-abelian anyons is primarily due to their potential to realize fault-tolerant quantum information processing. Up until now, the main focus has been on fermionic systems, and in particular systems that may host Majorana fermion quasi-particles. Spinor BECs, on the other hand, appear to have been overlooked in this context, which prompted us to carry out this work with the aim of shedding light on their potential for TQC. Theoretically, as we discussed in Section~\ref{quaternion}, one of the main hurdles in the development of a spinor BEC-based TQC platform is that the fusion spaces generally are large due to the high dimensionality. To construct a qubit, only two states are needed, which implies that information leakage \cite{2003PhRvA..67b2315M,PhysRevB.75.165310} may occur into the states not belonging to the computational space. Another issue is the possibility of multiplicities in the fusion rules, which complicates severely the calculation of the Clebsch--Gordan coefficients \cite{2010JPhA...43M5205A} and thus the braid group matrices. Similar issues are, of course, also present in the $\rm{SU}(2)_k$ anyon models but mainly for larger values of $k$. To reduce the chance of having to face such problems, it is useful to consider smaller groups since that will restrict the dimensionality of the Hilbert spaces spanned by the anyons of the theory. For instance, the quaternionic subgroup $\mathds{Q}_8$ which was considered in a separate work \cite{2022QuIP...21...31G}, can be obtained by breaking further the rotational part of the $D_4^*$ subgroup to a 4-cycle. For this group it is possible to define multiplicity free qubits that are not accompanied by additional non-computational states that generally cause leakage.

Finally, although a scalable fault tolerant topological quantum computer based on non-abelian vortex anyons remains out of reach, a neutral atom superfluid prepared in a non-abelian ground state could certainly facilitate a beautiful demonstration of the fundamental principles of topological quantum computation utilising a few topologically protected logical qubits.

\section*{Competing interests}
The authors declare that there are no competing interests.

\section*{Author's contributions}
T.S. conceived the idea for the work and contributed to the writing of the manuscript. E.G.-J. carried out the analytical studies and composed the first draft of the manuscript. C.V. provided expertise on the experimental aspects. All authors contributed substantially to the work.

\begin{acknowledgements}
 This research was supported by the Australian Research Council Future Fellowship FT180100020,
and was funded by the Australian Government.
\end{acknowledgements}

\section*{Data availability}
All essential data is included in the article.

\begin{appendix}

\section{Appendix A}
\subsection{The fundamental group}
\label{FG}
A detailed description of the fundamental group can be found in standard literature on topology, see e.g. \cite{AT}. Here we merely provide a brief exposition of its rudimentary elements. Consider a space $X$ of some topology, and let $x_0 \in X$ be an arbitrary point in $X$. Let $f_1$ and $f_2$ be two loop functions 
\begin{equation}
    f_1,f_2: \ [0,1] \rightarrow X
\end{equation}
belonging to different homotopy classes and let $f_1$ start at $x_0$ and $f_2$ end at $x_0$, such that $f_1(1)=f_2(0)=x_0$. Then, one may compose $f_1$ and $f_2$ according to the parametrization 
$$
            f_1 \circ f_2(t)= \left\{
                \begin{array}{ll}
                 f_1(2t), \ 0\leq t \leq \frac{1}{2}  \\
                 
                 f_2(2t-1), \ \frac{1}{2} \leq t \leq 1
                \end{array}\right.
$$
Given this composition rule, the set of homotopy classes $\{[f_1],[f_2],...,[f_n]\}$ forms a group of loops
$$\pi_1 (x_0,X)=\{[f_1],[f_2],...,[f_n]\},$$
which all start and end at $x_0\in X$. This group is known as the first homotopy group or the fundamental group.

\section{Appendix B}
\subsection{Representation theory and characters}

\subsubsection{Representation theory of a group $G$}

Here we outline the elements of representation theory that are relevant to this work. For a more thorough treatment we recommend e.g. Ref.~\cite{alma9936959566101361}. In physics, the main mathematical object of interest is that of a vector space. In particular, in quantum physics, the states of a system are eigenstates of a Hamiltonian, which spans a Hilbert space $\mathcal{H}$. In general, $\mathcal{H}$ has the structure of a direct sum since each subspace, labeled by some quantum number, is orthogonal to all other subspaces labelled by other quantum numbers. A symmetry $G$ of the system necessarily commutes with $H$ so that the quantum numbers are invariant under the action of $G$. Thus, in order to study this action we need to find a set of matrices that implement the elements of $G$ acting on $\mathcal{H}$. The map $\Gamma$ from $G$ to this set of matrices 
$$\Gamma: \ \ G \longrightarrow \rm{GL}(N),$$
must preserve the structure of $G$ (homomorphic), i.e.
$$\Gamma(ab) = \Gamma(a)\Gamma(b), \ \ \forall a,b \in G,$$
and is known as a representation, where $\rm{GL}(N)$ is the general linear group of matrices of dimension $N$. If the matrices in the image of $\Gamma$ are block diagonal, they may be decomposed as a direct sum into matrices that are not block diagonal. Hence, the representations that map onto matrices that cannot be decomposed in this way are called irreducible representations. In quantum physics we are, since normalization must be preserved, primarily concerned with unitary irreducible transformations so in this context $\Gamma(a) \in \rm{U}(N)$ for all $a \in G$, where $\Gamma$ is a unitary irreducible representation (UIR).

\subsubsection{Representation theory of the quantum double $\mathfrak{D}(H)$ of $H$}\label{QDrep}

A quantum double Hilbert space is spanned by the charge-flux states 

$$\{| h_i^{\rm CC},u^{\Gamma}_j \rangle \}_{0 \leq i \leq |\rm CC|}^{0 \leq j \leq dim(\Gamma_j)} \subset \mathds{C}[H] \otimes \mathfrak{F}[H],$$ 

where $h_i^{\rm CC} \in \rm CC$ is defined as $h_i^{\rm CC} = a_i h^{\rm CC}_* a_i^{-1}$ for $a_i \in H$ and $h^{\rm CC}_*$ is the representative element of the conjugacy class $\rm CC$. Note that we have a lot of freedom in our choice here. Any element of $h^{\rm CC}_* \in \rm CC$ can be chosen as the representative, and similarly, there might be several choices of $a_i$, all of which belong to the same equivalence class, which maps $h^{\rm CC}_*$ to $h_i$ under conjugation. The quantum double action, which performs a gauge transformation followed by a flux measurement, is implemented as
\begin{equation*}
    \Pi_{\rm CC}^{\Gamma}(P_h g) | h_i^{\rm CC}, u^{\Gamma}_j \rangle = \delta_{h, g h_i g^{-1}} | g h_i^{\rm CC} g^{-1}, \Gamma(g^*) u^{\Gamma}_j \rangle,
\end{equation*}
where $ g h_i^{\rm CC} g^{-1} = h^{\rm CC}_k$ and $g^*=a_k^{-1} g a_i$ is defined as the gauge transformation that commutes with the representative element $h^{\rm CC}_* \in \rm CC$. Thus, finding the representation $\Pi_{\rm CC}^{\Gamma}$ is equivalent to finding the matrices that are implementing this action on $\mathds{C}^d$. Here we shall outline the steps to achieve this. First, find all conjugacy classes $\rm CC_i$ of the group $H$ and their centralizers $Z(\rm CC_i)$, which are defined as the subgroups of elements that are commuting with $\rm CC_i$, and pick one arbitrary element $h^{\rm CC_i}_*$ from each conjugacy class $\rm CC_i$ as the representative element and work out the UIRs $\Gamma_j$ of each $Z(\rm CC_i)$. Then find the set of elements $\{a_j\}_{j=1}^{|\rm CC_i|}$ such that $h_k = a_k h_*^{\rm CC_i} a_k^{-1}$ for each $h_k \in \rm CC_i$. These elements belong to the various cosets formed by $H/Z(\rm CC_i)$. Next, explicitly define a basis to work with for the Hilbert space spanned by the states $| h_i^{\rm CC},u^{\Gamma}_j \rangle$. The dimension $d$ of this Hilbert space equals $d=dim(\Gamma) \times |\rm CC|$ so any set $\{\vec{v}_k \}_{k=1}^d$ of orthonormal vectors $\vec{v}_k \in \mathds{C}^d$ constitutes an orthonormal basis. Given such a basis, the action $\Pi_{\rm CC}^{\Gamma}(P_h g) | h_i^{\rm CC}, u^{\Gamma}_j \rangle$ is implemented as $\pi(h,g h^{\rm CC}_i g^{-1}) \Gamma(g^*)$ on the space spanned by $\{\vec{v}_k \}_{k=1}^d$, where $\pi(h,g h^{\rm CC}_i g^{-1})$ is a $d$-dimensional matrix projecting $g h^{\rm CC}_i g^{-1}$ onto $h$ and $\Gamma(g^*)$ is a $d$-dimensional matrix implementing the gauge transformation $g^*$. Note that in the case when $H$ is non-abelian, it may be possible to further decompose $g^*$ since the group has more than one generator. For instance, in the case of a dihedral group $D_n$, an arbitrary element is a combination of a rotation $r$ and a reflection $t$, such that $\Gamma(g^*) = \Gamma(t^m) \Gamma(r^n)$, where $m \in \{0,1 \}$ and $n \in \mathds{Z}_n$. Splitting the action up in this way makes it easier to construct the quantum double matrices since one can simply multiply the projection with the gauge transformation, whose matrix representations are easier to find.

\subsubsection{Characters}

The character $\chi_{\Gamma}(a)$ is defined as the trace of $a$ in the representation $\Gamma$
$$\chi_{\Gamma}(a) = \rm{Tr}[\Gamma(a)].$$
Due to the cyclic property of the trace, any two elements belonging to the same conjugacy class must have the same character since $\rm{Tr}[b a b^{-1}] = \rm{Tr}[b^{-1} b a] = \rm{Tr}[a]$ for all $a,b \in G$. Consequently, when computing the characters, it is only necessary to pick one representative from each conjugacy class. A fundamental theorem in representation theory, known as Burnside's theorem \cite{burnside}, states that the number of irreducible representations of a group equals the number of conjugacy classes in the group. Thus, the characters of the conjugacy classes in the various UIRs can be summarized in a table $\chi_{\Gamma_i,\rm CC_j}$, where the rows correspond to the UIRs $\Gamma_i$ and the columns to the conjugacy classes $\rm CC_j$. An important feature of the rows $\bar{\chi}_{\Gamma_m}$ in such a table is that they are orthonormal with respect to the inner product
$$\frac{1}{|G|}\langle \bar{\chi}_{\Gamma_m}, \bar{\chi}_{\Gamma_n} \rangle = \delta_{m,n}$$
which is nothing but the scalar multiplication of the two sets, where $\delta_{m,n}$ is the Kronecker delta. Moreover, since the size of the character sets always coincide with the Hilbert space dimension, they constitute an orthonormal basis for this space.

\section{Appendix C}
\subsection{Hopf algebras --- a unified framework}
\label{HA}
Here we supplement the physical picture of quantum doubles described in this work with a more detailed exposition of the underlying algebra. In particular, we did not discuss the generalization to multi-particle systems in much detail. Here we will mainly follow \cite{hopf} and \cite{bais1980flux}. A Hopf algebra over a field $\mathds{K}$ is an algebraic structure consisting of the tuple $(A,m,\eta,\Delta,\epsilon,\mathcal{S})$, where $(A,m,\eta)$ forms an algebra, and $(A,\Delta,\epsilon,\mathcal{S})$ a coalgebra, over $\mathds{K}$. For our purposes we are primarily interested in the coalgebra and the field $\mathds{K}$ is the field of quantum double elements, i.e. flux projections and gauge transformations $P_h g \in \mathfrak{D}(H)$. We adopt the following notation for multi-particle Hilber spaces. Consider two generic dyons with flux and charge $(\rm {CC}_i,\Lambda_i)$ and charge $(\rm {CC}_j,\Lambda_j)$, respectively, forming the Hilbert spaces $[V_{\rm {CC}}\otimes V_{\Lambda}]_i$ and $[V_{CC}\otimes V_{\Lambda}]_j$, then their combined fusion space is given by
$$[V_{\rm{CC}}\otimes V_{\Lambda}]_i \otimes [V_{\rm{CC}}\otimes V_{\Lambda}]_j.$$
We now need to know how the image of a quantum double representation acts on such a multi-particle Hilbert space. The extension to multiple particles can be achieved by incorporating the so-called coproduct map $\Delta$ of the Hopf algebra.

\subsubsection{Coproduct (fusion)}

The coproduct $\Delta$ of a Hopf algebra is a morphism
$$\Delta: \ \ \ \ \mathfrak{D}(H) \longrightarrow \mathfrak{D}(H) \otimes \mathfrak{D}(H),$$
so if we consider the quantum double action defined in Eq.~\eqref{QDaction}, it can be extended under the map by $\Delta$, i.e.
$$\Pi_{(\rm{CC}_i,\Lambda_i)} \otimes \Pi_{(\rm{CC}_j,\Lambda_j)}(P_h g) = \sum_{h_i h_j = h} \Pi_{(\rm{CC}_i,\Lambda_i)}(P_{h_i} g) \otimes \Pi_{(\rm{CC}_j,\Lambda_i)}(P_{h_j} g),$$
where the sum is carried out over all $h_i \in \rm{CC}_i$ and $h_j \in \rm{CC}_j$ satisfying $h_i h_j = h$. Hence, the two-particle quantum double action is implementing a gauge transformation $g$ on the individual Hilbert spaces and projects out the total flux $h$, so that the topological charges are conserved globally under fusion. This action satisfies the important coassociativity property under composition $\circ$
$$(I \otimes \Delta) \circ \Delta(P_h g) = (\Delta \otimes I) \circ \Delta(P_h g) = \\ \sum_{h_i h_j h_k = h} P_{h_i}g \otimes P_{h_j}g \otimes P_{h_k}g.$$
Hence, the coproduct of the Hopf algebra allows for a treatment of systems comprised of multiple generic dyons. We will now apply this map to construct the braid matrix $\mathcal{R}$.

\subsubsection{Braiding}

When interchanging two generic dyons, the charge is gauge transformed and the flux is projected out. Hence, the interchange operator $R$ must be an element of $\mathfrak{D}(H) \otimes \mathfrak{D}(H)$, acting on the flux-charge Hilbert space. More specifically, the operator $R$ implements a gauge transformation $g$ on the first dyon by the flux of the second dyon such that

\begin{equation}
    R = \sum_{h} \sum_g P_g \otimes P_h g.
\end{equation}

Thus, for a system comprised of three dyons which is required to form a qubit, we have two $R$ operators $R_1$ and $R_2$ which must satisfy the so-called quasi-triangularity conditions
\begin{equation*}
   R\Delta (P_h g)=\Delta(P_h a) R
\end{equation*}
\begin{equation*}
   (I \otimes \Delta)(R)=R_1 R_2
\end{equation*}
\begin{equation*}
   (\Delta \otimes I)(R)=R_2 R_1.
\end{equation*}
The first of these conditions assures that the global topological charge is conserved since it commutes with the quantum double elements. The other two relations state that if the fusion outcome of two generic dyons is independent of whether they were braided by a third dyon, before and after the fusion, respectively. These properties are essential since the Yang--Baxter equation \cite{1989IJMPA...4.3759J} can not be satisfied if they are not enforced. It is straight forward to check that indeed $R_1 R_2 R_1 = R_2 R_1 R_2$, which is required for any algebraically consistent anyon model. Thus, given a two-dyon representation, the action on the corresponding Hilbert space $V_{j_i}\otimes V_{j_j}$ is given by a matrix $\mathcal{R}_{j_i , j_j}^{j_k}$  
$$ \mathcal{R}_{j_i,j_j}^{j_k} = \sum_{m_i,m_j} \sum_{m_p,m_q} \sigma_{m_i,m_q}^{m_j,m_p} \circ [\Pi_{j_i} \otimes \Pi_{j_j} ](R),$$
where $j_i$ and $j_j$ are labelling the generalized angular momenta of the dyons that are being interchanged and $j_k$ that of their fusion channel. Further, $\sigma_{m_i,m_q}^{m_j,m_p}$ are the components of the of the permutation operator, which simply correspoinds to de-coupling followed by a re-coupling where the two dyons are swapped
$$\sigma_{m_i,m_q}^{m_j,m_p} = \left[ \begin{array}{cc|c}
    j_{i} & j_{j} & j_{k} \\
    m_{i} & m_{j} & m_{k} \\
    \end{array}
    \right] \left[ \begin{array}{c|cc}
    j_{k} & j_{j} & j_{i} \\
    m_{k} & m_{p} & m_{q} \\
    \end{array}
    \right],$$
and $m_i$ is the generalized topological magnetization. The elements denoted by the brackets are nothing but the Clebch--Gordan coefficients for which in Appendix~\ref{CGderivation} we shall derive an analytical expression. Now, the braid matrices $\sigma_1$ and $\sigma_2$ at single qubit level are given by $\sigma_1 = \mathcal{R}$ and
$\sigma_2 = \mathcal{F} \mathcal{R} \mathcal{F}^{-1}$, where $\mathcal{F}$ is the change of basis operator \cite{pachos2012introduction,preskill1999lecture}. The action of $\mathcal{F}$ is simply changing the order of fusion such that if $\ket{(j_i j_j) j_k;j_l}$ represents the states corresponding to the fusion channels $j_l$, formed when fusing $j_i$, $j_j$ and $j_k$, then $\mathcal{F}\ket{(j_i j_j) j_k;j_l} = \ket{j_i (j_j j_k);j_l}$, where the anyons within the parenthesis are fused first. The matrix elements of this operator can also be expressed as a series of de-couplings and re-couplings given the Clebsch--Gordan coefficients of the theory, such that
\begin{align}
\noindent
 &[\mathcal{F}^{j_q}_{j_i,j_j,j_k}]_{j_l}^{j_p} = \sum_{m_i,m_j,m_k,m_q,m_p} \left[ \begin{array}{cc|c}
    j_{i} & j_{j} & j_{l} \\
    m_{i} & m_{j} & m_{l} \\
    \end{array}
    \right]\left[ \begin{array}{cc|c}
    j_{l} & j_{k} & j_{q} \\
    m_{l} & m_{k} & m_{q} \\ 
    \end{array}  
    \right] 
    \times
    \nonumber \\ & \left[ \begin{array}{c|cc}
    j_{q} & j_{p} & j_{i} \\
    m_{q} & m_{p} & m_{i} \\
    \end{array}
    \right]\left[ \begin{array}{c|cc}
    j_{p} & j_{j} & j_{k} \\
    m_{p} & m_{j} & m_{k} \\
    \end{array}
    \right].
\end{align}
Note that the coproduct $\Delta$ can be applied iteratively, which allows for multiple n-dyonic representations to be constructed.

\subsubsection{Counit (vacuum)}

Next, let us consider the counit $\epsilon$ which is an algebra morphism
$$\epsilon: \ \ \ \mathfrak{D}(H) \longrightarrow \mathds{C}$$
such that
$$\epsilon \otimes I: \mathfrak{D}(H) \otimes \mathfrak{D}(H) \longrightarrow \mathds{C} \otimes \mathfrak{D}(H) \simeq \mathfrak{D}(H).$$
This statement is equivalent to
$$(\epsilon \otimes I)\circ \Delta = (I \otimes \epsilon)\circ \Delta = I \otimes I.$$
Hence, the counit maps $\mathfrak{D}(H)$ to the vacuum sector so that the fusion is trivial.

\subsubsection{Antipod (anti-particles)}

Finally, we shall discuss the antipod $\mathcal{S}$ which is another morphism
$$\mathcal{S}: \ \ \ \mathfrak{D}(H) \longrightarrow \mathfrak{D}(H)$$
such that
$$I \otimes \mathcal{S}: \ \ \ \mathfrak{D}(H) \otimes \mathfrak{D}(H) \longrightarrow \mathds{C}$$
or equivalently
$$(\mathcal{S} \otimes I)\circ \Delta = (I \otimes \mathcal{S})\circ \Delta = I \otimes \epsilon = \epsilon \otimes I$$
which shows that the antipode can interpreted as anti-particles. Next we shall derive analytically an expression for the quantum double Clebsch--Gordan coefficients used in this section.

\section{Appendix D}
\subsection{Derivation of the quantum double Clebsch-Gordan coefficients}\label{CGderivation}

Consider the expansion of the coupled basis $\ket{j_{k},m_k}$  in the product basis $\ket{j_i, m_i} \otimes \ket{j_j, m_j}$
\begin{equation}
    \ket{j_k, m_k} = \sum\limits_{m_i m_j} \sum_n \left[
    \begin{array}{cc|c}
    j_{i} & j_{j} & j_{k} \\
    m_{i} & m_{j} & m_{k} \\
    \end{array}
    \right]_n \ket{j_i, m_i} \otimes \ket{j_j, m_j},
\end{equation}
where $n$ denotes the multiplicity and let us define a projection operator $\hat{P}^{j_k}_{m_k m_l}$ that is projecting out the $m_k'$s component of the tensored basis
\begin{equation}
\label{project}
    \hat{P}^{j_k}_{m_k m_l} \ket{j_i, m_i} \otimes \ket{j_j, m_j} = \left[ \begin{array}{cc|c}
    j_{i} & j_{j} & j_{k} \\
    m_{i} & m_{j} & m_{l} \\
    \end{array}
    \right]_n ^* \ket{j_k, m_k}.
\end{equation}
The explicit form of the projection operator is given by the inner product 
\begin{equation}
    \hat{P}^{j_k}_{m_k m_l} = \frac{d_{j_k}}{|H|} \sum\limits_{h, g} \Lambda^{j_k}_{m_k m_l} (P_h g)^* \Lambda^{j_i \otimes j_j}(P_h g),
\end{equation}
where $P_h g$ is a generic quantum double element, that is, a gauge transformation $g \in Z(h)$ followed by a flux measurement $P_h$. $H$ is the residual symmetry group and $d_{j_k}$ is the dimension of the representation $j_k$. Moreover, $\Lambda^{j_i \otimes j_j}$ represents the reducible tensored representation and $\Lambda^{j_k}$ the irreducible component we wish to project onto. Now, in order compute the action of this operator on the tensored state, we need to expand the quantum double element with the coproduct 
\begin{equation}
\label{coproduct}
    \Delta:\     \ \mathfrak{D}(H) \longrightarrow \mathfrak{D}(H) \otimes \mathfrak{D}(H)
\end{equation}
such that
\begin{equation}
    \Delta(P_h g) = \sum\limits_{h' h'' = h} P_{h'} g \otimes P_{h''}g,
\end{equation}
which results in
\begin{equation}
  \Lambda^{j_i \otimes j_j}(\Delta(P_h g)) = \sum\limits_{h' h'' =h} \Lambda^{j_i}(P_{h'}g) \otimes \Lambda^{j_j}(P_{h''} g) .
\end{equation}
The action on $\ket{j_i, m_i} \otimes \ket{j_j, m_j}$ can now be expressed as
\begin{align}
\label{expand}
    &\hat{P}^{j_k}_{m_k m_l} \ket{j_i, m_i} \otimes \ket{j_j, m_j}=\\
    &\frac{d_{j_k}}{|H|} \sum\limits_{h, g} \Lambda^{j_k}_{m_k, m_l} (P_h g)^* \sum\limits_{h' h'' = h} \sum\limits_{m_q, m_p} \Lambda^{j_i}_{m_i m_q}(P_{h'}g) \Lambda^{j_j}_{m_j m_q}(P_{h''} g) \ket{j_i, m_q}\otimes\ket{j_j, m_p}. \nonumber
\end{align}
We may further expand Eq. \eqref{project} in the $\ket{j_i, m_q}\otimes\ket{j_j, m_p}$ basis. Going from the coupled basis $\ket{j_k, m_k}$ to the product basis $\ket{j_i, m_q}\otimes\ket{j_j, m_p}$ can be achieved through
\begin{equation}
    \ket{j_k, m_k} = \sum\limits_{m_q, m_p} \sum_n \left[ \begin{array}{cc|c}
    j_{q} & j_{p} & j_{k} \\
    m_{q} & m_{p} & m_{k} \\
    \end{array}
    \right]_n \ket{j_i, m_q} \otimes \ket{j_j, m_p},
\end{equation}
and substituting this into Eq.~\eqref{project} we obtain
\begin{equation}
\label{twoCG}
    \hat{P}^{j_k}_{m_k m_l} \ket{j_i, m_i} \otimes \ket{j_j, m_j} = \sum\limits_{m_q, m_p} \sum_n \left[ \begin{array}{cc|c}
    j_{i} & j_{j} & j_{k} \\
    m_{i} & m_{j} & m_{l} \\
    \end{array}
    \right]_n ^* \left[ \begin{array}{cc|c}
    j_{q} & j_{p} & j_{k} \\
    m_{q} & m_{p} & m_{k} \\
    \end{array}
    \right]_n \ket{j_i, m_q} \otimes \ket{j_j, m_p}.
\end{equation}

Setting Eq.~\eqref{expand} and Eq.~\eqref{twoCG} equal to one another yields 
\begin{equation}
\label{solution}
   \sum_n \left[ \begin{array}{cc|c}
    j_{i} & j_{j} & j_{k} \\
    m_{i} & m_{j} & m_{l} \\
    \end{array}
    \right]_ n^* \left[ \begin{array}{cc|c}
    j_{q} & j_{p} & j_{k} \\
    m_{q} & m_{p} & m_{k} \\
    \end{array}
    \right]_n = \frac{d_{j_k}}{|H|} \sum\limits_{h, g} \Lambda^{j_k}_{m_k m_l} (P_h g)^* \sum\limits_{h' h'' = h} \Lambda^{j_i}_{m_i m_q}(P_{h'}g) \Lambda^{j_j}_{m_j m_q}(P_{h''} g).
\end{equation}
As evident by the above equation, solutions are far more accessible in the multiplicity free case since we only have one term on the left hand side. Therefore we let $n=1$ to provide an analytical formula for the case with no multiplicities. Finally, by letting $m_i = m_p$, $m_j = m_p$ and $m_k = m_l$, and taking the square root, the Clebsch--Gordan coefficients can be expressed as
\begin{equation}
\label{diagelem}
    \left[ \begin{array}{cc|c}
    j_{i} & j_{j} & j_{k} \\
    m_{i} & m_{j} & m_{k} \\
    \end{array}
    \right] = \sqrt{\frac{d_{j_k}}{|H|} \sum\limits_{h, g} \Lambda^{j_k}_{m_k m_k} (P_h g)^* \sum\limits_{h' h'' = h} \Lambda^{j_i}_{m_i m_i}(P_{h'}g) \Lambda^{j_j}_{m_j m_j}(P_{h''} g)}.
\end{equation}
The remaining solutions not pertaining to the diagonal elements are obtained by simply dividing Eq.~\eqref{solution} by Eq.~\eqref{diagelem}, which brings us to the final result
\begin{equation}
    \left[ \begin{array}{cc|c}
    j_{q} & j_{p} & j_{k} \\
    m_{q} & m_{p} & m_{k} \\
    \end{array}
    \right]_{(m_i,m_j,m_k)} = \sqrt{\frac{d_{j_k}}{|H|}}\frac{ \sum\limits_{h, g} \Lambda^{j_k}_{m_k m_l} (P_h g)^* \sum\limits_{h' h'' = h} \Lambda^{j_i}_{m_i m_q}(P_{h'}g) \Lambda^{j_j}_{m_j m_q}(P_{h''} g)}{\sqrt{ \sum\limits_{h, g} \Lambda^{j_k}_{m_k m_k} (P_h g)^* \sum\limits_{h' h'' = h} \Lambda^{j_i}_{m_i m_i}(P_{h'}g) \Lambda^{j_j}_{m_j m_j}(P_{h''} g)}}
\end{equation}

This result is the same as the authors in \cite{2006CoPhC.174..903R} arrived at for regular (non-quantum double) finite groups. In order to obtain the quantum double Clebsch--Gordan coefficients from their derivation one simply has to expand the tensored representation according to the coproduct defined in Eq.~\eqref{coproduct}.

\section{Appendix E}
\subsection{Mean-field theory for spin-$S$ BECs}\label{mft}

In topological phases of matter excitations are classified according to the fundamental group over the some order parameter manifold. In a spinor BEC, within the mean-field approximation, the dynamics is governed by the Hamiltonian \cite{2010arXiv1001.2072K}
\begin{align}\label{spinGPE}
    H = \underbrace{\int dr \bigg[\sum_{m,m'=-S}^S \psi_m^*(r,t)\Big(-\frac{\hbar^2}{2M}\nabla^2 + V(r) - q[f_z]_{m m'} + p[f_z^2]_{m m'}\Big)\psi_{m'}(r,t)}_{\text{single particle energy}}+\\
    \underbrace{\frac{c_0}{2}n^2(r,t)+\frac{c_1}{2}|\vec{F}(r,t)|^2+\frac{c_2}{2}|A(r,t)|^2}_{\text{interaction energy}}\bigg],
\end{align}
where $m,m'$ denote the spin components, $p,q$ are external magnetic fields coupling linearly and quadratically, respectively, to the wavefunction via the z-component of the spin-$F$ matrix $[f_z]_{m m'}$ and $c_i$ ($i=1,2$) are couplings to the spin density $\vec{F} = \sum_{m,m'=-S}^S \psi^*_m [f_a]_{m m'} \psi_{m'}$ and spin-singlet pair amplitude $A(r)$. The Hamiltonian describing the various phases can thus be obtained by tuning the parameters to the desired values \cite{2010arXiv1001.2072K,mawson2018braiding}.
Here particles belonging to different states of spin participate in the current thus resulting in a multi-component velocity
\begin{equation}
    [{\boldsymbol v}_s]_{a} = \frac{\hbar}{2Mni} \sum_{m,m'=-S}^S [f_a]_{m m'}\Big(\psi_m^*(r,t)(\nabla \psi_{m'}(r,t))-(\nabla \psi_{m}^*(r,t))\psi_{m'}(r,t)\Big),
\end{equation}
where $a$ denotes the components if the spin, which serves as a non-abelian gauge field in contrast to the simple abelian phase gradient $v_s = \frac{\hbar}{2m} \nabla \theta$ of a scalar BEC. When computing the monodromy of such an object the gauge field $A^a_{\mu}$ implements a non-abelian transformation from which the particles spin can be deduces. In particular, in a low-energy phase with residual symmetry group $H \subset \rm{SO}(3)$, the first homotopy group is isomorphic to the subgroup itself \cite{1995hep.th...11201D} $\pi_1(\rm{U}(1) \times \rm{SO}(3)/H) \simeq H$, which is why the non-abelian vortices are labelled by the conjugacy classes of $H$. 

\section{Appendix F}
\subsection{Fourier analysis over arbitrary groups and the S-matrix}

Here we provide the basics elements of Fourier transforms on finite groups and algebras. For a more detailed analysis, the reader may consult e.g. the book \cite{terras1999fourier}. Let $G$ be a finite group and let $\Lambda$ be a representation of $G$, then the Fourier transform $\Tilde{F}_{\Lambda}$ of a function $\psi \in L^2 (G)$, with respect to $\Lambda$, is given by the inner product projection
$$\Tilde{F}_{\Lambda}(\psi) = \sum_{g\in G} \psi(g) \Lambda(g).$$
Note that in the case when $\Lambda$ is one-dimensional (abelian), we have the equality  $\chi_{\Lambda}(g)={\Lambda}(g)$, where $\chi_{\Lambda}(g)$ denotes the character of $g$ in the image of $\Lambda$, since the trace is trivial. As a concrete example, let us consider a system with discrete $N$-fold cyclic translational symmetry $T_a$, generated by $a$, such as in a one-dimensional crystal with periodic boundary conditions. A representation $\Lambda$ in this case is a map $\Lambda: \ \ T_a \longrightarrow \rm{U}(1)$, which implies that ${\chi}_{\Lambda}(g)= \Lambda (g)\in \rm{U}(1)$, which is nothing but the conventional discrete Fourier transform where the set $\{ e^{i\frac{2 \pi n}{N}}\}_{n=0}^{N-1}$ forms an orthornormal basis for the Hilbert space. Thus, given a system with a Hamiltonian which is invariant under $G$, the set of characters $\{\chi_{\Lambda_i}(g)\}_{g\in G}$ with respect to a representation $\Lambda_i$ of $G$, constitutes a basis vector corresponding to the subspace $\mathcal{H}_i \subset \mathcal{H}$, where $\mathcal{H} = \bigoplus_i \mathcal{H}_i$ is the full Hilbert space, and $i$ denotes the invariant label (quantum number) of the subspace $\mathcal{H}_i$. The sets of characters are always orthonormal and have a size equivalent to $\dim(L^2 (G))$ meaning that they always furnish a basis for the Hilbert space of the system.

\subsubsection{Modular S-matrix}

Now let us consider the quantum double $\mathfrak{D}(H)$ of some discrete group $H$. The matrices in the image of a representation $\Pi$ of $\mathfrak{D}(H)$ act on a Hilbert space $\mathcal{H}$ spanned by the sets of characters corresponding to those matrices. Here $\mathcal{H}=\mathds{C}[H] \otimes \mathfrak{F}[H]$ where $\mathds{C}[H]$ is spanned by the pure fluxon states and $\mathcal{F}[H]$ by the pure chargeon states. The Fourier transform relating these two vector spaces is defined by the modular S-matrix \cite{1988NuPhB.300..360V}
\begin{equation*}
    S^{\Gamma \Lambda}_{AB} = \frac{1}{|H|} \sum_{h_A \in A, h_B \in B} {\rm{Tr}[\Gamma(g_A^{-1} h_B g_A)]}^* \rm{Tr}[\Lambda(g_B^{-1} h_A g_B)]^*,
\end{equation*}
where the capital letters denote the conjugacy classes and $\Gamma$ and $\Lambda$ the centralizer UIRs. This formula can be interpreted as an inner product between two generic dyonic states, whose matrix action is expanding one state in the basis of the other. Note that this expression is symmetric under the interchange of labels implying that it is implementing a duality (a particle-vortex duality) swapping the fluxon part and the chargeon part. In this respect, the modular S-matrix can be viewed as a generalized character table for the quantum double where the rows and columns simultaneously represent the conjugacy classes and UIRs.

\end{appendix}

\bibliographystyle{unsrt}
\bibliography{main}

\begin{thebibliography}{100}

\bibitem{landau1937theory}
L.~D. Landau.
\newblock On the theory of phase transitions.
\newblock {\em Zh. Eksp. Teor. Fiz.}, 11:19--32, 1937.

\bibitem{arovas1985statistical}
D.P. Arovas, R.~Schrieffer, F.~Wilczek, and A.~Zee.
\newblock Statistical mechanics of anyons.
\newblock {\em Nucl. Phys. B}, 251:117--126, 1985.

\bibitem{wilczek1990fractional}
F.~Wilczek.
\newblock {\em Fractional statistics and anyon superconductivity}, volume~5.
\newblock World scientific, Singapore, 1990.

\bibitem{chen1989anyon}
Y.H. Chen, F.~Wilczek, E.~Witten, and B.~Halperin.
\newblock On anyon superconductivity.
\newblock {\em Int. J. Mod. Phys. B}, 3(07):1001--1067, 1989.

\bibitem{leinaas1977theory}
J.~M. {Leinaas} and J.~{Myrheim}.
\newblock {On the theory of identical particles}.
\newblock {\em Nuovo Cimento B Serie}, 37(1):1--23, 1977.

\bibitem{1982IJTP...21..467F}
R.P. {Feynman}.
\newblock {Simulating Physics with Computers}.
\newblock {\em Int. J. Theor. Phys.}, 21(6-7):467--488, 1982.

\bibitem{2010qcqi.book.....N}
M.A. {Nielsen} and I.L. {Chuang}.
\newblock {\em {Quantum computation and quantum information}}.
\newblock Cambridge University Press, 2010.

\bibitem{kitaev2009topological}
A.~{Kitaev} and C.~{Laumann}.
\newblock {\em {Topological phases and quantum computation}}.
\newblock {Oxford University Press}, 2009.

\bibitem{field2018introduction}
B.~Field and T.~Simula.
\newblock Introduction to topological quantum computation with non-abelian
  anyons.
\newblock {\em Quantum Sci. Tech.}, 3(4):045004, 2018.

\bibitem{pachos2012introduction}
J.K Pachos.
\newblock {\em Introduction to topological quantum computation}.
\newblock Cambridge University Press, 2012.

\bibitem{nayak2008non}
C.~Nayak, S.H. Simon, A.~Stern, M.~Freedman, and S.D. Sarma.
\newblock Non-abelian anyons and topological quantum computation.
\newblock {\em Rev. Mod. Phys.}, 80(3):1083, 2008.

\bibitem{wang2010topological}
E.C. {Rowell} and Z.~{Wang}.
\newblock {Mathematics of topological quantum computing}.
\newblock {\em Bull. Amer. Math. Soc.}, 55:183, 2018.

\bibitem{PhysRevB.61.10267}
N.~Read and D.~Green.
\newblock Paired states of fermions in two dimensions with breaking of parity
  and time-reversal symmetries and the fractional quantum {H}all effect.
\newblock {\em Phys. Rev. B}, 61:10267--10297, 2000.

\bibitem{witten1989quantum}
E.~Witten.
\newblock Quantum field theory and the {J}ones polynomial.
\newblock {\em Commun. Math. Phys.}, 121(3):351--399, 1989.

\bibitem{1992PhR...213..179I}
R.~{Iengo} and K.~{Lechne}.
\newblock {Anyon quantum mechanics and Chern-Simons theory}.
\newblock {\em Phys. Rep.}, 213(4):179--269, 1992.

\bibitem{1999tald.conf..177D}
G.~V. {Dunne}.
\newblock {Course 3: Aspects of Chern-Simons Theory}.
\newblock In {\em Topological Aspects of Low Dimensional Systems}, volume~69,
  page 177, 1999.

\bibitem{aharonov1959significance}
Y.~Aharonov and D.~Bohm.
\newblock Significance of electromagnetic potentials in the quantum theory.
\newblock {\em Phys. Rev.}, 115(3):485, 1959.

\bibitem{berry1984quantal}
M.V. Berry.
\newblock Quantal phase factors accompanying adiabatic changes.
\newblock {\em Proc. R. Soc. Lond.}, 392(1802):45--57, 1984.

\bibitem{2013PhRvB..87w5120G}
C.~{Gils}, E.~{Ardonne}, S.~{Trebst}, D.~A. {Huse}, A.~W.~W. {Ludwig},
  M.~{Troyer}, and Z.~{Wang}.
\newblock {Anyonic quantum spin chains: Spin-1 generalizations and topological
  stability}.
\newblock {\em Phys. Rev. B}, 87(23):235120, 2013.

\bibitem{2020arXiv200810790G}
E.~G\'enetay~Johansen and T.~Simula.
\newblock Fibonacci anyons versus {M}ajorana fermions: A {M}onte {C}arlo
  approach to the compilation of braid circuits in $\mathrm{SU}(2{)}_{k}$ anyon
  models.
\newblock {\em PRX Quantum}, 2:010334, 2021.

\bibitem{fan2010braid}
Z.~Fan and H.~de~Garis.
\newblock Braid matrices and quantum gates for {I}sing anyons topological
  quantum computation.
\newblock {\em Eur. Phys. J. B}, 74(3):419--427, 2010.

\bibitem{bombin2010topological}
H.~Bomb{\'\i}n.
\newblock Topological order with a twist: Ising anyons from an abelian model.
\newblock {\em Phys. Rev. Lett.}, 105(3):030403, 2010.

\bibitem{2007PhRvL..98a0506T}
S.~{Tewari}, S.~{Das Sarma}, C.~{Nayak}, C.~{Zhang}, and P.~{Zoller}.
\newblock {Quantum Computation using Vortices and {M}ajorana Zero Modes of a
  p$_{x}$+ip$_{y}$ Superfluid of Fermionic Cold Atoms}.
\newblock {\em Phys. Rev. Lett.}, 98(1):010506, 2007.

\bibitem{2001PhRvL..86..268I}
D.~A. {Ivanov}.
\newblock {Non-Abelian Statistics of Half-Quantum Vortices in p-Wave
  Superconductors}.
\newblock {\em Phys. Rev. B}, 86(2):268--271, 2001.

\bibitem{PhysRevB.97.104501}
Z.~Wang and K.R.A. Hazzard.
\newblock Analytic ground state wave functions of mean-field ${p}_{x}+i{p}_{y}$
  superconductors with vortices and boundaries.
\newblock {\em Phys. Rev. B}, 97:104501, 2018.

\bibitem{Gurarie2007a}
V.~Gurarie and L.~Radzihovsky.
\newblock Zero modes of two-dimensional chiral $p$-wave superconductors.
\newblock {\em Phys. Rev. B}, 75:212509, 2007.

\bibitem{Mizushima2008a}
T.~Mizushima, M.~Ichioka, and K.~Machida.
\newblock Role of the {M}ajorana fermion and the edge mode in chiral
  superfluidity near a $p$-wave {F}eshbach resonance.
\newblock {\em Phys. Rev. Lett.}, 101:150409, 2008.

\bibitem{kasahara2018majorana}
Y.~Kasahara, T.~Ohnishi, Y.~Mizukami, O.~Tanaka, S.~Ma, K.~Sugii, N.~Kurita,
  H.~Tanaka, J.~Nasu, and Y.~Motome.
\newblock {M}ajorana quantization and half-integer thermal quantum {H}all
  effect in a {K}itaev spin liquid.
\newblock {\em Nat. Phys.}, 559(7713):227, 2018.

\bibitem{zuo2016detecting}
Z.-W. {Zuo}, H.~{Li}, L.~{Li}, L.~{Sheng}, R.~{Shen}, and D.~Y. {Xing}.
\newblock {Detecting {M}ajorana fermions by use of superconductor-quantum Hall
  liquid junctions}.
\newblock {\em Europhys. Lett.}, 114(2):27001, 2016.

\bibitem{PhysRevLett.94.166802}
S.~Das~Sarma, M.~Freedman, and C.~Nayak.
\newblock Topologically protected qubits from a possible non-abelian fractional
  quantum {H}all state.
\newblock {\em Phys. Rev. Lett.}, 94:166802, 2005.

\bibitem{2019NatCo..10.5128Z}
H.~{Zhang}, D.E. {Liu}, M.~{Wimmer}, and L.P. {Kouwenhoven}.
\newblock {Next steps of quantum transport in Majorana nanowire devices}.
\newblock {\em Nat. Commun.}, 10:5128, 2019.

\bibitem{2019arXiv191104512P}
E.~{Prada}, P.~{San-Jose}, M.~W.~A. {de Moor}, A.~{Geresdi}, E.J.~H. {Lee},
  J.~{Klinovaja}, D.~{Loss}, J.~{Nyg{å}rd}, R.~{Aguado}, and L.P.
  {Kouwenhoven}.
\newblock {From Andreev to Majorana bound states in hybrid
  superconductor-semiconductor nanowires}.
\newblock {\em Nat. Rev. Phys.}, 2(10):575--594, 2020.

\bibitem{2019arXiv190706497B}
C.~W.~J. Beenakker.
\newblock {Search for non-Abelian Majorana braiding statistics in
  superconductors}.
\newblock {\em SciPost Phys. Lect. Notes}, 15, 2020.

\bibitem{stanescu2013majorana}
T.D. Stanescu and S.~Tewari.
\newblock {M}ajorana fermions in semiconductor nanowires: fundamentals,
  modeling, and experiment.
\newblock {\em J. Phys. Condens. Matter}, 25(23):233201, 2013.

\bibitem{Jiang2011a}
L.~Jiang, T.~Kitagawa, J.~Alicea, A.~R. Akhmerov, D.~Pekker, G.~Refael, J.I
  Cirac, E.~Demler, M.D. Lukin, and P.~Zoller.
\newblock {{M}ajorana fermions in equilibrium and in driven cold-atom quantum
  wires}.
\newblock {\em Phys. Rev. Lett.}, 106:220402, 2011.

\bibitem{2018NatRM...3...52L}
R.~M. {Lutchyn}, E.~P.~A.~M. {Bakkers}, L.~P. {Kouwenhoven}, P.~{Krogstrup},
  C.~M. {Marcus}, and Y.~{Oreg}.
\newblock {Majorana zero modes in superconductor-semiconductor
  heterostructures}.
\newblock {\em Nat. Rev. Mater.}, 3(5):52--68, 2018.

\bibitem{livanas2019alternative}
G.~{Livanas}, M.~{Sigrist}, and G.~{Varelogiannis}.
\newblock {Alternative paths to realize {M}ajorana fermions in
  superconductor-ferromagnet heterostructures}.
\newblock {\em Sci. Rep.}, 9:6259, 2019.

\bibitem{sato2016majorana}
M.~Sato and S.~Fujimoto.
\newblock {M}ajorana fermions and topology in superconductors.
\newblock {\em J. Phys. Soc. Jpn}, 85(7):072001, 2016.

\bibitem{Sang2022a}
Lina Sang, Zhi Li, Guangsai Yang, Muhammad Nadeem, Lan Wang, Qikun Xue,
  Alexander~R. Hamilton, and Xiaolin Wang.
\newblock Majorana zero modes in iron-based superconductors.
\newblock {\em Matter}, 5(6):1734--1759, 2022.

\bibitem{1995Sci...269..198A}
M.~H. {Anderson}, J.~R. {Ensher}, M.~R. {Matthews}, C.~E. {Wieman}, and E.~A.
  {Cornell}.
\newblock {Observation of Bose-Einstein Condensation in a Dilute Atomic Vapor}.
\newblock {\em Science}, 269(5221):198--201, 1995.

\bibitem{1995hep.th...11201D}
M.~de~Wild~Propitius and F.A Bais.
\newblock {\em {Discrete gauge theories}}.
\newblock {Springer, Berlin}, 1995.

\bibitem{gould1993quantum}
M.D. Gould.
\newblock Quantum double finite group algebras and their representations.
\newblock {\em Bulletin of the Australian Mathematical Society},
  48(2):275--301, 1993.

\bibitem{kassel1995drinfeld}
C.~Kassel.
\newblock Drinfeld’s quantum double.
\newblock In {\em Quantum Groups}, pages 199--238. Springer, Berlin, 1995.

\bibitem{2006JMP....47j3511D}
K.~A. {Dancer}, P.~S. {Isac}, and J.~{Links}.
\newblock {Representations of the quantum doubles of finite group algebras and
  spectral parameter dependent solutions of the Yang--Baxter equation}.
\newblock {\em J. Math. Phys.}, 47(10):103511--103511, 2006.

\bibitem{Drinfeld1988QuantumG}
V.G. Drinfel'd.
\newblock Quantum groups.
\newblock {\em Journal of Soviet Mathematics}, 41:898--915, 1988.

\bibitem{2007PhRvL..98j0401S}
G.W. {Semenoff} and F.~{Zhou}.
\newblock {Discrete Symmetries and 1/3 Quantum Vortices in Condensates of F=2
  Cold Atoms}.
\newblock {\em Phys Rev. Lett.}, 98(10):100401, 2007.

\bibitem{2010arXiv1001.2072K}
Y.~Kawaguchi and M.~Ueda.
\newblock Spinor {B}ose--{E}instein condensates.
\newblock {\em Phys. Rep.}, 520(5):253--381, 2012.

\bibitem{2012PhRvA..85e1606L}
B.~{Lian}, T.-L. {Ho}, and H.~{Zhai}.
\newblock {Searching for non-Abelian phases in the {B}ose--Einstein condensate
  of dysprosium}.
\newblock {\em Phys. Rev. A}, 85(5):051606, 2012.

\bibitem{makela2003topological}
H.~M{\"a}kel{\"a}, Y.~Zhang, and K.-A. Suominen.
\newblock Topological defects in spinor condensates.
\newblock {\em J. Phys. A Math. Theor.}, 36(32):8555, 2003.

\bibitem{hall2016tying}
D.~S. Hall, M.~W. Ray, K.~Tiurev, E.~Ruokokoski, A.~H. Gheorghe, and
  M.~M{ö}tt{ö}nen.
\newblock Tying quantum knots.
\newblock {\em Nat. Phys.}, 12(5):478--483, 2016.

\bibitem{liu2020interlocked}
Y.-K. Liu, S.-J. Yang, G.-H. Yang, and J.-J. Zhang.
\newblock Interlocked knot in spinor {B}ose-{E}instein condensates.
\newblock {\em Chaos, Solitons \& Fractals}, 140:110209, 2020.

\bibitem{PhysRevLett.100.180403}
Y.~Kawaguchi, M.~Nitta, and M.~Ueda.
\newblock Knots in a spinor {B}ose--{E}instein condensate.
\newblock {\em Phys. Rev. Lett.}, 100:180403, 2008.

\bibitem{1956PhRv..102.1217D}
F.J. {Dyson}.
\newblock {General Theory of Spin-Wave Interactions}.
\newblock {\em Phys. Rev.}, 102(5):1217--1230, 1956.

\bibitem{1940PhRv...58.1098H}
T.~{Holstein} and H.~{Primakoff}.
\newblock {Field Dependence of the Intrinsic Domain Magnetization of a
  Ferromagnet}.
\newblock {\em Phys. Rev.}, 58(12):1098--1113, 1940.

\bibitem{RevModPhys.30.1}
J.~van Krankedonk and J.H. van Vleck.
\newblock Spin waves.
\newblock {\em Rev. Mod. Phys.}, 30:1--23, 1958.

\bibitem{2009PhRvB..80b4420B}
R.~{Barnett}, D.~{Podolsky}, and G.~{Refael}.
\newblock {Geometrical approach to hydrodynamics and low-energy excitations of
  spinor condensates}.
\newblock {\em Phys. Rev. B.}, 80(2):024420, 2009.

\bibitem{mawson2018braiding}
T.~Mawson, T.~Petersen, J.~Slingerland, and T.~Simula.
\newblock Braiding and fusion of non-{A}belian vortex anyons.
\newblock {\em Phys. Rev. Lett.}, 123, 2018.

\bibitem{PhysRevA.64.053602}
N.N. Klausen, J.L. Bohn, and C.H. Greene.
\newblock Nature of spinor {B}ose--{E}instein condensates in {R}ubidium.
\newblock {\em Phys. Rev. A}, 64:053602, 2001.

\bibitem{PhysRevLett.81.5109}
J.~L. Roberts, N.~R. Claussen, James~P. Burke, Chris~H. Greene, E.~A. Cornell,
  and C.~E. Wieman.
\newblock Resonant magnetic field control of elastic scattering in cold
  $^{85}${R}b.
\newblock {\em Phys. Rev. Lett.}, 81:5109--5112, 1998.

\bibitem{2001PhRvA..64b4702R}
J.~L. {Roberts}, James~P. {Burke}, N.~R. {Claussen}, S.~L. {Cornish}, E.~A.
  {Donley}, and C.~E. {Wieman}.
\newblock {Improved characterization of elastic scattering near a {F}eshbach
  resonance in $^{85}$Rb}.
\newblock {\em Phys. Rev. A}, 64(2):024702, 2001.

\bibitem{PhysRevLett.92.040402}
H.~Schmaljohann, M.~Erhard, J.~Kronj\"ager, M.~Kottke, S.~van Staa,
  L.~Cacciapuoti, J.~J. Arlt, K.~Bongs, and K.~Sengstock.
\newblock Dynamics of $f=2$ spinor {B}ose--{E}instein condensates.
\newblock {\em Phys. Rev. Lett.}, 92:040402, 2004.

\bibitem{2006NJPh....8..152W}
A.~{Widera}, F.~{Gerbier}, S.~{F{\"o}lling}, T.~{Gericke}, O.~{Mandel}, and
  I.~{Bloch}.
\newblock {Precision measurement of spin-dependent interaction strengths for
  spin-1 and spin-2 $^{87}$Rb atoms}.
\newblock {\em New J. Phys.}, 8(8):152, 2006.

\bibitem{2006AnPhy.321....2K}
A.~{Kitaev}.
\newblock {Anyons in an exactly solved model and beyond}.
\newblock {\em Ann. of Phys.}, 321(1):2--111, 2006.

\bibitem{1988NuPhB.311...46W}
E.~{Witten}.
\newblock {2 + 1 dimensional gravity as an exactly soluble system}.
\newblock {\em Nucl. Phys. B.}, 311(1):46--78, 1988.

\bibitem{2023arXiv230516016G}
E.~{Génetay Johansen} and T.~{Simula}.
\newblock {Vortex spin in a superfluid}.
\newblock {\em arXiv e-prints}, page arXiv:2305.16016, 2023.

\bibitem{singer1982differential}
I.M. Singer.
\newblock Differential geometry, fiber bundles and physical theories.
\newblock {\em Physics Today}, 35:41--44, 1982.

\bibitem{trautman1980fiber}
A.~Trautman.
\newblock Fiber bundles, gauge fields, and gravitation.
\newblock {\em Gen. Relativ. Gravit.}, 1:287--308, 1980.

\bibitem{witten1988topological}
E.~Witten.
\newblock Topological quantum field theory.
\newblock {\em Commun. in Math. Phys.}, 117(3):353--386, 1988.

\bibitem{atiyah1988topological}
M.F. Atiyah.
\newblock Topological quantum field theory.
\newblock {\em Publications Math{\'e}matiques de l'IH{\'E}S}, 68:175--186,
  1988.

\bibitem{dunne1999aspects}
G.V. Dunne.
\newblock Aspects of {C}hern-{S}imons theory.
\newblock In {\em Aspects topologiques de la physique en basse dimension.
  Topological aspects of low dimensional systems}, pages 177--263. Springer,
  Berlin, 1999.

\bibitem{1987RvMP...59..781Y}
D.~R. {Yennie}.
\newblock {Integral quantum Hall effect for nonspecialists}.
\newblock {\em Rev. of Mod. Phys.}, 59(3):781--824, 1987.

\bibitem{2003PhT....56h..38A}
J.E. {Avron}, D.~{Osadchy}, and R.~{Seiler}.
\newblock {A Topological Look at the Quantum Hall Effect}.
\newblock {\em Physics Today}, 56(8):38--42, 2003.

\bibitem{2017PhRvB..95k5136M}
Roger S.~K. {Mong}, M.P. {Zaletel}, F.~{Pollmann}, and Z.~{Papi{\'c}}.
\newblock {{F}ibonacci anyons and charge density order in the 12/5 and 13/5
  quantum Hall plateaus}.
\newblock {\em Phys. Rev. B}, 95(11):115136, 2017.

\bibitem{2014PhRvL.113w6804V}
A.~{Vaezi} and M.~{Barkeshli}.
\newblock {Fibonacci Anyons From Abelian Bilayer Quantum Hall States}.
\newblock {\em Phys. Rev. Lett.}, 113(23):236804, 2014.

\bibitem{xia2004electron}
J.~S. {Xia}, W.~{Pan}, C.~L. {Vicente}, E.~D. {Adams}, N.~S. {Sullivan}, H.~L.
  {Stormer}, D.~C. {Tsui}, L.~N. {Pfeiffer}, K.~W. {Baldwin}, and K.~W. {West}.
\newblock {Electron correlation in the second Landau level: a competition
  between many nearly degenerate quantum phases}.
\newblock {\em Phys. Rev. B}, 93(17):176809, 2004.

\bibitem{komijani2019kondo}
Y.~{Komijani}.
\newblock {Kondo-based qubits for topological quantum computation}.
\newblock {\em Phys. Rev. B}, 101:235131, 2019.

\bibitem{hu1959homotopy}
S.-T. Hu.
\newblock {\em Homotopy theory}, volume~8.
\newblock Academic press, New York, 1959.

\bibitem{whitehead2012elements}
G.W. Whitehead.
\newblock {\em Elements of homotopy theory}, volume~61.
\newblock Springer Science \& Business Media, Berlin, 2012.

\bibitem{PhysRevD.10.2445}
K.G. Wilson.
\newblock Confinement of quarks.
\newblock {\em Phys. Rev. D}, 10:2445--2459, 1974.

\bibitem{Gurarie2007aa}
V.~{Gurarie} and L.~{Radzihovsky}.
\newblock {Resonantly paired fermionic superfluids}.
\newblock {\em Ann. Phys.}, 322(1):2--119, 2007.

\bibitem{2015PhRvA..92a2301L}
C.~{Levaillant}, B.~{Bauer}, M.~{Freedman}, Z.~{Wang}, and P.~{Bonderson}.
\newblock {Universal gates via fusion and measurement operations on $SU(2)_4$
  anyons}.
\newblock {\em Phys. Rev. A}, 92(1):012301, 2015.

\bibitem{2003PhRvA..67b2315M}
C.~{Mochon}.
\newblock {Anyons from nonsolvable finite groups are sufficient for universal
  quantum computation}.
\newblock {\em Phys. Rev. A}, 67(2):022315, 2003.

\bibitem{bais1980flux}
F.A. Bais.
\newblock Flux metamorphosis.
\newblock {\em Nucl. Phys. B}, 170(1):32--43, 1980.

\bibitem{1992PhLB..280...63A}
F.~{Alexander Bais}, P.~{van Driel}, and M.~{de Wild Propitius}.
\newblock {Quantum symmetries in discrete gauge theories}.
\newblock {\em Phys. Rev. Lett. B}, 280(1-2):63--70, 1992.

\bibitem{1993PhRvD..48.4821L}
H.-K. {Lo} and J.~{Preskill}.
\newblock {Non-Abelian vortices and non-Abelian statistics}.
\newblock {\em Phys. Rev. B.}, 48(10):4821--4834, 1993.

\bibitem{2012PhRvB..86p1107Y}
Y.-Z. {You} and X.-G. {Wen}.
\newblock {Projective non-Abelian statistics of dislocation defects in a
  Z$_{N}$ rotor model}.
\newblock {\em Phys. Rev. B}, 86(16):161107, 2012.

\bibitem{1992PhRvB..45.5737D}
M.C. {Diamantini} and P.~{Sodano}.
\newblock {Anyon representation of the ground-state degeneracy of the quantum
  frustrated XY model}.
\newblock {\em Phys. Rev. B}, 45(10):5737--5739, 1992.

\bibitem{2013JSMTE..10..024B}
R.~{Bondesan} and T.~{Quella}.
\newblock {Topological and symmetry broken phases of Z$_{N}$ parafermions in
  one dimension}.
\newblock {\em J. Stat. Mech.}, 2013(10):10024, 2013.

\bibitem{1988NuPhB.300..360V}
E.~{Verlinde}.
\newblock {Fusion rules and modular transformations in 2D conformal field
  theory}.
\newblock {\em Nucl. Phys. B}, 300:360--376, 1988.

\bibitem{burnside}
Gordon J.
\newblock {\em Representations and Characters of Groups}.
\newblock Cambridge University Press, 2001.

\bibitem{1982PhRvL..49..957W}
F.~{Wilczek}.
\newblock {Quantum Mechanics of Fractional-Spin Particles}.
\newblock {\em Phys. Rev. Lett.}, 49(14):957--959, 1982.

\bibitem{1988CMaPh.116..127F}
J.~{Fröhlich} and P.~{Marchetti}.
\newblock {Bosonization, topological solitons and fractional charges in
  two-dimensional quantum field theory}.
\newblock {\em Commun. Math. Phys.}, 116(1):127--173, 1988.

\bibitem{popov1973quantum}
V.N. Popov.
\newblock Quantum vortices and phase transitions in {Bose} systems.
\newblock {\em Sov. Phys. JETP}, 37(341):2--33, 1973.

\bibitem{2020PhRvA.101f3616S}
T.~{Simula}.
\newblock {Gravitational vortex mass in a superfluid}.
\newblock {\em Phys. Rev. A}, 101(6):063616, 2020.

\bibitem{peskin2018introduction}
M.E. Peskin.
\newblock {\em An introduction to quantum field theory}.
\newblock CRC press, 2018.

\bibitem{2018PhRvA..97b3613E}
M.~{Eto} and M.~{Nitta}.
\newblock {Confinement of half-quantized vortices in coherently coupled
  Bose--Einstein condensates: Simulating quark confinement in a QCD-like
  theory}.
\newblock {\em Phys. Rev. A}, 97(2):023613, 2018.

\bibitem{tylutki2016confinement}
M.~Tylutki, L.P. Pitaevskii, A.~Recati, and S.~Stringari.
\newblock Confinement and precession of vortex pairs in coherently coupled
  bose--einstein condensates.
\newblock {\em Phys. Rev. A}, 93(4):043623, 2016.

\bibitem{cirac2010cold}
J.I. Cirac, P.~Maraner, and J.K. Pachos.
\newblock Cold atom simulation of interacting relativistic quantum field
  theories.
\newblock {\em Phys. Rev. Lett.}, 105(19):190403, 2010.

\bibitem{2003AnPhy.308..692Z}
F.~{Zhou} and M.~{Snoek}.
\newblock {Spin singlet mott states and evidence for spin singlet quantum
  condensates of spin-one bosons in lattices}.
\newblock {\em Ann. Phys.}, 308(2):692--738, 2003.

\bibitem{genetay2023topological}
E.~Génetay~Johansen and T.~Simula.
\newblock Topological quantum computation using analog gravitational holonomy
  and time dilation.
\newblock {\em SciPost Phys. Core}, 6(1):005, 2023.

\bibitem{1999JPhA...32.8539K}
T.~H. {Koornwinder}, B.~J. {Schroers}, J.~K. {Slingerland}, and F.~A. {Bais}.
\newblock {Fourier transform and the Verlinde formula for the quantum double of
  a finite group}.
\newblock {\em J. Phys. A: Math. Gen.}, 32(48):8539--8549, 1999.

\bibitem{2008arXiv0810.3225L}
M.~{Lehn} and C.~{Sorger}.
\newblock {A symplectic resolution for the binary tetrahedral group}.
\newblock {\em arXiv e-prints}, pages 429--435, 2008.

\bibitem{gt}
R.~Steven.
\newblock {\em Fundamentals of Group Theory: An Advanced Approach}.
\newblock Birkhäuser Basel, 2012.

\bibitem{2022QuIP...21...31G}
E.~{G{\'e}netay Johansen} and T.~{Simula}.
\newblock {Prime number factorization using a spinor {B}ose--{E}instein
  condensate-inspired topological quantum computer}.
\newblock {\em QIP}, 21(1):31, 2022.

\bibitem{PhysRevB.75.165310}
L.~Hormozi, G.~Zikos, N.~E. Bonesteel, and S.~H. Simon.
\newblock Topological quantum compiling.
\newblock {\em Phys. Rev. B}, 75:165310, 2007.

\bibitem{2010JPhA...43M5205A}
E.~{Ardonne} and J.~{Slingerland}.
\newblock {Clebsch-Gordan and 6j-coefficients for rank 2 quantum groups}.
\newblock {\em J. Phys. A: Math. Gen.}, 43(39):395205, 2010.

\bibitem{Matthews1999a}
M.~R. Matthews, B.~P. Anderson, P.~C. Haljan, D.~S. Hall, C.~E. Wieman, and
  E.~A. Cornell.
\newblock Vortices in a {B}ose--{E}instein condensate.
\newblock {\em Phys. Rev. Lett.}, 83:2498--2501, 1999.

\bibitem{Zwierlein2005a}
M.~W. {Zwierlein}, J.~R. {Abo-Shaeer}, A.~{Schirotzek}, C.~H. {Schunck}, and
  W.~{Ketterle}.
\newblock {Vortices and superfluidity in a strongly interacting {F}ermi gas}.
\newblock {\em Nat. Phys.}, 435(7045):1047--1051, 2005.

\bibitem{Anderson2010a}
B.~P. Anderson.
\newblock Experiments with vortices in superfluid atomic gases.
\newblock {\em J. Low Temp. Phys.}, 161:574, 2010.

\bibitem{Fetter2009a}
A.~L. {Fetter}.
\newblock {Rotating trapped {B}ose--{E}instein condensates}.
\newblock {\em Rev. Mod. Phys.}, 81(2):647--691, 2009.

\bibitem{Simula2019a}
T.~Simula.
\newblock {\em Quantised Vortices}.
\newblock 2053-2571. Morgan \& Claypool Publishers, 2019.

\bibitem{Kopnin1991a}
N.~B. Kopnin and M.~M. Salomaa.
\newblock Mutual friction in superfluid $^{3}\mathrm{He}$: Effects of bound
  states in the vortex core.
\newblock {\em Phys. Rev. B}, 44:9667--9677, 1991.

\bibitem{Li2022a}
M.~{Li}, G.~{Li}, L.~{Cao}, X.~{Zhou}, X.~{Wang}, C.~{Jin}, C.-K. {Chiu}, S.~J.
  {Pennycook}, Z.~{Wang}, and H.-J. {Gao}.
\newblock {Ordered and tunable {M}ajorana-zero-mode lattice in naturally
  strained LiFeAs}.
\newblock {\em Nat. Phys.}, 606(7916):890--895, 2022.

\bibitem{Masaki2023a}
Y.~{Masaki}, T.~{Mizushima}, and M.~{Nitta}.
\newblock {Non-Abelian Anyons and Non-Abelian Vortices in Topological
  Superconductors}.
\newblock {\em arXiv e-prints}, page arXiv:2301.11614, 2023.

\bibitem{Mizushima2010a}
T.~Mizushima and K.~Machida.
\newblock Splitting and oscillation of {M}ajorana zero modes in the $p$-wave
  {BCS}-{BEC} evolution with plural vortices.
\newblock {\em Phys. Rev. A}, 82:023624, 2010.

\bibitem{Zhang2004}
J.~Zhang, E.~G.~M. van Kempen, T.~Bourdel, L.~Khaykovich, J.~Cubizolles,
  F.~Chevy, M.~Teichmann, L.~Tarruell, S.~J. J. M.~F. Kokkelmans, and
  C.~Salomon.
\newblock $p$-wave {F}eshbach resonances of ultracold $^{6}\mathrm{Li}$.
\newblock {\em Phys. Rev. A}, 70:030702, 2004.

\bibitem{Schunck2005}
C.~H. Schunck, M.~W. Zwierlein, C.~A. Stan, S.~M.~F. Raupach, W.~Ketterle,
  A.~Simoni, E.~Tiesinga, C.~J. Williams, and P.~S. Julienne.
\newblock {F}eshbach resonances in fermionic $^{6}\mathrm{Li}$.
\newblock {\em Phys. Rev. A}, 71:045601, 2005.

\bibitem{Gaebler2007}
J.~P. Gaebler, J.~T. Stewart, J.~L. Bohn, and D.~S. Jin.
\newblock $p$-wave {F}eshbach molecules.
\newblock {\em Phys. Rev. Lett.}, 98:200403, 2007.

\bibitem{Fuchs2008}
J.~Fuchs, C.~Ticknor, P.~Dyke, G.~Veeravalli, E.~Kuhnle, W.~Rowlands,
  P.~Hannaford, and C.~J. Vale.
\newblock Binding energies of $^{6}\text{L}\text{i}$ $p$-wave {F}eshbach
  molecules.
\newblock {\em Phys. Rev. A}, 77:053616, 2008.

\bibitem{Inada2008}
Yasuhisa Inada, Munekazu Horikoshi, Shuta Nakajima, Makoto Kuwata-Gonokami,
  Masahito Ueda, and Takashi Mukaiyama.
\newblock Collisional properties of $p$-wave {F}eshbach molecules.
\newblock {\em Phys. Rev. Lett.}, 101:100401, 2008.

\bibitem{Maier2010}
R.~A.~W. Maier, C.~Marzok, C.~Zimmermann, and Ph.~W. Courteille.
\newblock Radio-frequency spectroscopy of $^{6}\mathrm{Li}$ $p$-wave molecules:
  Towards photoemission spectroscopy of a $p$-wave superfluid.
\newblock {\em Phys. Rev. A}, 81:064701, 2010.

\bibitem{Luciuk2016}
Christopher Luciuk, Stefan Trotzky, Scott Smale, Zhenhua Yu, Shizhong Zhang,
  and Joseph~H. Thywissen.
\newblock Evidence for universal relations describing a gas with p-wave
  interactions.
\newblock {\em Nat. Phys.}, 12(6):599--605, feb 2016.

\bibitem{Gerken2019}
Manuel Gerken, Binh Tran, Stephan H\"afner, Eberhard Tiemann, Bing Zhu, and
  Matthias Weidem\"uller.
\newblock Observation of dipolar splittings in high-resolution atom-loss
  spectroscopy of $^{6}\mathrm{Li}$ $p$-wave {F}eshbach resonances.
\newblock {\em Phys. Rev. A}, 100:050701, 2019.

\bibitem{Waseem2019}
Muhammad Waseem, Jun Yoshida, Taketo Saito, and Takashi Mukaiyama.
\newblock Quantitative analysis of $p$-wave three-body losses via a cascade
  process.
\newblock {\em Phys. Rev. A}, 99:052704, May 2019.

\bibitem{Levinsen2008}
J.~Levinsen, N.~R. Cooper, and V.~Gurarie.
\newblock Stability of fermionic gases close to a $p$-wave {F}eshbach
  resonance.
\newblock {\em Phys. Rev. A}, 78:063616, 2008.

\bibitem{Kurlov2017}
D.~V. Kurlov and G.~V. Shlyapnikov.
\newblock Two-body relaxation of spin-polarized fermions in reduced
  dimensionalities near a $p$-wave {F}eshbach resonance.
\newblock {\em Phys. Rev. A}, 95:032710, 2017.

\bibitem{Muhammed2018}
Muhammad Waseem, Jun Yoshida, Taketo Saito, and Takashi Mukaiyama.
\newblock Unitarity-limited behavior of three-body collisions in a $p$-wave
  interacting {F}ermi gas.
\newblock {\em Phys. Rev. A}, 98:020702, 2018.

\bibitem{Fonta2020}
Francisco Fonta and Kenneth~M. O'Hara.
\newblock Experimental conditions for obtaining halo $p$-wave dimers in
  quasi-one-dimension.
\newblock {\em Phys. Rev. A}, 102:043319, 2020.

\bibitem{Chang2020}
Ya-Ting Chang, Ruwan Senaratne, Danyel Cavazos-Cavazos, and Randall~G. Hulet.
\newblock Collisional loss of one-dimensional fermions near a $p$-wave
  {F}eshbach resonance.
\newblock {\em Phys. Rev. Lett.}, 125:263402, Dec 2020.

\bibitem{Venu2023}
Vijin Venu, Peihang Xu, Mikhail Mamaev, Frank Corapi, Thomas Bilitewski,
  Jose~P. D'Incao, Cora~J. Fujiwara, Ana~Maria Rey, and Joseph~H. Thywissen.
\newblock Unitary p-wave interactions between fermions in an optical lattice.
\newblock {\em Nat. Phys.}, 613(7943):262--267, 2023.

\bibitem{Foster2013}
Matthew~S. Foster, Victor Gurarie, Maxim Dzero, and Emil~A. Yuzbashyan.
\newblock Quench-induced floquet topological $p$-wave superfluids.
\newblock {\em Phys. Rev. Lett.}, 113:076403, Aug 2014.

\bibitem{Kobayashi2009a}
M.~Kobayashi, Y.~Kawaguchi, M.~Nitta, and M.~Ueda.
\newblock Collision dynamics and rung formation of non-abelian vortices.
\newblock {\em Phys. Rev. Lett.}, 103:115301, 2009.

\bibitem{UedaRMP}
D.~M. Stamper-Kurn and M.~Ueda.
\newblock Spinor {B}ose gases: Symmetries, magnetism, and quantum dynamics.
\newblock {\em Rev. Mod. Phys.}, 85:1191--1244, 2013.

\bibitem{2015arXiv151101575M}
G.~E. {Marti} and D.~M. {Stamper-Kurn}.
\newblock {Spinor {B}ose-{E}instein gases}.
\newblock {\em arXiv e-prints}, page arXiv:1511.01575, 2015.

\bibitem{Samson2016a}
E.~C. {Samson}, K.~E. {Wilson}, Z.~L. {Newman}, and B.~P. {Anderson}.
\newblock {Deterministic creation, pinning, and manipulation of quantized
  vortices in a {B}ose--{E}instein condensate}.
\newblock {\em Phys. Rev. A}, 93(2):023603, 2016.

\bibitem{Kwon2021a}
W.~J. {Kwon}, G.~{Del Pace}, K.~{Xhani}, L.~{Galantucci}, A.~{Muzi Falconi},
  M.~{Inguscio}, F.~{Scazza}, and G.~{Roati}.
\newblock {Sound emission and annihilations in a programmable quantum vortex
  collider}.
\newblock {\em Nat. Phys.}, 600(7887):64--69, 2021.

\bibitem{SaffmanRMP}
M.~Saffman, T.~G. Walker, and K.~M\o{}lmer.
\newblock Quantum information with {R}ydberg atoms.
\newblock {\em Rev. Mod. Phys.}, 82:2313--2363, 2010.

\bibitem{Graham2022a}
T.~M. {Graham}, Y.~{Song}, J.~{Scott}, C.~{Poole}, L.~{Phuttitarn}, K.~{Jooya},
  P.~{Eichler}, X.~{Jiang}, A.~{Marra}, B.~{Grinkemeyer}, M.~{Kwon},
  M.~{Ebert}, J.~{Cherek}, M.~T. {Lichtman}, M.~{Gillette}, J.~{Gilbert},
  D.~{Bowman}, T.~{Ballance}, C.~{Campbell}, E.~D. {Dahl}, O.~{Crawford}, N.~S.
  {Blunt}, B.~{Rogers}, T.~{Noel}, and M.~{Saffman}.
\newblock {Multi-qubit entanglement and algorithms on a neutral-atom quantum
  computer}.
\newblock {\em Nat. Phys.}, 604(7906):457--462, 2022.

\bibitem{Bluvstein2022a}
D.~{Bluvstein}, H.~{Levine}, G.~{Semeghini}, T.~T. {Wang}, S.~{Ebadi},
  M.~{Kalinowski}, A.~{Keesling}, N.~{Maskara}, H.~{Pichler}, M.~{Greiner},
  V.~{Vuleti{\'c}}, and M.~D. {Lukin}.
\newblock {A quantum processor based on coherent transport of entangled atom
  arrays}.
\newblock {\em Nat. Phys.}, 604(7906):451--456, 2022.

\bibitem{PhysRevA.104.L061301}
R.~E.~S. Polkinghorne and T.~P. Simula.
\newblock Two {N}ambu-{G}oldstone zero modes for rotating {B}ose--{E}instein
  condensates.
\newblock {\em Phys. Rev. A}, 104:L061301, 2021.

\bibitem{PhysRevLett.101.020402}
T.~P. Simula, T.~Mizushima, and K.~Machida.
\newblock Kelvin waves of quantized vortex lines in trapped {B}ose--{E}instein
  condensates.
\newblock {\em Phys. {R}ev. {Lett.}}, 101:020402.

\bibitem{PhysRevA.69.043617}
A.~L. Fetter.
\newblock Kelvin mode of a vortex in a nonuniform {B}ose--{E}instein
  condensate.
\newblock {\em Phys. {R}ev. {A}}, 69:043617, 2004.

\bibitem{doi:10.1080/14786448008626897}
Sir~W. Thomson.
\newblock {XVI}. {O}n gravitational oscillations of rotating water.
\newblock {\em Lond. Edinb. Dublin Philos. Mag. J. Sci.}, 10(60):109--116,
  1880.

\bibitem{PhysRevLett.90.100403}
V.~Bretin, P.~Rosenbusch, F.~Chevy, G.~V. Shlyapnikov, and J.~Dalibard.
\newblock Quadrupole oscillation of a single-vortex bose-einstein condensate:
  Evidence for kelvin modes.
\newblock {\em Phys. Rev. Lett.}, 90:100403, 2003.

\bibitem{AT}
A.~Hatcher.
\newblock {\em Algebraic Topology}.
\newblock Cambridge University Press, 2002.

\bibitem{alma9936959566101361}
M.~Burrow.
\newblock {\em Representation theory of finite groups}.
\newblock Ebook Central Library. Academic Press, New York, 1965.

\bibitem{hopf}
D.E. Radford.
\newblock {\em Hopf algebras}, volume~49.
\newblock World Scientific, Singapore, 2012.

\bibitem{1989IJMPA...4.3759J}
M.~{Jimbo}.
\newblock {Introduction to the {Y}ang--{B}axter equation}.
\newblock {\em Int. J. Mod. Phys. A}, 4(15):3759--3777, 1989.

\bibitem{preskill1999lecture}
John Preskill.
\newblock Lecture notes for physics 219: Quantum computation.
\newblock {\em Caltech Lecture Notes}, 1999.

\bibitem{2006CoPhC.174..903R}
K.~{Rykhlinskaya} and S.~{Fritzsche}.
\newblock {Generation of Clebsch Gordan coefficients for the point and double
  groups}.
\newblock {\em Comput. Phys. Commun.}, 174(11):903--913, 2006.

\bibitem{terras1999fourier}
A.~Terras.
\newblock {\em Fourier analysis on finite groups and applications}.
\newblock Number~43. Cambridge University Press, 1999.

\end{thebibliography}
\end{document}